\documentclass[12pt]{article}
\usepackage{xr}
\usepackage{fullpage}
\usepackage{amsmath}
\usepackage{amsfonts}
\usepackage{amssymb}
\usepackage{graphicx}
\usepackage{subfigure}
\usepackage{psfrag}
\usepackage{rotating}
\usepackage{lscape}
\usepackage{geometry}
\usepackage{color}
\usepackage{float}
\usepackage{bm,enumerate, graphics, multirow, booktabs, url,cite,ctable,subfigure,epstopdf}
\usepackage{natbib}
\usepackage[colorlinks, citecolor=blue]{hyperref}
\usepackage{mathrsfs}
\usepackage{verbatim}
\usepackage{multirow}
\usepackage{threeparttable}
\usepackage[section]{placeins}
\usepackage{array}

\graphicspath{{figures/}}

\numberwithin{equation}{section}
\numberwithin{figure}{section}
\allowdisplaybreaks[3]

\newtheorem{theorem}{Theorem}[section]

\newtheorem{lemma}{Lemma}[section]

\newtheorem{assumption}{Assumption}
\newtheorem{example}{Example}[section]
\newtheorem{remark}{Remark}[section]

\def\calplongarrow{\stackrel{{\mathcal{P}}}{\longrightarrow}}
\def\darrow{\stackrel{\mathcal{D}}{\longrightarrow}}

\renewcommand{\baselinestretch}{1}

\begin{document}

\title{Factors in Fashion: Factor Analysis towards the Mode}
\author{Zhe Sun and Yundong Tu\thanks{Corresponding author. Address: Guanghua
 		School of Management, Peking University,
 		Beijing, 100871, China. E-mail: yundong.tu@gsm.pku.edu.cn.}\\
 	Guanghua School of Management\\
 	Peking University\\
 }
 	\maketitle
	\begin{abstract}

The modal factor model represents a new factor model for dimension reduction in high dimensional panel data. Unlike the approximate factor model that targets for the mean factors, it captures factors that influence the conditional mode of the distribution of the observables. Statistical inference is developed with the aid of mode estimation, where the modal factors and the loadings are estimated through maximizing a kernel-type objective function. An easy-to-implement alternating maximization algorithm is designed to obtain the estimators numerically. Two model selection criteria are further proposed to determine the number of factors. The asymptotic properties of the proposed estimators are established under some regularity conditions. Simulations demonstrate the nice finite sample performance of our proposed estimators, even in the presence of heavy-tailed and asymmetric idiosyncratic error distributions. Finally, the application to inflation forecasting illustrates the practical merits of modal factors.
	

	\end{abstract}

	\pagenumbering{Alph}
	\thispagestyle{empty}
	\newpage
	
	\pagenumbering{arabic}
	\setcounter{page}{1}
	\renewcommand{\baselinestretch}{1.25} \normalsize

\section{Introduction}

Factor model has become one of the most important tools in analyzing high dimensional time series, due to its capability of dimension reduction and feature extraction through a small number of common factors, especially in the era of big data. Theoretical advancements in factor analysis have been made using principal component analysis \citep[PCA]{BN2002,Bai2003,FHLR2000,FHLR2005,AH2013,HKYZ2022} and maximum likelihood approach \citep{BL2016,Wang2022}, and so on. In the meanwhile, high dimensional factor models have found practical applications in a wide range of financial and economic studies, such as modeling monetary policy \citep{BB2003}, break and threshold detection \citep{MT2023a,MT2023b}, group structure identification \citep{AB2017,AGP2020,ZPYZ2023}, forecasting excess stock returns \citep{LN2007}, bond returns \citep{LN2009} and macroeconomic time series \citep{SW2002a,SW2002b,BN2006,CH2015,TL2019,GKP2016},
and among many others. 
See \citet{FLL2022} for a recent review and references in the above studies for more related literature.

The majority of the above theoretical contributions to factor analysis have been confined to extracting common features that explain the (conditional) mean of the observed high dimensional time series, the factors obtained from which may be referred to as mean factors. While there is no dispute that mean is one of the most commonly used location parameters for a random variable, other location measures, such mode, median, quantiles, and so on,  are also frequently seen in empirical studies as they could contain alternative unique distributional information as well. To enable factor analysis across the whole (conditional) distribution, \citet{CDG2021} recently put forward quantile factors that are derived under a quantile factor model. The quantile factors are allowed to vary across the quantile level, and can completely characterize features that shift any part of the conditional distribution. The empirical evidence provided by \citet{CDG2021} demonstrates that quantile factors are very informative for density forecasting of the inflation rate and real GDP growth.

This paper advocates a modal factor model (MFM) in order to capture common features that explain the mode of the (conditional) densities of the observed high dimensional time series. This leads to modal factors that are the most likely to appear in the (conditional) densities, which are thus referred to as ``fashionable factors''. It is worth noting that the modal factors are in nature quantile factors corresponding to the quantile level at which the densities of the observed time series reach their peak,  just like that mode is a specific quantile at which the density achieves its maximum. In this sense, the modal factor model is nested in the quantile factor model as a special case. However, the latter does not automatically produce the former, because the quantile factors are defined only for a given quantile level $\tau$, whose value at the mode is unknown unless in certain (impractical) scenarios such as that the density is symmetric ($\tau=0.5$) or is fully known. Consequently, the quantile factor model could fail to reveal how the conditional mode of the high dimensional time series depends on the modal factors directly to detect the ``most likely'' effect and may produce low density point predictions, as similarly argued by \citet{UWY2023} in the regression setting. As a result, the modal factor model is potentially a very useful tool that can be of interest in itself, or used to complement the PCA and quantile factors in dimension reduction for high dimensional data.

This paper contributes to the literature in several aspects. First, a modal factor analysis (MFA) procedure, called ``alternating modal expectation-maximization'', is proposed to provide a basis on which statistical inference on modal factor model can be conducted.
The loss function we use to derive the modal factor and loading estimators involves a kernel function with vanishing bandwidth, which adapts that designed to obtain modal regression estimators \citep{YL2014,KS2012}.
The estimation procedure marries the alternating maximization algorithm used in factor estimation \citep{CDG2021} and the modal expectation-maximization algorithm adopted in estimating modal regressions \citep{YL2014}. The resulting algorithm is computationally efficient and easy to implement with the choice of a normal kernel function, which largely alleviates the practical challenge that there is no analytical closed-form solution for the MFA estimators.

Second, asymptotic properties of the proposed MFA estimators are established. We derive the average convergence rate of the MFA estimators, establish their asymptotic normality, and obtain consistent estimators for the associated asymptotic variances. The results are obtained under the condition that, given the factors, the errors are independent cross-sectionally but follow an $\alpha$-mixing process time serially, without any restriction imposed on the existence of the error moments. The time serial dependence allowed largely relaxes the independence requirement made by \citet{CDG2021}. We show that the MFA estimators converge at the fastest possible rate $L_{NT}^{2/7}$, where $L_{NT}=\min\{N,T\}$, with $N$ and $T$ being the cross-sectional dimension and the time length, respectively. This rate is slower than the convergence rate $L_{NT}^{1/2}$ for PCA factor estimators \citep{Bai2003} and quantile factor estimators \citep{CDG2021}. The slower convergence stems from the nature of nonparametric inference due to the use of a vanishing bandwidth, and is the cost we have to pay for estimating the conditional mode without the knowledge of the density functional form \citep{Parzen1962}. The optimal order of bandwidth choice is also discussed.

Third, two data-driven model selection methods, based on the rank of a certain matrix and the information criterion, respectively, are proposed to determine the number of modal factors. We characterize the conditions for the tuning parameters under which the selection for the factor number can achieve consistency. Although these selection criteria bear similarity to those used by \citet{CDG2021},
there are important distinctions in the theoretical development in current modal factor models, which are outlined in the remarks. Examples of tuning parameter choices that meet the consistency requirement are also provided.

Fourth, numerical evidences are provided to demonstrate the nice finite sample performance of the proposed estimators. In particular, simulated examples show that, in spite of the reduced convergence rate, the MFA estimators can effectively capture the true factor space in a variety of parameter settings, and tend to outperform the PCA and quantile factor estimators (at $\tau=0.5$), especially when the errors are heavy-tailed. Moreover, the two factor number selection criteria can select the correct number of factors with high probability. The above simulation findings are robust to the presence of heavy-tailed or skewed errors. Finally, empirical applications illustrate that MFA factors contain valuable information in enhancing the predictive accuracy of U.S. inflation rate.

This paper is also related to the growing literature on modal regressions. \citet{KS2012} and \citet{YL2014} consider the linear modal regression through maximizing a kernel-based objective function with a vanishing bandwidth, and largely extend the pioneer work of \citet{Lee1989, Lee1993} by developing asymptotic results under skewed error distributions. For nonparametric modal regressions, \citet{YLL2012} estimate the global conditional mode by local polynomial smoothing, while \citet{CGTW2016} estimate the collection of all conditional modes based on a kernel density estimate. Recently, \citet{UWY2021} study the fixed effects modal regression for panel data, \citet{UWY2023} consider a semiparametric partially linear varying coefficient modal regression, and \citet{Wang2024} investigates the nonlinear modal regression for dependent data.
It is worth emphasizing that the above studies only involve observed regressors, while both the factors and loadings are unknown and need to be estimated in the current setup. There has been no study that considers the inference on the conditional mode in factor models so far. We note that \citet{SH2020} propose a modal PCA that uses the probability density value of the mode as a measure of concentration, the direction maximizes which is regarded as the minor component (factor) direction.
In that way, their formulation is notably different from ours, and they do not consider the asymptotic properties of the estimators. These differences highlight the new contribution of current paper.

The outline of the rest of this paper is as follows. Section \ref{Sec2} introduces MFM, provides a list of illustrative examples of MFM,  presents the MFA estimators and the computational algorithm, and proposes two methods for selecting the number of factors. Section \ref{Sec3} establishes the asymptotic properties of the proposed estimators. Section \ref{Sec4} evaluates the finite sample performance of the estimators using Monte Carlo simulations. Section \ref{Sec5} assesses the predictive power of the MFA factors in forecasting U.S. inflation rate. The proofs of Theorems \ref{Thm1}, \ref{Thm2} and \ref{Thm3} are contained in the Appendix,
while the proofs of Theorems \ref{Thm4} and \ref{Thm5}, together with some additional simulation results, are relegated to the Supplementary Material.


\emph{Notations}. For any real number $a$, $\operatorname{sgn}(a)=1$ if $a\geq 0$ and $-1$ if $a< 0$. For any matrix $\mathbf A$, let $\operatorname{rank}(\mathbf A)$,
$\operatorname{tr}(\mathbf A)$, $\mathbf A^{\prime}$, $\|A\|=[\operatorname{tr}(A^{\prime}A)]^{1/2}$ and $\operatorname{vech}(\mathbf A)$ denote its rank,
trace, transpose, Frobenius norm and the vectorization of $A$, respectively. For any square matrix $\mathbf B$ with real eigenvalues, denote $\rho_{\mathrm{min}}(\mathbf B)$ (resp. $\rho_{\mathrm{max}}(\mathbf B)$) as its minimum (resp. maximum) eigenvalue, $\mathbf B_{jj}$ as its $j$-th diagonal element, and $\operatorname{sgn}(\mathbf B)$ as a diagonal matrix whose $j$-th diagonal element equals $\operatorname{sgn}({\mathbf{B}}_{jj})$. We use $\mathbf B>0$ (resp. $\mathbf B<0$) to signify that $\mathbf B$ is positive (resp. negative) definite.

\section{The model and estimation}\label{Sec2}
Section \ref{Sec2-1} presents the modal factor models. Section \ref{Sec2-2} defines the estimators for the
factors and the loadings, and introduces an algorithm to obtain the estimators. Section \ref{Sec2-3} proposes two criteria for selecting the number of factors.
\subsection{The modal factor model}\label{Sec2-1}
Suppose that the observed variables $\{X_{it};i=1,\ldots,N,t=1,\ldots,T\}$ satisfy the following modal factor model (MFM):
\begin{equation}\label{Sec21-1}
    \operatorname{Mode}(X_{it}|{\bf{f}}_{0t})=\bm{{\bm{\lambda}}}_{0i}^{\prime}{\bf{f}}_{0t}, \quad \text{for}\;\;i=1,\cdots,N; t=1,\cdots, T,
\end{equation}
where ${\bf{f}}_{0t}$ is an $r\times1$ vector of random common factors, $\bm{{\bm{\lambda}}}_{0i}$ is an $r\times1$ vector of non-random factor loadings, with the conditional mode function of $X_{it}$ given ${\bf{f}}_{0t}$ denoted as
\begin{equation*}
    \operatorname{Mode}(X_{it}|{\bf{f}}_{0t})=\arg \underset{x}{\mathrm{max}} \,\,
    f_{X_{it}}(x|{\bf{f}}_{0t}),
\end{equation*}
and $f_{X_{it}}(\cdot|{\bf{f}}_{0t})$ being the conditional distribution of $X_{it}$ given ${\bf{f}}_{0t}$.
Let $e_{it}^0=X_{it}-\bm{{\bm{\lambda}}}_{0i}^{\prime} \mathbf{f}_{0t}$ denote the idiosyncratic error, then \eqref{Sec21-1} can be equivalently represented as
\begin{equation*}
\begin{split}
    &X_{it}={\bm{{\bm{\lambda}}}}_{0i}^{\prime}{\bf{f}}_{0t}+e_{it}^0,\\
    &\operatorname{Mode}(e_{it}^0|{\bf{f}}_{0t})=\arg\underset{e}{\mathrm{max}}\,\, g_{e_{it}^0}(e|{\bf{f}}_{0t})=0, \quad \text{for}\;\;
    i=1\cdots,N; t=1,\cdots, T, \\
\end{split}
\end{equation*}
where $g_{e_{it}^0}(\cdot|{\bf{f}}_{0t})$ is the conditional density of $e_{it}^0$ given ${\bf{f}}_{0t}$.

If the $g_{e_{it}^0}(\cdot|{\bf{f}}_{0t})$ is symmetric about 0 across both $i$ and $t$, then $\operatorname{Mode}(X_{it}|{\bf{f}}_{0t})=E(X_{it}|{\bf{f}}_{0t})$. In this case, the factors and loadings from the above model are the same as those of the approximate factor models (AFMs), and thus can be estimated by principal component analysis (PCA) as studied by \citet{BN2002} and \citet{Bai2003}, provided that certain moment conditions hold for the error terms. However, if $g_{e_{it}^0}(\cdot|{\bf{f}}_{0t})$ is skewed for some $i$ and $t$, then these components from the two models become different. The following example provides more illustrations on the relationship among our MFM, the AFM \citep{BN2002} and the quantile factor model \citep[QFM]{CDG2021}.

\begin{example}\label{exm0}
Let $\{e_{it}\}$ be independent and identically distributed (i.i.d.) errors with density function $g_e(\cdot)$ and cumulative distribution function $G_e(\cdot)$. Let $Q_e(\tau)=G_e^{-1}(\tau)=\inf\{c: G_e(c)\geq \tau)$ be the quantile function of $e_{it}$. Additionally, let ${\bf{f}}_{1t}\in \mathbb{R}^{r_1}, {\bf{f}}_{1t}\in \mathbb{R}^{r_2}$, where $r_1$ and $r_2$ are positive constants,
and $\{e_{it}\}$ is independent of $\{{\bf{f}}_{1t}\}$ and $\{{\bf{f}}_{2t}\}$. Without loss of generality, suppose that $E(e_{it})=0$, $Q_{e}(\tau_0)=0$ for some $\tau_0\in [0,1]$, and $\operatorname{Mode}(e_{it})=e_m$, which corresponds to the $\tau_m$-th quantile, i.e., $Q_{e}(\tau_m)=e_m$ for some $\tau_m\in [0,1]$.

Consider the following location-scale-shift factor model:
$$X_{it}={\bm{\lambda}}_{i}^{\prime}{\bf{f}}_{1t}+{\bm{\alpha}}_i^{\prime}{\bf{f}}_{2t}e_{it},$$
where ${\bm{\lambda}}_i\in\mathbb{R}^{r_1}, {\bm{\alpha}}_i\in\mathbb{R}^{r_2}, {\bm{\alpha}}^{\prime}_{i}{\bf{f}}_{2t}>0$. Let ${\bf{f}}_t=[{\bf{f}}^{\prime}_{1t},{\bf{f}}^{\prime}_{2t}]^{\prime}.$
\begin{enumerate}
\item[(a)] If ${\bf{f}}_{2t}$ and ${\bf{f}}_{1t}$ do not share any common element, then
\begin{equation*}
    E(X_{it}|{\bf{f}}_{t})={\bm{\lambda}}_{i}^{\prime}{\bf{f}}_{1t},\qquad \operatorname{Mode}(X_{it}|{\bf{f}}_{t})={{\bm{\lambda}}_i^M}^{\prime}{\bf{f}}_{t},\qquad Q_{X_{it}}(\tau|{\bf{f}}_{t})={{\bm{\lambda}}_i(\tau)}^{\prime}{\bf{f}}_{t},
\end{equation*}
where ${\bm{\lambda}}^M_i=[{\bm{\lambda}}_i^{\prime}, e_m{\bm{\alpha}}_{i}^{\prime}]^{\prime}$, ${\bm{\lambda}}_i(\tau)=[{\bm{\lambda}}_{i}^{\prime},Q_e(\tau){\bm{\alpha}}_i^{\prime}]^{\prime}$.
\item[(b)]
If ${\bf{f}}_{2t}={\bf{f}}_{1t}$, then
\begin{equation*}
    E(X_{it}|{\bf{f}}_{t})={\bm{\lambda}}_{i}^{\prime}{\bf{f}}_{1t},\qquad \operatorname{Mode}(X_{it}|{\bf{f}}_{t})={{\bm{\lambda}}_i^M}^{\prime}{\bf{f}}_{1t},\qquad Q_{X_{it}}(\tau|{\bf{f}}_{t})={{\bm{\lambda}}_i(\tau)}^{\prime}{\bf{f}}_{1t},
\end{equation*}
where ${\bm{\lambda}}^M_i={\bm{\lambda}}_i+e_m{\bm{\alpha}}_i, {\bm{\lambda}}_i(\tau)={\bm{\lambda}}_i+Q_e(\tau){\bm{\alpha}}_i$.
\item[(c)]
If $e_m=0$ and $\tau=\tau_0$, then
\begin{equation*}\label{exm-eq1}
    E(X_{it}|{\bf{f}}_{t})= \operatorname{Mode}(X_{it}|{\bf{f}}_{t})= Q_{X_{it}}(\tau|{\bf{f}}_{t})={\bm{\lambda}}_{i}^{\prime}{\bf{f}}_{1t}.
\end{equation*}
\end{enumerate}
\end{example}
The above example illustrates that only in case (c) where $e_m=0$ and $\tau=\tau_0$, AFM, MFM and QFM have the same representation, while in general the three models are different. In particular, the above example demonstrates that different characteristics of $X_{it}$ can be driven by different common factors. For instance, in case (a), the conditional mean of $X_{it}$ is only affected by ${\bf{f}}_{1t}$, while the conditional mode and quantiles of $X_{it}$ are affected by both ${\bf{f}}_{1t}$ and ${\bf{f}}_{2t}$ if the location and scale factors are different. From this perspective, MFM can capture (scale) factors that are missed by AFM.

\subsection{Estimating the factors and loadings}\label{Sec2-2}

For the observed sample $\{X_{it}\}$, we take the fixed-effect approach that treats $\{\mathbf{f}_{0t}\}$ and $\{\bm{\lambda}_{0i}\}$ as unknown parameters to be estimated, while the asymptotic analysis is conditional on $\{\mathbf{f}_{0t}\}$. We first assume that the factor number $r_0$ is known in this subsection. The data-driven selection of $r_0$ will be presented in Section \ref{Sec2-3}.

Let  $M=(N+T) r_0, \bm{{\bm{\theta}}}=\left({\bm{{\bm{\lambda}}}}_{1}^{\prime}, \ldots, {\bm{{\bm{\lambda}}}}_{N}^{\prime}, {\bf{f}}_{1}^{\prime}, \ldots, {\bf{f}}_{T}^{\prime}\right)^{\prime}$. Denote  $\bm{{\bm{\theta}}}_{0}=\left(\bm{{\bm{\lambda}}}_{01}^{\prime}, \ldots, \bm{{\bm{\lambda}}}_{0 N}^{\prime}, \mathbf{f}_{01}^{\prime}, \ldots, \mathbf{f}_{0 T}^{\prime}\right)^{\prime} $ as the vector of true parameters, and write ${\bm{{\bm{\Lambda}}}}_0=(\bm{{\bm{\lambda}}}_{01},\cdots,{\bm{{\bm{\lambda}}}}_{0N})^{\prime}$, ${\bf{F}}_0=({\bf{f}}_{01},\cdots,{\bf{f}}_{0T})^{\prime}.$ The dependence of  $\bm{{\bm{\theta}}} $ and  $\bm{{\bm{\theta}}}_0 $ on  $M$  is suppressed for notational convenience.

A well-known fact in factor models \citep{BN2002} is that $\left\{\bm{{\bm{\lambda}}}_{0i}\right\} $ and $ \left\{{\bf{f}}_{0t}\right\} $ cannot be separately identified without imposing certain normalizations. Without loss of generality, we adopt the following normalizations:
\begin{align}\label{31}
		&\frac{1}{T}\sum\limits_{t=1}^{T} {\bf{f}}_{t}{\bf{f}}_{t}^{\prime}=\mathbb{I}_{r_0}, \nonumber\\ 
		&\frac{1}{N} \sum\limits_{i=1}^{N} \bm{{\bm{\lambda}}}_{i} \bm{{\bm{\lambda}}}_{i}^{\prime}, \quad \text { is diagonal with non-increasing diagonal elements. }
\end{align}
Let  $\mathcal{A}, \mathcal{F} \subset \mathbb{R}^{r_0}$  and define the parameter space as
\begin{equation*}
	{\bm{\Theta}}^{r_0}=\left\{\bm{{\bm{\theta}}} \in \mathbb{R}^{M}: \bm{{\bm{\lambda}}}_i\in \mathcal{A}, \mathbf{f}_{t} \in \mathcal{F} \text{ for all }   i,t,\left\{\bm{{\bm{\lambda}}}_{i}\right\} \text{and}  \left\{\mathbf{f}_{t}\right\}  \text{satisfy the normalizations in }  (\ref{31}) \right\}.
\end{equation*}

To estimate the modal factor model parameter $\bm{{\bm{\theta}}}_0$, we follow the strategy of \citet{KS2012} and \citet{YL2014} to use a kernel-based objective function in the regression setting. Specifically,
define
\begin{equation}\label{61}
	\mathbb{M}_{NT}(\bm{{\bm{\theta}}})=\frac{1}{NTh}\sum\limits_{i=1}^{N}\sum\limits_{t=1}^{T}K\left(\frac{ X_{it}-\bm{{\bm{\lambda}}}_{i}^{\prime}\mathbf{f}_t}{h}\right)=\frac{1}{NT}\sum\limits_{i=1}^{N}\sum\limits_{t=1}^{T}K_h\left(X_{it}-\bm{{\bm{\lambda}}}_{i}^{\prime}{\mathbf{f}}_t\right),
\end{equation}
where $K_h(u)=\frac{1}{h}K(\frac{u}{h})$, $K(\cdot)$ is a smooth kernel function, and $h$ is a bandwidth diminishing towards $0$ as $N, T\to \infty$.
The estimator of $\bm{{\bm{\theta}}}_0$ is then defined as
\begin{equation}\label{51}
\bm{\hat{{\bm{\theta}}}}=\left(\bm{\hat{{\bm{\lambda}}}}_{1}^{\prime}, \ldots, \bm{\hat{{\bm{\lambda}}}}_{N}^{\prime}, \mathbf{\hat{f}}_{1}^{\prime}, \ldots,\mathbf{\hat{f}}_{T}^{\prime}\right)^{\prime}=\underset{\bm{{\bm{\theta}}} \in {\bm{\Theta}}^{r_0}}{\arg \max } \,\, \mathbb{M}_{N T}(\bm{{\bm{\theta}}}) .
\end{equation}
The reason why we estimate ${\bm{{\bm{\theta}}}}_0$ in this way is related to the fact that, for any fixed $\bm{{\bm{\theta}}}$, $\mathbb{M}_{NT}(\bm{{\bm{\theta}}})$ can be seen as a kernel density estimator for the residuals $e_{it}=X_{it}-{\bm{{\bm{\lambda}}}}_i^{\prime}{\bf{f}}_t$ at $0$ (i.e. $g_{e_{it}}(0)$).
Note that $g_{e_{it}}(0)=E\big[g_{e_{it}}(0|{\bf{f}}_{0t})\big]=E\big[f_{X_{it}}({\bm{{\bm{\lambda}}}}_i^{\prime}{\bf{f}}_t|{\bf{f}}_{0t})\big]\leq E\big[f_{X_{it}}({\bm{{\bm{\lambda}}}}_{0i}^{\prime}{\bf{f}}_{0t}|{\bf{f}}_{0t})\big]=E\big[g_{e_{it}^0}(0|{\bf{f}}_{0t})\big]$,  provided that $f_{X_{it}}({\bm{{\bm{\lambda}}}}_i^{\prime}{\bf{f}}_t|{\bf{f}}_{0t})\leq f_{X_{it}}({\bm{{\bm{\lambda}}}}_{0i}^{\prime}{\bf{f}}_{0t}|{\bf{f}}_{0t})$ for all $\bm{\theta}\in{\bm{\Theta}}^{r_0}$, with a strict inequality when ${\bm{{\bm{\lambda}}}}_i^{\prime}{\bf{f}}_t\neq {\bm{{\bm{\lambda}}}}_{0i}^{\prime}{\bf{f}}_{0t}$. This suggests that $\bm{{\bm{\theta}}}_0$ can be well estimated by the vector at which the kernel density function $\mathbb{M}_{NT}(\bm{{\bm{\theta}}})$ reaches the peak. See \citet{KS2012} and \citet{YL2014} for illustrations of such criterion function choices in the regression setting.

As typical in modal regressions, the maximization in \eqref{51} does not yield an analytical closed form for $\hat{\bm{{\bm{\theta}}}}$. For practical implementation, we propose an algorithm, which we call alternating modal expectation-maximization (AMEM), to obtain the numerical estimate for an observed sample $\{X_{it}\}$. As both the factor and loading vectors are unknown and need to be estimated, the algorithm estimates one by maximizing the objective function in \eqref{51} given the other and then alternates, until some stopping rule is satisfied. Each maximization involved in the iterated process is essentially a linear modal regression estimation, which can be conveniently solved by the modal expectation-maximization algorithm proposed by \citet{YL2014}.

To precisely describe the algorithm, let $\bm{{\bm{\Lambda}}}=({\bm{{\bm{\lambda}}}}_1,\cdots, {\bm{{\bm{\lambda}}}}_N)^{\prime}, {\bf{F}}=({\bf{f}}_1,\cdots, {\bf{f}}_T)^{\prime}$, and write
\begin{equation*}	\mathbb{M}_{i,T}(\bm{{\bm{\lambda}}},\mathbf{F})=\frac{1}{T}\sum\limits_{t=1}^{T}K_h(X_{it}-{\bf{f}}_t^{\prime}{\bm{{\bm{\lambda}}}}) \quad \text{and} \quad \mathbb{M}_{t,N}(\bm{{\bm{\Lambda}}},{\bf{f}})=\frac{1}{N}\sum\limits_{i=1}^{N}K_h(X_{it}-{\bm{{\bm{\lambda}}}}_i^{\prime}\bf{f}).
\end{equation*}
Then $\mathbb{M}_{NT}(\bm{{\bm{\theta}}})=N^{-1}\sum_{i=1}^N\mathbb{M}_{i,T}(\bm{{\bm{\lambda}}}_i,\mathbf{F})=T^{-1}\sum_{t=1}^T\mathbb{M}_{t,N}(\bm{{\bm{\Lambda}}},{\bf{f}}_t)$.

\bigskip

\noindent\textbf{The AMEM algorithm}:

\begin{enumerate}[{\bf Step 1}]
\item Choose random starting parameters: $\mathbf{F}^{(0)}, \bm{\Lambda}^{(0)}$.
\item For $l\geq1$, with given $\mathbf{F}^{(l-1)}$, solve $\bm{{\bm{\lambda}}}_i^{(l)}=\arg\underset{\bm{{\bm{\lambda}}}}{\max}\,\,\mathbb{M}_{i,T}(\bm{{\bm{\lambda}}},\mathbf{F}^{(l-1)})$ for $i=1,\cdots,N$. With initial value $\bm{{\bm{\lambda}}}_i(0)={\bm{\lambda}}_i^{(l-1)}$, the solution $\bm{{\bm{\lambda}}}_i^{(l)}$ can be found by repeating
the following two steps until convergence:

\hspace{0.5cm}\emph{E-Step}: Calculate weights $\pi(t|\bm{{\bm{\lambda}}}_i(k),\mathbf{F}^{(l-1)}),$ for $t=1,\cdots,T$ as
\begin{equation}\label{AMEM-eq1}
\pi(t|\bm{{\bm{\lambda}}}_i (k),\mathbf{F}^{(l-1)})=\frac{K_h(X_{it}-\bm{{\bm{\lambda}}}_i (k)^{\prime} \mathbf{f}_t^{(l-1)})}{\sum_{t=1}^{T}K_h(X_{it}-\bm{{\bm{\lambda}}}_i (k)^{\prime} \mathbf{f}_t^{(l-1)})}\propto K_h(X_{it}-\bm{{\bm{\lambda}}}_i (k)^{\prime} \mathbf{f}_t^{(l-1)}).
\end{equation}
\hspace{0.5cm}\emph{M-Step}: Update $\bm{{\bm{\lambda}}}_i (k+1)$ as
\begin{align}\label{AMEM-eq2}
\bm{{\bm{\lambda}}}_i (k+1)&=\arg\underset{\bm{{\bm{\lambda}}}}{\max}\,\,\sum\limits_{t=1}^{T}\big\{\pi(t|\bm{{\bm{\lambda}}}_i (k),\mathbf{F}^{(l-1)})\log K_h(X_{it}-\bm{{\bm{\lambda}}}^{\prime} \mathbf{f}_t^{(l-1)})\big\}\nonumber\\
&=(\mathbf{F}^{(l-1)^{T}}\mathbf{W}_k\mathbf{F}^{(l-1)})^{-1}\mathbf{F}^{(l-1)^{T}}\mathbf{W}_k\mathbf{X}_i,
\end{align}
where $\mathbf{X}_i=(X_{i1},\cdots,X_{iT})^{\prime}$ and $\mathbf{W}_k$ is a $T\times T$ diagonal matrix with $t$-th diagonal
element $\pi(t|\bm{{\bm{\lambda}}}_i (k),\mathbf{F}^{(l-1)})$.
\item Given $\bm{{\bm{\Lambda}}}^{(l)}$, solve $\mathbf{f}_t^{(l)}=\arg\underset{\mathbf{f}}{\max}\,\,\mathbb{M}_{t,N}(\bm{{\bm{\Lambda}}}^{(l)}, \mathbf{f})$ for $t=1,\cdots,T$ following a similar procedure as outlined in Step 2, with initial value $\bm{{\bf{f}}}_t(0)=\bm{{\bf{f}}}_t^{(l-1)}$.
\item For $l=1,2,\cdots$, iterate Steps 2-3 until $\mathbb{M}_{NT}(\bm{{\bm{\theta}}}^{(L)})$ is close to
$\mathbb{M}_{NT}(\bm{{\bm{\theta}}}^{(L-1)})$ for some $L$, i.e., $|\mathbb{M}_{NT}(\bm{{\bm{\theta}}}^{(L)})-\mathbb{M}_{NT}(\bm{{\bm{\theta}}}^{(L-1)})|<\epsilon$, for some small positive $\epsilon$, where $\bm{{\bm{\theta}}}^{(l)}=\operatorname{vech}((\bm{{\bm{\Lambda}}}^{(l)})^{\prime},(\mathbf{F}^{(l)})^{\prime})$.
\item Normalize $\bm{{\bm{\Lambda}}}^{(L)}$ and $\mathbf{F}^{(L)}$ to  satisfy the normalizations in \eqref{31}.
\end{enumerate}


Note that the closed-form solution as shown in \eqref{AMEM-eq2} above during the \emph{M-Step} is only obtained for the standard normal kernel choice $K(u)=\phi(u)$, which largely enhances the computational efficiency. The asymptotic results obtained in this article remain valid for other kernel choices as well, even though in general they do not produce explicit solution as in \eqref{AMEM-eq2}, in which cases numerical optimization becomes necessary.

\begin{remark}
\begin{enumerate}[(i)]
\item The AMEM algorithm is a hybrid procedure that combines the alternating maximization (AM) algorithm and the modal expectation-maximization (MEM) algorithm. The former refers to the process of alternately maximizing $\mathbb{M}_{NT}(\bm{{\bm{\theta}}})$ with respect to $\bm{{\bm{\Lambda}}}$ or $\bf{F}$ given the other, while the latter algorithm is borrowed from \citet{YL2014} and solves each maximization as a linear modal regression.
    The idea of AM algorithm has been utilized in a variety of factor models in which no explicit solutions are available for the factor and loading estimators, including the quantile factor model of \citet{CDG2021} and the generalized factor model of \citet{Wang2022}, among others.

\item Following the arguments for Theorem 2.1 of \citet{YL2014}, we can establish the monotonically ascending property of the objective function in Steps 2 and 3, i.e., $\mathbb{M}_{i,T}(\bm{{\bm{\lambda}}}_i(k+1),{\bf{F}}^{(l-1)})\geq \mathbb{M}_{i,T}({\bm{{\bm{\lambda}}}_i}(k),{\bf{F}}^{(l-1)})$, $ \mathbb{M}_{t,N}({\bm{{\bm{\Lambda}}}}^{(l)},{\bf{f}}_t(k+1))\geq \mathbb{M}_{t,N}({\bm{{\bm{\Lambda}}}}^{(l)},{\bf{f}}_t(k))$ for $i=1,\cdots,N; t=1,\cdots,T$, and for each $l$ and $k$. Consequently, we have that $\mathbb{M}_{NT}(\bm{{\bm{\theta}}}|{\bm{{\bm{\Lambda}}}}^{(l+1)},{\bf{F}}^{(l+1)})\geq \mathbb{M}_{NT}(\bm{{\bm{\theta}}}|{\bm{{\bm{\Lambda}}}}^{(l+1)}, {\bf{F}}^{(l)})$ $\geq \mathbb{M}_{NT}(\bm{{\bm{\theta}}}|{\bm{{\bm{\Lambda}}}}^{(l)},{\bf{F}}^{(l)})$, for $l=0,1,\cdots, L$, which guarantees the convergence of the AMEM algorithm.

\item For both the AM algorithm and the MEM algorithm, the solutions obtained upon convergence are necessarily local maxima. As a result, the converged value obtained by the AMEM algorithm depends on the starting points. To find the global optimum, it is necessary to run the algorithm multiple times with different initial values of ${\bf{F}}^{(0)}$, $\bm{{\bm{\Lambda}}}^{(0)}$, and then choose the best local maximum.
\end{enumerate}
\end{remark}

\subsection{Selecting the number of factors}\label{Sec2-3}
The previous subsection assumes that the number of factors $r_0$ is known, which is needed to outline the estimation algorithm for the factors and loadings. In practice, data-driven procedures are desired to select the correct $r_0$. In the following, we propose two methods that can consistently determine $r_0$ based on the observed sample.

For the sake of exposition, we first introduce some notations. Let $\bar{r}$ be a large positive integer such that $r_0<\bar{r}<\infty$. For any $r=1,\ldots, \bar{r}$, let $\mathcal{A}^r$ and $\mathcal{F}^r$ be compact subsets of $\mathbb{R}^r$. Let ${\bm{{\bm{\lambda}}}}_i^r, {\bf{f}}_t^r\in\mathbb{R}^r$ for  $i=1,\cdots, N, t=1,\cdots,T$, and write ${\bm{{\bm{\theta}}}}^r=({{\bm{{\bm{\lambda}}}}_1^r}^{\prime},\cdots,{{\bm{{\bm{\lambda}}}}_N^r}^{\prime}, {{\bf{f}}_1^r}^{\prime}, \cdots, {{\bf{f}}_T^r}^{\prime})^{\prime}, {\bm{{\bm{\Lambda}}}}^r=({{\bm{{\bm{\lambda}}}}_1^r},\cdots,{{\bm{{\bm{\lambda}}}}_N^r})^{\prime}, {\bf{F}}^r=({{\bf{f}}_1^r}, \cdots, {{\bf{f}}_T^r})^{\prime}$.
Similar to \eqref{31}, we adopt the following normalizations:
\begin{align}\label{81}
		&\frac{1}{T}\sum\limits_{t=1}^{T} \mathbf{f}^r_{t} {\mathbf{f}^r_{t}}^{\prime}=\mathbb{I}_{r}, \nonumber \\ 
		&\frac{1}{N}\sum\limits_{i=1}^{N} \bm{{\bm{\lambda}}}^r_{i} {\bm{{\bm{\lambda}}}^r_{i}}^{\prime}, \quad \text { is diagonal with non-increasing diagonal elements.}
\end{align}
Define ${\bm{\Theta}}^{r}=\left\{\bm{{\bm{\theta}}}^r : \bm{{\bm{\lambda}}}^r_i\in \mathcal{A}^r, \mathbf{f}^r_{t} \in \mathcal{F}^r \text{ for all }   i,t,\left\{\bm{{\bm{\lambda}}}^r_{i}\right\} \text{and}  \left\{\mathbf{f}^r_{t}\right\}  \text{satisfy } (\ref{81}) \right\}$,  and denote
\begin{equation}
    {\hat{\bm{{\bm{\theta}}}}}^r=({\hat{\bm{{\bm{\lambda}}}}_1}^{r^{\prime}},\cdots, {\hat{\bm{{\bm{\lambda}}}}_N}^{r^{\prime}}, {\hat{\mathbf{f}}_1}^{r^{\prime}},\cdots,{\hat{\mathbf{f}}_T}^{r^{\prime}})^{\prime}=
    \underset{\bm{{\bm{\theta}}}^r\in{\bm{\Theta}}^r}{\arg\max}\frac{1}{NT}\sum_{i=1}^N\sum_{t=1}^T K_h(X_{it}-{\bm{{\bm{\lambda}}}_i^r}^{\prime}\mathbf{f}_t^r).
\end{equation}
Further, let $\hat{\mathbf{{\bm{\Lambda}}}}^{\bar{r}}=(\hat{\bm{{\bm{\lambda}}}}^{\bar{r}}_1,\cdots,\hat{\bm{{\bm{\lambda}}}}^{\bar{r}}_N)^{\prime}$ and write
	\begin{equation*}
\left(\hat{\mathbf{{\bm{\Lambda}}}}^{\bar{r}}\right)^{\prime} \hat{\mathbf{{\bm{\Lambda}}}}^{\bar{r}} / N=\operatorname{diag}\left(\hat{\sigma}_{N,1}^{\bar{r}}, \ldots, \hat{\sigma}_{N,\bar{r}}^{\bar{r}}\right) .
	\end{equation*}

Our first estimator of $r_0$ is related to the rank of the above matrix. In particular, the rank estimator $\hat{r}_{\mathrm{rank}}$ is defined as
		\begin{equation*}
			\hat{r}_{\mathrm{rank }}=\sum_{j=1}^{\bar{r}} \mathbf{1}\left\{\hat{\sigma}_{N,j}^{\bar{r}}>P_{1,N T}\right\},
		\end{equation*}
where  $P_{1,N T}$  is a sequence that approaches 0 as  $N, T \rightarrow \infty$. It is easily seen that $\hat{r}_{\mathrm{rank }}$ equals the number of diagonal elements in $\left(\hat{\bm{{\bm{\Lambda}}}}^{\bar{r}}\right)^{\prime} \hat{\bm{{\bm{\Lambda}}}}^{\bar{r}}/N$ that exceed the threshold  $P_{1,N T}$. Consequently, $\hat{r}_{\mathrm{rank}}$ can be interpreted as the rank estimator of $\left(\hat{\bm{{\bm{\Lambda}}}}^{\bar{r}}\right)^{\prime} \hat{\bm{{\bm{\Lambda}}}}^{\bar{r}} / N$. It will be shown that $\hat{r}_{\text {rank }}$ approaches $r_0$ as $N, T\to \infty$, provided that the tuning parameter $P_{1,N T}$ diminishes at certain rates.

Our second estimator of $r_0$ is derived from the information criterion (IC), inspired from \citet{BN2002}. To be specific, we consider the following IC:
\begin{equation}
    \mathbb{\mathrm{IC}}(r)=-\mathbb{ M}_{NT}(\hat{\bm{{\bm{\theta}}}}^r)+r\cdot P_{2,NT}, \quad \text{for}\;\; 1\leq r\leq \bar{r},
\end{equation}
where $P_{2,NT}$ is a sequence that approaches 0 as  $N, T \rightarrow \infty$. The IC-based estimator $\hat{r}_{\mathrm{IC}}$ of $r_0$ is defined as
\begin{equation*}
		\hat{r}_{\mathrm{IC}}=\arg \underset{1\leq r\leq \bar{r}}{\min} \mathbb{\mathrm{IC}} (r).
\end{equation*}

It is worth noting that the computation for $\hat{r}_{\mathrm{rank}}$ is significantly less demanding than that for $\hat{r}_{\mathrm{IC}}$. This is because for $\hat{r}_{\mathrm{rank}}$ the MFM only needs to be estimated once with the number of factors setting as $\bar{r}$, while for $\hat{r}_{\mathrm{IC}}$ the MFM needs to be estimated $\bar{r}$ times, with the number of factors specified as $r=1,\ldots,\bar{r}$, respectively. This indicates that $\hat{r}_{\mathrm{rank}}$ is practical preferred in terms of computational time, particularly with large sample sizes.

\section{Asymptotic properties}\label{Sec3}
This section establishes the consistency and asymptotic normality of the proposed estimators for the factors and loadings, and the selection consistency of the two criteria for selecting the factor number.

\subsection{Consistency}
The following assumptions are needed to facilitate the theoretical development.

\begin{assumption}\label{Ass1} (Factors and factor loadings) Suppose that the parameter spaces $\mathcal{A}$ and $\mathcal{F}$ are compact, and ${\bm{{\bm{\theta}}}}_{0}$ is an interior point of ${\bm{\Theta}}^{r_0}$. Further, it holds that
	\begin{enumerate}[(i)]
\item $T^{-1}\sum_{t=1}^T{\bf{f}}_{0t}^{\prime}{\bf{f}}_{0t}\overset{p}{\to}\mathbb{I}_{r_0}$, and there exists a finite positive constant $M_1$, such that $\|\mathbf{f}_{0t}\|\leq M_1$ for all $t=1,\ldots,T$.
\item $N^{-1}\sum_{i=1}^{N}\bm{{\bm{\lambda}}}_{0i}^{\prime}{\bm{\lambda}}_{0i}=\operatorname{diag}(\sigma_{N1},\cdots,\sigma_{Nr_0})$, $\sigma_{N1}\geq\sigma_{N2}\cdots\geq \sigma_{Nr_0},$ and $\sigma_{Nj} \to \sigma_{j}$ as $N \to \infty$ for $j=1,\cdots, r_0$, with $\infty>\sigma_{1}>\sigma_{2}\cdots >\sigma_{r_0}>0$. In addition, there exists a finite positive constant $M_2$ such that $\|\bm{{\bm{\lambda}}}_{0i} \|\leq M_2 $ for all $i=1,\ldots,N$.
\end{enumerate}
\end{assumption}

\begin{assumption}\label{Ass2} (Cross-section dependence and heteroskedasticity)
\begin{enumerate}[(i)]
\item Given $\{\mathbf{f}_{0t}, 1\leq t\leq T\}, \{e^0_{it}, i=1,\cdots,N; t=1, \cdots, T\} $ are independent across $i$.
\item
The sequence $\{{\bf{f}}_{0t},{\bf{e}}^0_t\}$ is $\alpha$-mixing, with mixing coefficients $\alpha(k)\leq B\rho^k$, for $\rho\in (0,1)$ and $B>0$, where ${\bf{e}}^0_t=(e^0_{1t},\cdots,e_{Nt}^0)^{\prime}.$
\end{enumerate}
\end{assumption}

\begin{assumption}\label{Ass3} (Conditional density)
For notational convenience,
let $g_{it}(\cdot)=g_{e^0_{it}}(\cdot|{\bf{f}}_{0t})$, 
\begin{enumerate}[(i)]
\item  $g_{it}(\cdot)$ is continuous and uniformly bounded for all $i,t$.
\item For any positive real number $C$, there exists $\underline{g}>0$ (depending on $C$) such that $g_{it}(0)-\sup_{|u|\geq C}g_{it}(u)\geq \underline{g}$ for all $i,t$.
\item $g_{it}(\cdot)$ is three times continuously differentiable, with $g_{it}^{(v)}(u)={\partial}^vg_{it}(u)/\partial u^v$ being uniformly bounded for $v=1,2,3$ and for all $i,t$.
\item $g_{it}^{(1)}(0)=0$, $-g_{it}^{(2)}(0)\geq \underline{g_1}$  for some $\underline{g_1}>0$, and for all $i,t$.
\end{enumerate}
\end{assumption}


\begin{assumption}(Joint density)\label{Ass6}
 Let $g_{it,js}(\cdot, \cdot)$ denote the joint conditional density function of $e^0_{it}$ and $e^0_{js}, (i,t)\neq (j,s)$, given ${\bf{f}}_{0t}$ and ${\bf{f}}_{0s}$. $g_{it,js}(\cdot, \cdot)$ is uniformly bounded for all $i,t,j,s, (i,t)\neq (j,s)$.
\end{assumption}

\begin{assumption}\label{Ass4} (Kernel function)
$K(\cdot): \mathbb{R}\to \mathbb{R}$ is a twice continuously differentiable  kernel function such that
(i) $\int_{-\infty}^{\infty}K(u)du=1$; (ii) $K(\cdot)$ is symmetric about $0$. (iii) $\lim_{u\to \pm \infty} K(u)=0$; (iv) $\sup_{u\in \mathbb{R}}|K(u)|=c_0<\infty$; (v) $\sup_{u\in \mathbb{R}}|K^{(1)}(u)|=c_1<\infty$, where $K^{(1)}(u)=dK(u)/du$; (vi) $\sup_{u\in \mathbb{R}}|K^{(2)}(u)|=c_2<\infty$, where $K^{(2)}(u)=d^2K(u)/du^2$;
(vii) $\int_{-\infty}^{\infty}|K(u)|u^2du=L_0<\infty$; (viii) $\int_{-\infty}^{\infty}|K^{(1)}(u)|du=L_1<\infty$. 
\end{assumption}

\begin{assumption}\label{Ass5}(Bandwidth) Let $L_{NT}=\min\{N,T\},U_{NT}=\max\{N,T\}.$ As $N, T \to \infty$, we have $ \log L_{NT}/(L_{NT}h^3)\to 0$ and $U_{NT}h^{\gamma}\to 0$ for some $\gamma>3$.\footnote{Assumption \ref{Ass5} requires $U_{NT}\leq L_{NT}^{\gamma/3}$ for some $\gamma>3$, which is quite weak when $\gamma$ is large.}
\end{assumption}

Assumption \ref{Ass1} places conditions on the factors and loadings following \citet[Assumptions A, B, G]{Bai2003}. Assumption \ref{Ass1} (i) requires the factors to be non-degenerate, while Assumption \ref{Ass1} (ii) ensures that each factor has a nontrivial contribution and can be ordered according to their contributions.
The requirement that true parameter is an interior point of the compact parameter space has been similarly made in \citet{BL2016} and  \citet{Wang2022}, and is common in nonlinear models where the estimators have no explicit forms.

Assumption \ref{Ass2} imposes restrictions on the dependence of the idiosyncratic errors. In particular, Assumption \ref{Ass2} (i) requires the errors to be cross-sectionally independent, and Assumption \ref{Ass2} (ii) assumes that the errors follow an $\alpha$-mixing process over time. The mixing property resembles that allowed in \citet{UWY2021} and largely relaxes the independence requirement imposed by \citet{KS2012} and \citet{YL2014} for modal regressions and by \citet{CDG2021} for quantile factor models. It is essential for establishing the bound for the sum of dependent variables. Additionally, both cross-sectional and time-series heteroscedasticity are allowed.

Assumption \ref{Ass3} imposes smoothness conditions on the conditional density of the error term. Assumption \ref{Ass3} (ii) requires that the idiosyncratic error has a well defined unique global mode at $0$ \citep{KS2012,UWY2021}, which is necessary for parameter identification. Note that $g_{it}(\cdot)$ is allowed to be heterogeneous over $i$ and $t$, and it is not required to be unimodal.
Assumption \ref{Ass3} (iii) assumes the derivatives of the conditional density function up to three order are uniformly bounded \citep{KS2012}, which is needed to control the remainder term in the Taylor expansion. Assumption \ref{Ass3} (iv) supposes the conditional density function is concave over a neighbourhood of the mode \citep{KPS2020}, which is required to determine the sign of a specific term in the Taylor expansion. It is worth emphasizing that the existence of moments of the errors is not needed, in contrast to \citet{Bai2003}.
Assumption \ref{Ass6} stipulates that the joint conditional density of the error terms given the factors should be bounded, which is needed to control the sum of covariances when calculating the variance for the sum of dependent variables \citep{Masry1996}.

The kernel function under Assumption \ref{Ass4} is a bounded density function that is symmetric about the mode \citep{YL2014}, with tail approaching $0$ and has bounded first two derivatives \citep{KS2012,KPS2020}. Assumptions \ref{Ass4} (vii) and (viii) are needed when calculating the moments of some random quantities \citep{KPS2020}.
Assumption \ref{Ass5} specifies the convergence rate for the bandwidth required for asymptotic analysis, which is similar to those made by \citet{Romano1988} and \citet{KS2012}.

Write $\hat{\bm{{\bm{\Lambda}}}}=(\hat{\bm{{\bm{\lambda}}}}_1,\cdots,\hat{\bm{{\bm{\lambda}}}}_N)^{\prime}$, $\hat{\bf{F}}=(\hat{\bf{f}}_{1},\cdots,\hat{\bf{f}}_{T})^{\prime}.$
For any $\bm{\theta}_{a},\bm{\theta}_{b} \in \bm{\Theta}^{r_0},$ let $\bm{\theta}_{a}=({\bm{\lambda}}^{\prime}_{a1}, \cdots, {\bm{\lambda}}^{\prime}_{aN},$ $ {\bf{f}}^{\prime}_{a1},\cdots,{\bf{f}}^{\prime}_{aT})^{\prime}$, $\bm{\theta}_{b}=({\bm{\lambda}}^{\prime}_{b1}, \cdots, {\bm{\lambda}}^{\prime}_{bN}, {\bf{f}}^{\prime}_{b1},\cdots,{\bf{f}}^{\prime}_{bT})^{\prime}$, and define
$$d({\bm{\theta}}_a,\bm{\theta}_{b})=\frac{1}{NT}\sum_{i=1}^N\sum_{t=1}^T({\bm{\lambda}}^{\prime}_{ai}{\bf{f}}_{at}-{\bm{\lambda}}^{\prime}_{bi}{\bf{f}}_{bt})^2,$$
which measures the distance between the common components of ${\bm{\theta}}_a$ and ${\bm{\theta}}_b$. The following theorem establishes the average convergence rate of  $\hat{\bf{F}}$ and $\hat{\bm{\Lambda}}$.

\begin{theorem}\label{Thm1}
Under Assumptions \ref{Ass1}-\ref{Ass5}, as $N,T \to \infty$, it holds that
\begin{enumerate}[(a)]
\item $\|\hat{{\bm{\Lambda}}}-{\bm{\Lambda}}_0 \hat{\bf{S}}\|/\sqrt{N}=O_p\left(\frac{1}{\sqrt{L_{NT}h^3}}+h^2\right)$;
\item $\|\hat{\mathbf{F}}-\mathbf{F}_0 \hat{\bf{S}}\|/\sqrt{T}=O_p\left(\frac{1}{\sqrt{L_{NT}h^3}}+h^2\right)$;
\item $d(\hat{\bm{\theta}},{\bm{\theta}}_0)=O_p\left(\frac{1}{\sqrt{L_{NT}h^3}}+h^2\right)$,
\end{enumerate}
with $\hat{\bf{S}}=\operatorname{sgn}(\hat{\bf{F}}^{\prime}{\bf{F}}_0)$, which appears because the value of $\hat{\bm{{\bm{\lambda}}}}_i^{\prime}\hat{\bf{f}}_t$ remains unchanged if both $\hat{\bm{{\bm{\lambda}}}}_i$ and $\hat{\bf{f}}_t$ are multiplied by $-1$.
\end{theorem}

\begin{remark}
\begin{enumerate}[(i)]
\item
There are two notable differences between our theoretical study for MFM and those in the modal regression setting. First, unlike in the modal regressions \citep{KS2012,YL2014} where regressors are observed, both factors and loadings are unobserved in our setup. This calls for simultaneous inference on both the factor and loading estimators. Second, the number of parameters of interest in the regression setting is often of finite dimension, while the number of parameters in MFM diverges along with both $N$ and $T$. Such distinctions prevent the use of proof strategies designed for modal regression in our theoretical development.

\item
 The proof of Theorem \ref{Thm1} borrows asymptotic techniques developed by \citet{CDG2021} for analyzing QFM, given that both estimation procedures involve iterative estimation of diverging number of parameters. However, compared to \citet{CDG2021}, there exist at least two major innovative aspects in the theoretical development. First, the crucial inequality to bound $d^2(\hat{\bm{\theta}},{\bm{\theta}}_0)$ in
 \citet[(A.1) in Appendix A]{CDG2021} does not hold in our setup. Instead, Taylor expansion and properties of the error density functions are used to establish the bound. Second, we remove the restrictive time serial independence error assumption imposed by \citet{CDG2021} and replace it with a mixing condition in the asymptotic development, under which the exponential-type inequality \citet[Theorem 1.3]{Bosq2012} remains effective to control the tail probabilities.

\item Theorem \ref{Thm1} reveals that the optimal bandwidth order for estimation is $h_{opt}=O_p(L_{NT}^{-1/7})$, with which the fastest average convergence rates of $\hat{\bm{\Lambda}}$ and $\hat{\bf{F}}$ are both $L_{NT}^{2/7}$. This rate is slower than the typical rate $L_{NT}^{1/2}$ obtained for factor and loading estimators for the AFM \citep{BN2002}, the QFM \citep{CDG2021} or generalized factor model \citep{Wang2022}. This finding aligns with the result in the regression framework, where the fastest convergence rates for the modal regression estimator \citep{KS2012,YL2014} is $n^{2/7}$ and that for the mean regression estimator is $n^{1/2}$, with $n$ representing the sample size. As explained by \citet{UWY2022}, such reduced convergence rate is attributed to the use of a shrinking bandwidth, which makes the modal estimators rely only on the observations in a small neighbourhood of the mode. In spite of this, the simulations in Section \ref{Sec4} indicate that the proposed estimators enjoy desirable estimation accuracy compared to other alternatives.
\end{enumerate}
\end{remark}

\subsection{Asymptotic normality}
We next study the distributional properties of the factor and loading estimators. The following additional assumptions are needed.

\begin{assumption}(Kernel function)\label{Ass7} $K(\cdot)$ is three times continuously differentiable, such that (i) $\sup_{u\in \mathbb{R}}|K^{(3)}(u)|=c_3<\infty$, where $K^{(3)}(u)=d^3K(u)/du^3$;
(ii) $\int_{-\infty}^{\infty}|K(u)| |u|^5du=L_2<\infty$;
(iii) $\int_{-\infty}^{\infty}|K^{(1)}(u)|^2|u| du=L_3<\infty$;
(iv)$\int_{-\infty}^{\infty}|K^{(2)}(u)|^2du=L_4<\infty$.
\end{assumption}

\begin{assumption}(Conditional density)\label{Ass8}
\begin{enumerate}[(i)]
\item $g_{it}(\cdot)$ is six times continuously differentiable, with $g_{it}^{(v)}(u)={\partial}^vg_{it}(u)/\partial u^v$ being uniformly bounded for $v=4,5,6$ and for all $i,t$.
\item $g_{it}^{(v)}(0)=0$ for $v=3,5$ and for all $i,t$.
\end{enumerate}
\end{assumption}

\begin{assumption}(Bandwidth)\label{Ass9}
As $N,T\to \infty, N \propto T$, $Th^{11}\to\infty$, $Th^{13}\to 0$.
\end{assumption}

Assumptions \ref{Ass7}-\ref{Ass9} strengthen the conditions used in establishing the consistency results earlier. Assumptions \ref{Ass7} and \ref{Ass8} require that the kernel function and the conditional density of the error terms should exhibit higher order smoothness. Assumption \ref{Ass7} is standard in modal regression for asymptotic normality \citep{KS2012,KPS2020}. Assumption \ref{Ass8} is stronger than Assumption B3 in \citet{KS2012}, and is required to control the higher order terms in the stochastic expansions of the estimators. Assumption \ref{Ass9} specifies a suitable rate for the bandwidth, which ensures the modal estimators to be asymptotically unbiased.

Define
\begin{equation*}
{\bm{{\bm{\Phi}}}}_i=\underset{T\to \infty}{\operatorname{plim}}\frac{1}{T}\sum_{t=1}^Tg^{(2)}_{it}(0){\bf{f}}_{0t}{\bf{f}}_{0t}^{\prime},\qquad {\bm{{\bm{\Psi}}}}_t=\underset{N\to \infty}{\lim}\frac{1}{N}\sum_{i=1}^Ng^{(2)}_{it}(0){\bm{\lambda}}_{0i}{\bm{\lambda}}_{0i}^{\prime},
\end{equation*}
and
\begin{equation*}
{\bm{{\bm{\Sigma}}}}_i=\underset{T\to \infty}{\operatorname{plim}}\frac{1}{T}\sum_{t=1}^TLg_{it}(0){\bf{f}}_{0t}{\bf{f}}_{0t}^{\prime},\qquad {\bm{{\bm{\Omega}}}}_t=\underset{N\to \infty}{\lim}\frac{1}{N}\sum_{i=1}^NLg_{it}(0){\bm{\lambda}}_{0i}{\bm{\lambda}}_{0i}^{\prime},
\end{equation*}
where $L=\int_{-\infty}^{\infty}|K^{(1)}(u)|^2du<\infty$.

\begin{assumption}(Matrix definiteness)\label{Ass10}
\begin{enumerate}[(i)]
\item ${\bm{{\bm{\Phi}}}}_i<0$ and ${\bm{{\bm{\Psi}}}}_t<0$ for all $i,t$.
\item ${\bm{{\bm{\Sigma}}}}_i>0$ and ${\bm{{\bm{\Omega}}}}_t>0$ for all $i,t$.
\end{enumerate}
\end{assumption}

Assumption \ref{Ass10} requires the matrices to be positive or negative definite, which is essential for establishing the asymptotic variances of the estimators \citep{CDG2021}.

\begin{theorem}\label{Thm4}
Let $\hat{\bf{S}}=\operatorname{sgn}(\hat{\bf{F}}^{\prime}{\bf{F}}_0/T)$. Under Assumptions \ref{Ass1}-\ref{Ass10}, as $N,T\to \infty$, 
\begin{equation*}
\sqrt{Th^3}(\hat{\bm{\lambda}}_i-\hat{\bf{S}}{\bm{\lambda}}_{0i})\overset{d}{\rightarrow}\mathcal{N}(0,{\bm{\Phi}}_i^{-1}{\bm{\Sigma}}_i{\bm{\Phi}}_i^{-1}), \qquad \sqrt{Nh^3}(\hat{\bf{f}}_t-\hat{\bf{S}}{\bf{f}}_{0t})\overset{d}{\rightarrow}\mathcal{N}(0,{\bm{\Psi}}_t^{-1}{\bm{\Omega}}_t{\bm{\Psi}}_t^{-1}).
\end{equation*}
\end{theorem}

\begin{remark}
\begin{enumerate}[(i)]
 \item Theorem \ref{Thm4} establishes the limiting distributions of $\hat{\bm{\lambda}}_i$ and $\hat{\bf{f}}_t$. The asymptotic variance for
$\hat{\bm{\lambda}}_i$ and $\hat{\bf{f}}_t$ are similar to those for regression coefficient estimator in \citet{KS2012} with observed regressors.

\item  The proof of this theorem relies on expanding the first order conditions around the true values of the factors and loadings. Taking the first result for example. Let $K_h^{(j)}(u)=d^jK_h(u)/du^j,j=1,2$. The first order condition $\sum_{t=1}^TK_h^{(1)}(X_{it}-\hat{\bm{\lambda}}_i^{\prime}\hat{\bf{f}}_t)\hat{\bf{f}}_t/T=0$ expanded around
$(\hat{\bf{S}}{\bm{\lambda}}_{0i}, {\bf{F}}_0\hat{\bf{S}})$
gives rise to, after some simple calculations,
\begin{align}\label{thm4-eq2}
&\frac{1}{T}\sum_{t=1}^TK^{(2)}_h(e_{it}^0){\bf{f}}_{0t}{\bf{f}}_{0t}^{\prime}(\hat{\bm{\lambda}}_i-\hat{\bf{S}}{\bm{\lambda}}_{0i})\nonumber\\
=&\frac{1}{T}\sum_{t=1}^TK^{(1)}_h(e_{it}^0)\hat{\bf{S}}{\bf{f}}_{0t}+\frac{1}{T}\sum_{t=1}^TK_{h}^{(1)}(e_{it}^{1})(\hat{\bf{f}}_t-\hat{\bf{S}}{\bf{f}}_{0t})-\frac{1}{T}\sum_{t=1}^TK_{h}^{(2)}(e_{it}^{2}){\bf{f}}_{0t}{\bm{\lambda}}_{0i}^{\prime}(\hat{\bf{f}}_t-\hat{\bf{S}}{\bf{f}}_{0t})\nonumber\\
+&o_p(\|\hat{\bm{\lambda}}_{i}-\hat{\bf{S}}{\bm{\lambda}}_{0i}\|),
\end{align}
where $e_{it}^0=X_{it}-{\bm{\lambda}}_{0i}^{\prime}{\bf{f}}_{0t},\hat{e}_{it}=X_{it}-\hat{\bm{\lambda}}_i^{\prime}\hat{\bf{f}}_t$, and $e_{it}^1, e_{it}^2$ lie between $e_{it}^0$ and $\hat{e}_{it}$.
With the above expansion, the proof 
then proceeds in two steps. The first step involves demonstrating that both
the second and the third terms on the right hand side of \eqref{thm4-eq2} are $o_p(1/\sqrt{Th^3})$. Since $\hat{\bf{f}}_t-\hat{\bf{S}}{\bf{f}}_{0t}$ does not yield an analytical form, we require a stochastic expansion for $ \hat{\bf{f}}_t-\hat{\bf{S}}{\bf{f}}_{0t}$. This technical challenge is solved by showing that the expected Hessian matrix is asymptotically block diagonal.
In this way, \eqref{thm4-eq2} can be simplified as
\begin{equation*}
\frac{1}{T}\sum_{t=1}^TK^{(2)}_h(e_{it}^0){\bf{f}}_{0t}{\bf{f}}_{0t}^{\prime}(\hat{\bm{\lambda}}_i-\hat{\bf{S}}{\bm{\lambda}}_{0i})=\frac{1}{T}\sum_{t=1}^T{K}_h^{(1)}(e^0_{it})\hat{\bf{S}}{\bf{f}}_{0t}+o_p(1/\sqrt{Th^3})+o_p(\|\hat{\bm{\lambda}}_i-\hat{\bf{S}}{\bm{\lambda}}_{0i}\|).
\end{equation*}
The second step is then to establish the asymptotic normality of $\sum_{t=1}^T{K}_h^{(1)}(e^0_{it})\hat{\bf{S}}{\bf{f}}_{0t}/T$ under the mixing condition, following the proof strategy used by \citet{Masry1996}. Consequently, the asymptotic normality of the loading estimators can be established.
\end{enumerate}
\end{remark}

Define
\begin{equation*}
\begin{split}
\hat{\bm{\Phi}}_i=\frac{1}{T}\sum_{t=1}^TK_h^{(2)}(X_{it}-\hat{\bm{\lambda}}_i^{\prime}\hat{\bf{f}}_t)\hat{\bf{f}}_t\hat{\bf{f}}_{t}^{\prime},\qquad \hat{\bm{\Psi}}_t=\frac{1}{N}\sum_{i=1}^NK_h^{(2)}(X_{it}-\hat{\bm{\lambda}}_i^{\prime}\hat{\bf{f}}_t)\hat{\bm{\lambda}}_i\hat{\bm{\lambda}}_{i}^{\prime},
\end{split}
\end{equation*}
and
\begin{equation*}
\begin{split}
\hat{\bm{\Sigma}}_i=\frac{1}{Th}\sum_{t=1}^T[K^{(1)}(\frac{X_{it}-\hat{\bm{\lambda}}_i^{\prime}\hat{\bf{f}}_t}{h})]^2\hat{\bf{f}}_t\hat{\bf{f}}_{t}^{\prime},\qquad \hat{\bm{\Omega}}_t=\frac{1}{Nh}\sum_{i=1}^N[K^{(1)}(\frac{X_{it}-\hat{\bm{\lambda}}_i^{\prime}\hat{\bf{f}}_t}{h})]^2\hat{\bm{\lambda}}_i\hat{\bm{\lambda}}_{i}^{\prime}.
\end{split}
\end{equation*}

The following theorem establishes the consistency of the variance estimators for those of the factor and loading estimators.

\begin{theorem}\label{Thm5}
Under Assumptions \ref{Ass1}-\ref{Ass10}, as $N,T\to \infty$, it holds that
\begin{equation*}
\hat{\bm{\Phi}}_i^{-1}\hat{\bm{\Sigma}}_i\hat{\bm{\Phi}}_i^{-1}\overset{p}{\rightarrow} {\bm{\Phi}}_i^{-1}{\bm{\Sigma}}_i{\bm{\Phi}}_i^{-1},
\qquad \hat{\bm{\Psi}}_t^{-1}\hat{\bm{\Omega}}_t\hat{\bm{\Psi}}^{-1}_t\overset{p}{\rightarrow} {\bm{\Psi}}_t^{-1}{\bm{\Omega}}_t{\bm{\Psi}}_t^{-1}.
\end{equation*}
\end{theorem}

\subsection{Selection consistency}
We next consider the consistency for the factor number estimators $\hat{r}_{\mathrm{IC}}$ and $\hat{r}_{\mathrm{rank}}$.

\begin{theorem}\label{Thm2}
Let $\delta_{NT}=(\sqrt{L_{NT}h^3})^{-1}+h^2$. Under Assumptions \ref{Ass1}-\ref{Ass5}, if $P_{1,NT} \to 0$, $P_{1,NT}\delta_{NT}^2\to \infty,$ then $ P[\hat{r}_{\mathrm{rank}}=r_0]\to 1 $, as $ N, T \to  \infty$.
\end{theorem}

\begin{remark}\label{Rem3-2}
\begin{enumerate}[(i)]
    \item 
Theorem \ref{Thm2} establishes the consistency of the rank estimator $\hat{r}_{\mathrm{rank}}$. The idea of the proof is as follows. For any $\mathbf{{\bm{\Lambda}}}^{r}$, $r_0<r\leq \bar{r}$, decompose
$\mathbf{{\bm{\Lambda}}}^{r}=[\mathbf{{\bm{\Lambda}}}^{r,r_0},\mathbf{{\bm{\Lambda}}}^{r,-r_0}] $, where the submatrix $\mathbf{{\bm{\Lambda}}}^{r,r_0}$ collects the first $r_0$ columns of $\mathbf{{\bm{\Lambda}}}^{r}$, and $\mathbf{{\bm{\Lambda}}}^{r,-r_0}$ collects the remaining $r-r_0$ columns. To prove Theorem \ref{Thm2}, we establish that up to sign, it holds that
\begin{equation}\label{thm2-eq1}
\|\hat{\bm{{\bm{\Lambda}}}}^{\bar{r}}-\bm{{\bm{\Lambda}}}^*_0 \|/\sqrt{N}=O_p(\delta_{NT}^{-1}),
\end{equation}
where ${\bm{\Lambda}}_0^*=[{\bm{\Lambda}}_0,\mathbf{0}_{N\times(\bar{r}-r_0)}]$.
Then $\hat{\sigma}^{\bar{r}}_{N,j}={\sigma}_{Nj}+o_p(1) \overset{p}{\to}\sigma_{j}> 0$, for $j=1,\cdots, r_0$ by Assumption \ref{Ass1} (ii), and $\hat{\sigma}^{\bar{r}}_{N,j}\leq \sum_{s=r_0+1}^{\bar{r}}\hat{\sigma}^{\bar{r}}_{N,s}=\|\hat{\bm{{\bm{\Lambda}}}}^{\bar{r},-r_0}\|^2/N=O_p(\delta_{NT}^{-2})$ for $j=r_0+1,\cdots,\bar{r}$.
Therefore, $\hat{{\bm{\Lambda}}}^{{\bar{r}}^{\prime}}\hat{{\bm{\Lambda}}}^{\bar{r}}/N
$ converges in probability to a matrix with rank $r_0$
 at rate $\delta_{NT}^2$, which leads to the consistency of $\hat{r}_{\mathrm{rank}}$ naturally.
 
 \item  The proof of \eqref{thm2-eq1} parallels that of Lemma 2 in \citet{CDG2021}. The difference lies in that their proof requires $\rho_{\mathrm{min}}[{\bm{\Lambda}}_0^{\prime}{\bm{\Lambda}}_0]=\sigma_{Nr_0} \overset{p}{\to} \sigma_{r_0}>0$ as $N\to \infty$, which is not satisfied here, as $\rho_{\mathrm{min}}[{{\bm{\Lambda}}^*_0}^{\prime}{\bm{\Lambda}}^*_0]=0$. We address this discrepancy by showing that, for $\|{\bm{\Lambda}}^{r}{\bf{F}}^{r^{\prime}}-{\bm{\Lambda}}_0{\bf{F}}_0^{\prime}\|\leq \delta$ with $\delta>0$ sufficiently small and $r>r_0$, $\rho_{\mathrm{min}}[{\bm{\Lambda}}^{{r,r_0}^{\prime}}{\bm{\Lambda}^{r,r_0}}]$ is positively bounded below, together with leveraging properties of
positive definite matrices. 
\end{enumerate}
\end{remark}

\begin{theorem}\label{Thm3}
Under Assumptions \ref{Ass1}-\ref{Ass5}, if $P_{2,NT}\to 0$, $P_{2,NT}\delta_{NT}^2\to \infty$, then $P[\hat{r}_{\mathrm{IC}}=r_0]\to 1$, as $ N, T \to  \infty$.
\end{theorem}

\begin{remark}\label{Rem3-3}
\begin{enumerate}[(i)]
\item
Theorem \ref{Thm3} establishes the consistency of the IC-based estimator $\hat{r}_{\mathrm{IC}}$. The proof of this result follows closely from \citet{BN2002}. In particular, we consider two cases (i) $0<r<r_0$ and (ii) $r_0< r\leq \bar{r}$, respectively.
    For case (i) $0<r<r_0$, we prove that there exists $C>0$, such that $\mathbb{M}_{NT}(\hat{\bm{{\bm{\theta}}}}^{r_0})-\mathbb{M}_{NT}(\hat{\bm{{\bm{\theta}}}}^{r})\geq C+o_p(1)$.
    For case (ii) $r_0<r\leq \bar{r}$, we demonstrate  $\mathbb{M}_{NT}(\hat{\bm{{\bm{\theta}}}}^{r_0})-\mathbb{M}_{NT}(\hat{\bm{{\bm{\theta}}}}^{r})=O_p(\delta_{NT}^{-2})$.
     Therefore, when $P_{2,NT}$
vanishes at a rate slower than $\delta_{NT}^2$, for any $r$ such that $0<r\neq r_0\leq \bar{r}$, IC$(r)$-IC$(r_0)$ will always be dominated by a positive term. Hence, $P[\hat{r}_{\mathrm{IC}}=r_0]\to 1$.
\item 
The conditions placed on the penalties in Theorem \ref{Thm2} and Theorem \ref{Thm3} are identical, therefore, the choices of $P_{1,NT}$ and $P_{2,NT}$ can be made at the same order.
With the optimal bandwidth $h=O(L_{NT}^{-1/7})$, we obtain $\delta_{NT}=O(L_{NT}^{2/7})$. For $\hat{r}_{\mathrm{rank}}$, the choice
\begin{equation*}
   P_{1,NT}=\hat{\sigma}^{\bar{r}}_{N,1}\cdot (L_{NT}^{4/7})^{-0.3},
\end{equation*}
meets the rate requirement, which leads to high probability of correctly selecting the factor number as long as $\min\{N,T\}\geq 100$ in finite sample simulations later on.
\end{enumerate}

For $\hat{r}_{\mathrm{IC}}$, let $\mathbb{M}_{NT}=\frac{1}{NT}\sum_{i=1}^N\sum_{t=1}^TK_h(X_{it})$,
then the choice
\begin{equation*}
    P_{2,NT}=3/7\cdot (\mathbb{M}_{NT}(\hat{\bm{\theta}}^{\bar{r}})-\mathbb{M}_{NT})\cdot (L_{NT}^{4/7})^{-0.4},
\end{equation*}
is desirable. Simulations indicate that $\hat{r}_{\mathrm{IC}}$ outperforms $\hat{r}_{\mathrm{rank}}$, and $\hat{r}_{\mathrm{IC}}$ can accurately select the number of factors in most settings even with sample size as small as $(N,T)=(60,60)$.
\end{remark}

\section{Simulation studies}\label{Sec4}
This section conducts a set of Monte Carlo experiments to evaluate the finite sample performance of our proposed estimators, with a comparison to that of the estimators derived under AFM \citep{BN2002} and QFM \citep{CDG2021}.

\subsection{Data generating processes}
\noindent The data are drawn from the following three-factor model:
\begin{equation*}
	X_{it}=\sum\limits_{j=1}^3{\lambda}_{ji}f_{jt}+ e_{it},\quad \text{for} \;\; i=1,\cdots,N; t=1,\cdots,T,
\end{equation*}
where $f_{it},{\lambda}_{it}$ are independent $\mathcal{N}(0,1)$ variates. For the error term,
we entertain three different error specifications.

\smallskip

\noindent\textbf{S1} (Heavy-tailed errors). The error term ${e_{it}}^{\prime}s$ are independently drawn from $ t_\nu$, the students $t$-distribution with $\nu$ degrees of freedom, for  $\nu=1,2,3$.

\medskip

\noindent\textbf{S2} (Dependent errors).
Following \citet{BN2002}, $e_{it}$ is generated according to
\begin{equation*}
	e_{it}=\rho e_{it-1}+v_{it}+\sum\limits_{j\neq0,j=-J}^{J}\beta v_{i-jt},
\end{equation*}
where ${v_{it}}^{\prime}s$ are independently drawn from $t_3$. The autoregressive coefficient $\rho$ reflects the serial correlation of $e_{it}$, while the parameters $\beta$ and $J$ reflect the cross-sectional dependence of $e_{it}$. In particular, the following three sets of error dependence parameters are considered.

(D1) Serially correlated errors: $\rho=0.2,\beta=0$.

(D2) Cross-sectionally correlated errors: $\rho=0,\beta=0.2,J=3$.

(D3) Serially and cross-sectionally correlated errors: $\rho=0.2,\beta=0.2,J=3$.

\medskip

\noindent\textbf{S3} (Skewed errors).
Following \citet{YL2014}, we entertain a mixture normal distribution for $e_{it}$,
\begin{equation*}
    e_{it}\sim 0.5 \mathcal{N}(0.8,0.6^2)+0.5 \mathcal{N}(-0.8,\sigma^2),
\end{equation*}
where $\sigma\in \{2.6,3,3.4\}$. This distribution is skewed left with E$(e_{it})=0$, Mode$(e_{it})\approx 0.8$, Median$(e_{it})\approx \{0.53,0.57,0.60\}$, corresponding to the three values of $\sigma$. Additionally, as discussed in Example \ref{exm0}, to ensure that AFM, MFM and QFM have the same representations, the error terms are accordingly normalized when obtaining the estimates for MFM ($e_{it}-\operatorname{Mode}(e_{it})$) and QFM ($e_{it}-\operatorname{Median}(e_{it})$).

Included in the comparison are the MFA estimator ${\hat{\bf{F}}}_M$ proposed in this paper,
the PCA estimator ${\hat{\bf{F}}}_{P}$ studied by \citet{BN2002}, and the QFA estimator at $\tau=0.5$, i.e., ${\hat{\bf{F}}}_{Q}^{0.5}$ proposed by \citet{CDG2021}. To obtain ${\hat{\bf{F}}}_M$, we run the AMEM algorithm with two different sets of random starting parameters,
and adopt the estimate that maximizes the objective function. For each error specification, we set $N,T \in \{60,100,200\}$. For the tuning parameters, we follow the discussion in Remark \ref{Rem3-3} to set $h=c\cdot L_{NT}^{-1/7}$, $P_{1,NT}=\hat{\sigma}^{\bar{r}}_{N,1}\cdot (L_{NT}^{4/7})^{-0.3}$ and $P_{2,NT}=3/7\cdot (\mathbb{M}_{NT}(\hat{\bm{\theta}}^{\bar{r}})-\mathbb{M}_{NT})\cdot(L_{NT}^{4/7})^{-0.4}$.
To evaluate the robustness of our MFA estimator to the bandwidth choice, we consider $c\in \{3,5,7\}$. Finally, we set the maximum number of factors $\bar{r}=8$ following \citet{BN2002}.

Several commonly used criteria are employed to evaluate the performance of the estimators. First, to assess the performance of the estimated factors in capturing the space of the true factors, we consider the trace-ratio statistic adopted by \citet{BN2006} and \citet{CKKK2018}, which measures the distance between the estimated factor space and the true factor space. Specifically, let $\hat{\mathbf{F}}$ and ${\mathbf{F}}_{0}$ denote the estimated and the true factor matrices, respectively. The trace-ratio statistic $\operatorname{tr}(\mathbf{\hat{F}})$ is defined as
\begin{equation*}
\operatorname{tr}(\hat{\bf{F}})=\frac{\operatorname{tr}[{{{\bf{F}}^{\prime}_{0}}\hat{\bf{F}}({\hat{\bf{F}}}^{\prime}\hat{\bf{F}})^{-1}{\hat{\bf{F}}}^{\prime}{\bf{F}}_{0}}]}{\operatorname{tr}({{{\bf{F}}^{\prime}_{0}}{\bf{F}}_{0}})}.
\end{equation*}
It is seen that $\operatorname{tr}(\mathbf{\hat{F}})\in [0,1]$, and a larger $\operatorname{tr}(\mathbf{\hat{F}})$ indicates a smaller distance between the space spanned by $\mathbf{\hat{F}}$ and ${\bf{F}}_{0}$.
Second, to evaluate the precision of model selection, we compute the average estimated number of factors and the frequency of correctly selecting the number of factors. All the results are obtained over $S=500$ replications.

\subsection{Results}
\renewcommand\arraystretch{0.85}
\begin{table}[t]
\centering
\caption{Factor estimation accuracy results for S1.}\label{table1}
\resizebox{1\linewidth}{!}{
\begin{tabular}{c l c c c c c c c c c c }
\toprule
&   & \multicolumn{3}{c}{$T=60$}& \multicolumn{3}{c}{$T=100$} & \multicolumn{3}{c}{$T=200$}\\
                \cmidrule(r){3-5}  \cmidrule(r){6-8} \cmidrule(r){9-11}
                  & &  \multicolumn{3}{c}{$N$} & \multicolumn{3}{c}{$N$} &\multicolumn{3}{c}{$N$}\\
                \cmidrule(r){3-5} \cmidrule(r){6-8} \cmidrule(r){9-11}
              &  & $60$    & $100$   & $200$   & $60$    & $100$   & $200$    & $60$    & $100$   & $200$  \\
                    \midrule
\multirow{5}{*}{$e_{it}\sim t_1$} & $\operatorname{tr}(\hat{\bf{F}}_M) \; c=3$ & 0.944 & 0.970 & 0.987 & 0.946 & 0.972 & 0.987 & 0.947 & 0.973 &0.987 \\
            & $\operatorname{tr}(\hat{\bf{F}}_M) \; c=5$& 0.940 & 0.968 & 0.985 & 0.942 & 0.970 & 0.985 & 0.948 & 0.970 &0.986 \\
         & $\operatorname{tr}(\hat{\bf{F}}_M) \; c=7$& 0.931 & 0.962 & 0.981 & 0.932 & 0.965 & 0.982 & 0.939 & 0.964 &0.984\\
         & $\operatorname{tr}(\hat{\bf{F}}_P)$& 0.048 & 0.055 & 0.040 & 0.030 & 0.031 & 0.026 & 0.019 & 0.016 & 0.012 \\
        & $\operatorname{tr}(\hat{\bf{F}}^{0.5}_Q)$& 0.928 & 0.964 & 0.983 & 0.934 & 0.966 & 0.985 & 0.943 & 0.968 & 0.985 \\
                   \midrule
\multirow{5}{*}{$e_{it}\sim t_2$} & $t r(\hat{\bf{F}}_M) \; c=3$ & 0.963 & 0.980 & 0.983 & 0.963 & 0.980 & 0.990 & 0.966 & 0.980 &0.990\\
        & $\operatorname{tr}(\hat{\bf{F}}_M) \; c=5$ & 0.961 & 0.980 & 0.990 & 0.962 & 0.980 & 0.990 & 0.965 & 0.979 &0.990 \\
        & $\operatorname{tr}(\hat{\bf{F}}_M) \; c=7$ & 0.956 & 0.977 & 0.989 & 0.957 & 0.978 & 0.989 & 0.961 & 0.977 & 0.989\\
     & $\operatorname{tr}(\hat{\bf{F}}_P)$ & 0.596 & 0.642 & 0.586 & 0.564 & 0.590 & 0.695 & 0.535 & 0.599 & 0.538\\
    & $\operatorname{tr}(\hat{\bf{F}}^{0.5}_Q)$ & 0.956 & 0.977 & 0.988 & 0.956 & 0.977 & 0.989 & 0.961 & 0.976 &0.989 \\
                   \midrule
\multirow{5}{*}{$e_{it}\sim t_3$} & $\operatorname{tr}(\hat{\bf{F}}_M) \; c=3$& 0.969 & 0.976 & 0.992 & 0.968 & 0.982 & 0.992 & 0.970 & 0.982 &0.991\\
    & $\operatorname{tr}(\hat{\bf{F}}_M) \; c=5$ & 0.969 & 0.983 & 0.992 & 0.968 & 0.983 & 0.992 & 0.971 & 0.982&0.992 \\
        & $\operatorname{tr}(\hat{\bf{F}}_M) \; c=7$& 0.966 & 0.982 & 0.991 & 0.965 & 0.982 & 0.991 & 0.969 & 0.981 &0.991 \\
     & $\operatorname{tr}(\hat{\bf{F}}_P)$& 0.934 & 0.955 & 0.984 & 0.929 & 0.919 & 0.983 & 0.949 & 0.967 & 0.981 \\
    & $\operatorname{tr}(\hat{\bf{F}}^{0.5}_Q)$& 0.962 & 0.979 & 0.990 & 0.962 & 0.979 & 0.990 & 0.965 & 0.979 & 0.990\\
                    \bottomrule
\end{tabular}}
\begin{tablenotes}
\item Simulation results over 500 repetitions. The DGP considered in this table: $X_{it}=\sum_{j=1}^3{\lambda}_{ji}f_{jt}+ e_{it}$, ${\lambda}_{ji},f_{jt}\sim i.i.d \; \mathcal{N}(0,1), e_{it}\sim t_\nu$, with $\nu=1$ (upper panel), $\nu=2$ (middle panel), and $\nu=3$ (lower panel). $\hat{\bf{F}}_M$, $\hat{\bf{F}}_P$ and $\hat{\bf{F}}^{0.5}_Q$ denote the estimates of the factors by MFA,  PCA \citep{BN2002} and QFA at $\tau=0.5$ \citep{CDG2021}, respectively. $\operatorname{tr}(\hat{\bf{F}}_M)$, $\operatorname{tr}(\hat{\bf{F}}_P)$ and $\operatorname{tr}(\hat{\bf{F}}^{0.5}_Q)$ represent the trace-ratio statistics of $\hat{\bf{F}}_M$, $\hat{\bf{F}}_P$ and $\hat{\bf{F}}^{0.5}_Q$, respectively, where $\hat{\bf{F}}_M$ is obtained with the bandwidth $h=c\cdot L_{NT}^{-1/7}$ for $c\in\{3,5,7\}$.
\end{tablenotes}
\end{table}
Table \ref{table1} presents the factor estimation accuracy results under heavy-tailed errors S1. The main findings are summarized as follows.
First, both the modal factor $\hat{\bf{F}}_M$ and the median factor ${\hat{\bf{F}}}_{Q}^{0.5}$ can effectively capture the true factor space for all three error choices. Second, our estimator ${\bf{\hat{F}}}_M$ is hardly affected by the choice of bandwidth and consistently outperforms the other two estimators across all parameter configurations. Third, ${\hat{\bf{F}}}_{P}$ is notably inferior to the other two estimators under $t_1$ and $t_2$ errors, but becomes comparable under $t_3$ errors. 
Overall, the performances of all three estimators tend to improve as the error distribution shifts from $t_1$ to $t_3$, and as both $N$ and $T$ increase.

The factor estimation accuracy results under dependent errors S2 and skewed errors S3 are presented in
the Supplementary Materials. 
Additional to those discovered from Table \ref{table1}, there are several new findings which are summarized as follows. First, as the dependence in the error terms increases in S2, the performances of all three estimators deteriorate, but our estimator ${\hat{\bf{F}}}_M$ continues to provide the most accurate estimates in almost all cases. Second, the accuracy
of ${\hat{\bf{F}}}_M$ and that of ${\hat{\bf{F}}}_Q^{0.5}$ remain nearly unchanged as the error terms become increasingly skewed in S3, while that of ${\hat{\bf{F}}}_P$ decreases notably. This finding demonstrates that ${\hat{\bf{F}}}_M$ and ${\hat{\bf{F}}}_Q^{0.5}$ exhibit robustness to the presence of skewness as compared to ${\hat{\bf{F}}}_P$. Third, the performance of ${\hat{\bf{F}}}_M$ is only slightly influenced by the choice of bandwidth in S2 and S3, even though the effect seems a bit larger than that in S1.

\begin{table}[t]
\caption{Factor number estimation results for S1.}\label{table4}
	\centering	
		\resizebox{1\linewidth}{!}{
			\begin{tabular}{l c c c c c c c c c c c c c c}
				\toprule
				&   &  &  \multicolumn{4}{c}{$e_{it}\sim  t_1$} &  \multicolumn{4}{c}{$e_{it}\sim t_2$} & \multicolumn{4}{c}{$e_{it}\sim t_3$}  \\
				\cmidrule(r){4-7}  \cmidrule(r){8-11} \cmidrule(r){12-15}
				 & $T$ & $N$   & $\bar{\hat{r}}_{\mathrm{rank}}$ & $\mathit{Freq}_1$  & $\bar{\hat{r}}_{\mathrm{IC}}$ & $\mathit{Freq}_2$  & $\bar{\hat{r}}_{\mathrm{rank}}$ & $\mathit{Freq}_1$ & $\bar{\hat{r}}_{\mathrm{IC}}$ & $\mathit{Freq}_2$ & $\bar{\hat{r}}_{\mathrm{rank}}$ & $\mathit{Freq}_1$&  $\bar{\hat{r}}_{\mathrm{IC}}$ & $\mathit{Freq}_2$  \\
				\midrule
\multirow{9}{*}{$c=3$} & 60  & 60  & 2.45 & 0.41 & 2.95 & 0.89 & 2.77 & 0.65 & 2.96 & 0.91 & 2.45 & 0.45 & 2.95 & 0.95 \\
                     & 60  & 100 & 2.97 & 0.74 & 2.98 & 0.97 & 2.61 & 0.65 & 2.98 & 0.98 & 2.58 & 0.58 & 2.97 & 0.97 \\
                     & 60  & 200 & 2.81 & 0.74 & 2.98 & 0.98 & 2.74 & 0.74 & 2.94 & 0.93 & 2.77 & 0.81 & 2.98 & 0.98 \\
                     & 100 & 60  & 3.02 & 0.65 & 2.97 & 0.97 & 2.68 & 0.71 & 2.98 & 0.98 & 2.45 & 0.48 & 2.99 & 0.99 \\
                     & 100 & 100 & 3.12 & 0.82 & 3.00 & 1.00 & 2.82 & 0.82 & 3.00 & 1.00 & 2.82 & 0.88 & 3.00 & 1.00 \\
                     & 100 & 200 & 3.00 & 1.00 & 3.00 & 1.00 & 3.00 & 1.00 & 3.00 & 1.00 & 2.94 & 0.94 & 3.00 & 1.00 \\
                     & 200 & 60  & 3.06 & 0.61 & 2.96 & 0.94 & 2.81 & 0.81 & 2.98 & 0.98 & 2.55 & 0.65 & 2.96 & 0.96 \\
                     & 200 & 100 & 2.94 & 0.94 & 3.00 & 1.00 & 3.00 & 1.00 & 3.00 & 1.00 & 2.94 & 0.94 & 3.00 & 1.00 \\
                     & 200 & 200 & 3.00 & 1.00 & 3.00 & 1.00 & 3.00 & 1.00 & 3.00 & 1.00 & 3.00 & 1.00 & 3.00 & 1.00 \\
                     \midrule
\multirow{9}{*}{$c=5$} & 60  & 60  & 2.52 & 0.44 & 2.90 & 0.87 & 2.58 & 0.58 & 2.93 & 0.89 & 2.45 & 0.48 & 2.93 & 0.93 \\
                     & 60  & 100 & 2.77 & 0.65 & 2.97 & 0.95 & 2.61 & 0.65 & 2.93 & 0.91 & 2.48 & 0.48 & 2.97 & 0.97 \\
                     & 60  & 200 & 2.90 & 0.71 & 2.96 & 0.94 & 2.68 & 0.68 & 2.93 & 0.93 & 2.81 & 0.81 & 2.96 & 0.96 \\
                     & 100 & 60  & 2.71 & 0.65 & 2.92 & 0.92 & 2.65 & 0.68 & 2.94 & 0.94 & 2.45 & 0.48 & 2.92 & 0.92 \\
                     & 100 & 100 & 2.88 & 0.76 & 3.00 & 1.00 & 2.82 & 0.82 & 3.00 & 1.00 & 2.82 & 0.88 & 3.00 & 1.00 \\
                     & 100 & 200 & 3.00 & 1.00 & 3.00 & 1.00 & 3.00 & 1.00 & 3.00 & 1.00 & 2.94 & 0.94 & 3.00 & 1.00 \\
                     & 200 & 60  & 2.84 & 0.84 & 2.93 & 0.91 & 2.74 & 0.74 & 2.97 & 0.97 & 2.55 & 0.65 & 2.97 & 0.97 \\
                     & 200 & 100 & 2.94 & 0.94 & 3.00 & 1.00 & 3.00 & 1.00 & 3.00 & 1.00 & 2.94 & 0.94 & 3.00 & 1.00 \\
                     & 200 & 200 & 3.00 & 1.00 & 3.00 & 1.00 & 3.00 & 1.00 & 3.00 & 1.00 & 3.00 & 1.00 & 3.00 & 1.00 \\
                     \midrule
\multirow{9}{*}{$c=7$} & 60  & 60  & 2.39 & 0.48 & 2.58 & 0.64 & 2.52 & 0.52 & 2.79 & 0.82 & 2.45 & 0.48 & 2.90 & 0.90 \\
                     & 60  & 100 & 2.77 & 0.65 & 2.88 & 0.92 & 2.55 & 0.61 & 2.81 & 0.88 & 2.45 & 0.45 & 2.94 & 0.94 \\
                     & 60  & 200 & 2.81 & 0.81 & 2.88 & 0.88 & 2.68 & 0.68 & 2.87 & 0.89 & 2.77 & 0.77 & 2.94 & 0.94 \\
                     & 100 & 60  & 2.87 & 0.74 & 2.80 & 0.83 & 2.68 & 0.71 & 2.90 & 0.90 & 2.45 & 0.48 & 2.84 & 0.85 \\
                     & 100 & 100 & 2.82 & 0.82 & 3.00 & 1.00 & 2.82 & 0.82 & 3.00 & 1.00 & 2.82 & 0.88 & 3.00 & 1.00 \\
                     & 100 & 200 & 3.00 & 1.00 & 3.00 & 1.00 & 3.00 & 1.00 & 3.00 & 1.00 & 2.94 & 0.94 & 3.00 & 1.00 \\
                     & 200 & 60  & 2.81 & 0.81 & 2.88 & 0.88 & 2.77 & 0.77 & 2.95 & 0.95 & 2.55 & 0.65 & 2.98 & 0.98 \\
                     & 200 & 100 & 2.94 & 0.94 & 3.00 & 1.00 & 3.00 & 1.00 & 3.00 & 1.00 & 2.94 & 0.94 & 3.00 & 1.00 \\
                     & 200 & 200 & 3.00 & 1.00 & 3.00 & 1.00 & 3.00 & 1.00 & 3.00 & 1.00 & 3.00 & 1.00 & 3.00 & 1.00 \\
				\bottomrule
		\end{tabular}}		
\begin{tablenotes}
\item Simulation results over 500 repetitions. The DGP considered in this table: $X_{it}=\sum_{j=1}^3{\lambda}_{ji}f_{jt}+ e_{it}$, ${\lambda}_{ji},f_{jt}\sim i.i.d \; \mathcal{N}(0,1), e_{it}\sim t_\nu$, with $\nu=1$ (columns 4-7), $\nu=2$ (columns 8-11), and $\nu=3$ (columns 12-15). For each error specification, $\bar{\hat{r}}_{\mathrm{rank}}$ and $\mathit{Freq}_1$ denote respectively the average number of factors selected and the frequency of selecting the correct number of factors by rank estimator $\hat{r}_{\mathrm{rank}}$, and  $\bar{\hat{r}}_{\mathrm{IC}}$, $\mathit{Freq}_2$ denote the corresponding estimates for the IC-based estimator $\hat{r}_{\mathrm{IC}}$. The estimates are obtained with the bandwidth $h=c\cdot L_{NT}^{-1/7}$, for $c=3$ (upper panel), $c=5$ (middle panel), $c=7$ (lower panel).
\end{tablenotes}
\end{table}

Table \ref{table4}
presents the results for model selection using the rank estimator $\hat{r}_{\mathrm{rank}}$ and the IC-based estimator $\hat{r}_{\mathrm{IC}}$ under S1, while those under S2-S3 are relegated to the Supplementary Material.
In these tables, $\bar{\hat{r}}_{\mathrm{rank}}$ and $\mathit{Freq}_1$ denote the average number of factors selected and the frequency of selecting the correct number of factors by  $\hat{r}_{\mathrm{rank}}$, respectively. $\bar{\hat{r}}_{\mathrm{IC}}$ and $\mathit{Freq}_2$ denote the corresponding estimates for ${\hat{r}}_{\mathrm{IC}}$.
The main findings from Table \ref{table4}
are summarized as follows.
First, although both $\hat{r}_{\mathrm{IC}}$ and  $\hat{r}_{\mathrm{rank}}$ tend to underestimate the factor number with small sample sizes, the probability of correct estimation increases along with $N$ and $T$.  In particular, both estimators can correctly estimate the true factor number when both $N$ and $T$ are as large as 200. Second, $\hat{r}_{\mathrm{IC}}$ is more precise with a smaller bandwidth across all scenarios, while $\hat{r}_{\mathrm{rank}}$ seems insensitive to the bandwidth setting. Third, $\hat{r}_{\mathrm{IC}}$ tends to outperform $\hat{r}_{\mathrm{rank}}$ in almost all cases. Fourth,  $\hat{r}_{\mathrm{IC}}$ and $\hat{r}_{\mathrm{rank}}$ can provide accurate estimates as long as $\min\{N,T\}\geq 100$, while $\hat{r}_{\mathrm{rank}}$ achieves high accuracy in most settings even for sample sizes as small as $(N,T)=(60,60)$.

We finally evaluate the finite sample distributional behavior of the MFA estimators and check how closely the asymptotic distributions derived in Theorem \ref{Thm4} approximate the finite sample distributions. Following \citet{Bai2003}, we consider the following factor specification:
\begin{equation*}
X_{it}=\lambda_if_t+e_{it},
\end{equation*}
where $\lambda_i, f_t, e_{it}$ are $i.i.d$ $\mathcal{N}(0,1)$ for all $i,t$. The number of factors is set to $1$, and $(N,T)=(60,60),(100,100),(200,200)$. According to Assumption \ref{Ass9}, we set the bandwidth as $h=c \cdot T^{-1/12}, c\in\{3,5,7\}$.

Figure S.1 in the Supplementary Material 
displays standard normal density curve and the histogram for the estimates
\begin{equation*}
\sqrt{Nh^3}[\hat{\bm{\Psi}}_t^{-1}\hat{\bm{\Omega}}_t\hat{\bm{\Psi}}^{-1}_t]^{-1/2}(\hat{f}_t-f_{0t}),
\end{equation*}
at $t=T/2$ obtained from $1000$ repetitions.\footnote{We choose the sign of $\hat{f}_t$ such that $\hat{\bf{S}}=\mathbb{I}_{r_0}$}
It can be seen that the standard normal density curve provides good approximations to the histograms in all cases. Therefore, the finite sample distribution of the standardized factor estimator is well approximated by standard normal distribution.

We further construct $95\%$ confidence intervals for the true factor process $\{f_{0t}, t=1,\cdots, 20\}$ 
in Figure S.2.
The confidence intervals for the remaining time points are not presented for the sake of clarity. The solid curve in the middle of each plot represents the true factor process, and the dashed curves signify the estimated confidence intervals. It can be observed that, the confidence intervals are hardly sensitive to the choice of bandwidth and contain the true factors in most cases. In addition, they become narrower as $N, T$ grow.

\section{Empirical applications}\label{Sec5}
This section examines the prediction of U.S. inflation rate
as studied by \citet{CDG2021}, and explores the usefulness of the MFA factors as predictors. Specifically, we extract a set of common factors by MFA, PCA and QFA from a large panel of macroeconomic data, and
evaluate the predictive power of each set of factors in forecasting the inflation rate.
The data set we use
is the FRED-QD dataset, which includes quarterly data for 211 U.S. macroeconomic time series from 1960Q1 to 2019Q2 ($N=211, T=238$),  and is available from the Federal Reserve Economic Data database at \url{https://research.stlouisfed.org/econ/mccracken/fred-databases/}.
Each series is first transformed to be stationary using MATLAB codes available on the FRED-QD data website. The transformed series is then demeaned and standardized to have zero mean and unit variance.

Let $y_{t}$ denote the realized value of U.S. inflation rate
at period $t$.\footnote{We use CPI to measure the inflation rate. Let $x_t$ denote CPI at period $t$, then $y_t=\ln{x_t}-\ln{x_{t-4}}$.}
The $s$-step-ahead forecasting model with factor-augmented predictors writes:
\begin{equation}\label{FAR}
y_{t+s}=\alpha+\sum_{j=0}^{p_{\mathrm{max}}}\beta_{j}y_{t-j}+{\bm{\gamma}}^{\prime}{\bf{F}}_t+\epsilon_{t+s},
\end{equation}
where ${\bf{F}}_t$ denotes the latent factors derived from the large data set. Based on the vector of estimated factors $\hat{\bf{F}}_t$, the least squares prediction of $y_{t+s}$ is obtained as:
\begin{equation*}
\hat{y}_{t+s}=\hat{\alpha}+\sum_{j=0}^{\hat{p}}\hat{\beta}_{j}y_{t-j}+\hat{\bm{\gamma}}^{\prime}\hat{\bf{F}}_t,
\end{equation*}
where $\hat{\alpha},\hat{\beta}_{j}, \hat{\bm{\gamma}}$ are least squares estimates of the coefficients, and $\hat{p}$ is the optimal lag length determined by BIC.

For each time point $t$, we extract the MFA factors $\hat{\bf{F}}_{M,t}$, PCA factors $\hat{\bf{F}}_{P,t}$, as well as the QFA factors $\hat{\bf{F}}_{Q,t}^{\tau}$ at the $\tau$-th quantile from the data set. Then consider the specifications of the forecasting model:
(i) $\hat{\bf{F}}_t=0$, which is the benchmark autoregressive (AR) model;
(ii) AR+$\hat{\bf{F}}_{M,t}$; 
(iii) AR+$\hat{\bf{F}}_{P,t}$;
(iv) AR+$\hat{\bf{F}}^{0.5}_{Q,t}$;
(v) AR+$\hat{\bf{F}}_{P,t}+\hat{\bf{F}}^{\tau_1}_{Q,t}, \tau_1=0.9$;
(vi) AR+$\hat{\bf{F}}_{P,t}+\hat{\bf{F}}^{\tau_2}_{Q,t}, \tau_2=0.99$.
The model specifications in (v)-(vi) are taken from \citet{CDG2021}.

A rolling window of $120$ quarters is adopted to estimate the coefficients and generate the rolling window forecasts. Within each window, the number of MFA, PCA, and QFA factors is determined by our IC-based estimator $\hat{r}_{\mathrm{IC}}$, the $PC_{p1}$ criterion of \citet{BN2002}, and the rank-minimization estimator proposed by \citet{CDG2021}, respectively. The maximum number of each kind of factor is set to $8$, and the maximum lag length is limited to 4, i.e.,  $p_{\mathrm{max}}=3$. Further, for the ``AR+$\hat{\bf{F}}_{M,t}$" model, we set the bandwidth $h=c\cdot 120^{-1/7}$, and consider $c\in\{3,5,7\}$, as done in the simulations. Following \citet{CDG2021}, the initial estimation period spans from 1960Q1 to 1989Q4 (120 quarters), and the forecast evaluation period is divided into two subperiods: the great moderation pre-crisis (1990Q1 to 2007Q2) and the financial crisis/recovery (2007Q3 to 2019Q2). The mean squared errors (MSEs) of forecasts from these models are calculated, and their relative MSEs (RMSEs) to that of the
AR model are reported in Table \ref{table7}
 for the full evaluation period and two subperiods.

\begin{table}[t]
\centering
\caption{ {RMSEs} of models (ii)-(vi) for inflation rate forecasting. }\label{table7}
\resizebox{1\linewidth}{!}{
\begin{tabular}{llccccccc}
\toprule
\multirow{2}{*}{s} & \multirow{2}{*}{Period} & \multicolumn{3}{c}{$\hat{\bm{F}}_{M,t}$}   & \multirow{2}{*}{$\hat{\bm{F}}_{P,t}$} & \multirow{2}{*}{$\hat{\bm{F}}^{0.5}_{Q,t}$} & \multicolumn{2}{c}
{$\hat{\bm{F}}_{P,t}+\hat{\bm{F}}^{\tau}_{Q,t}$} \\
\cmidrule(r){3-5}  \cmidrule(r){8-9}
                   &                         & $c=3$    & $c=5$    & $c=7$    &                     &                     & $\tau_1$         & $\tau_2$        \\
                   \midrule
                    & Pre-crisis      & 0.8742                        & 0.8656                        & 0.8699                        & 0.9739 & {\color{red} 0.8579} & 0.9839                        & 1.0012 \\
                    & Crisis/Recovery & {\color{red} 0.9385} & 0.9417                        & 0.9404                        & 0.9892 & 1.0693                        & 1.0483                        & 1.0354 \\
\multirow{-3}{*}{1} & Full            & 0.9173                        & {\color{red} 0.9167} & 0.9172                        & 0.9842 & 0.9997                        & 1.0271                        & 1.0241 \\
\midrule
                    & Pre-crisis      & 0.8220                        & {\color{red} 0.8219} & 0.8311                        & 1.0072 & 0.9421                        & 1.0083                        & 1.0323 \\
                    & Crisis/Recovery & 0.9355                        & {\color{red} 0.8935} & 0.9225                        & 0.9055 & 1.0853                        & 0.9126                        & 0.9526 \\
\multirow{-3}{*}{2} & Full            & 0.8998                        & {\color{red} 0.8710} & 0.8938                        & 0.9375 & 1.0402                        & 0.9427                        & 0.9777 \\
\midrule
                    & Pre-crisis      & {\color{red} 0.7588} & 0.7862                        & 0.7936                        & 1.0500 & 0.8547                        & 1.1006                        & 1.1100 \\
                    & Crisis/Recovery & 0.9585                        & 0.9169                        & 0.9119                        & 0.8666 & 0.9242                        & {\color{red} 0.8191} & 0.8456 \\
\multirow{-3}{*}{3} & Full            & 0.8868                        & 0.8699                        & {\color{red} 0.8694} & 0.9325 & 0.8992                        & 0.9202                        & 0.9406 \\
\midrule
                    & Pre-crisis      & 0.7770                        & 0.7795                        & {\color{red} 0.7744} & 1.1226 & 0.9073                        & 1.2221                        & 1.2438 \\
                    & Crisis/Recovery & 0.9619                        & 0.9515                        & 0.9659                        & 0.9088 & 1.0433                        & {\color{red} 0.8737} & 0.9579 \\
\multirow{-3}{*}{4} & Full            & 0.8792                        & {\color{red} 0.8745} & 0.8802                        & 1.0045 & 0.9824                        & 1.0296                        & 1.0859   \\
                   \bottomrule
\end{tabular}}
\begin{tablenotes}
\item This table reports the relative MSEs (RMSEs) of five ``AR $+\hat{\bf{F}}_{t}$" models compared with the benchmark AR model for forecasting one-quarter-ahead ($s=1$) to four-quarter-ahead ($s=4$) inflation rate. The forms of $\hat{\bf{F}}_{t}$ are specified in the top row, where $\hat{\bf{F}}_{M,t}, \hat{\bf{F}}_{P,t}, $ and $\hat{\bf{F}}^{\tau}_{Q,t}$ denote the estimates of the factors at period $t$ by MFA, PCA \citep{BN2002}, and QFA at $\tau$-th quantile \citep{CDG2021}, respectively. For the ``AR+$\hat{\bf{F}}_{M,t}$" model, $\hat{\bf{F}}_{M,t}$ is obtained with the bandwidth $h=c \cdot 120^{-1/7}$ for $c\in\{3,5,7\}$. The out-sample-forecasting is generated using rolling window with the most recent 120 observations. The full forecasting evaluation period is from 1990Q1to 2019Q2, the pre-crisis subperiod is from 1990Q1 to 2007Q2, and the crisis/recovery subperiod is from 2007Q3 to 2019Q2.   
\end{tablenotes}
\end{table}
The main findings in Table \ref{table7} are summarized as follows. First, incorporating factors as predictors can improve the forecast accuracy of the benchmark AR model, and the ``AR+$\hat{\bf{F}}_{M,t}$" model enjoys the best precision. Second, the ``AR+$\hat{\bf{F}}_{M,t}$" model produces more accurate predictions than the AR model for both subperiods, which demonstrates the robustness of the ``AR+$\hat{\bf{F}}_{M,t}$" model even in crisis scenarios. Third, the performance of the ``AR+$\hat{\bf{F}}_{M,t}$" model is only slightly affected by the choice of bandwidth, and achieves optimal when the bandwidth constant $c$ is set as $5$.

To overcome the limitation of point forecasts that little is known regarding its accuracy,
we next consider density forecasts of inflation rate
based on \eqref{FAR}. Following \citet{CDG2021}, we obtain these density forecasts using quantile regression (QR). Specifically, we first predict the conditional quantiles of the target variable by
\begin{equation*}
\hat{q}_{\tau,t+s}=\hat{\alpha}_{\tau}+\sum_{j=1}^p\hat{\beta}_{\tau,j}y_{t-j}+\hat{\bm{\gamma}}^{\prime}_{\tau}\hat{\bf{F}}_{t},
\end{equation*}
for $\tau\in\{0.05, 0.25, 0.75,0.95\}$, where $\hat{\alpha}_{\tau},\hat{\beta}_{\tau,j}, \hat{\bm{\gamma}}_{\tau}$ are estimated by running QR of $y_{t+s}$ on $[1,y_t,\cdots, y_{t-p}, \hat{\bf{F}}_{t}]$. Given the predicted qunatiles $[\hat{q}_{0.05,t+s},\hat{q}_{0.25,t+s},\hat{q}_{0.75,t+s},\hat{q}_{0.95,t+s}]$, the predicted density of $y_{t+s}$ is constructed as the density of a skewed $t$-distribution by matching the predicted quantiles. Finally, the predictive score, i.e., the predicted density evaluated at the observed value of $y_{t+s}$, which measures the accuracy of density forecasts is obtained.\footnote{We refer readers to \citet{ABG2019} for further details of doing density forecasts, and to \citet{AC2003} for the definition and properties of the skewed $t$-distribution.} Note that higher scores indicate more accurate predictions. In line with before, we consider factor specifications (i)-(vi), and set $h=c\cdot 120^{-1/7}, c\in \{3,5,7\}$, $p=3$. Additionally, a rolling window of the most recent 120 observations is adopted to generate the out-of-sample density forecasts, and the evaluation period is from 1990Q1 to 2019Q2.

\begin{table}[t]
\centering
\caption{{Average predictive scores} of models (i)-(vi) for inflation rate forecasting. }\label{table9}
\resizebox{1\linewidth}{!}{
\begin{tabular}{llcccccccc}
\toprule
\multirow{2}{*}{s} & \multirow{2}{*}{Period} & \multicolumn{3}{c}{$\hat{\bm{F}}_{M,t}$}   & \multirow{2}{*}{$\hat{\bm{F}}_{P,t}$} & \multirow{2}{*}{$\hat{\bm{F}}^{0.5}_{Q,t}$} & \multicolumn{2}{c}
{$\hat{\bm{F}}_{P,t}+\hat{\bm{F}}^{\tau}_{Q,t}$} & \multirow{2}{*}{AR}\\
\cmidrule(r){3-5}  \cmidrule(r){8-9}
                   &                         & $c=3$    & $c=5$    & $c=7$    &                     &                     & $\tau_1$         & $\tau_2$ &       \\
                   \midrule
                    & Pre-crisis      & 0.7356                        & {\color{red} 0.7887} & 0.7638                        & 0.6652 & 0.7548                        & 0.6822 & 0.6747                        & 0.6728 \\
                    & Crisis/Recovery & 0.5576                        & {\color{red} 0.5613} & 0.5033                        & 0.4464 & 0.5351                        & 0.5150 & 0.5174                        & 0.5099 \\
\multirow{-3}{*}{1} & Full            & 0.6632                        & {\color{red} 0.6962} & 0.6578                        & 0.5762 & 0.6654                        & 0.6142 & 0.6107                        & 0.6065 \\
\midrule
                    & Pre-crisis      & {\color{red} 0.4826} & 0.4455                        & 0.4388                        & 0.3793 & 0.4143                        & 0.4342 & 0.3613                        & 0.4033 \\
                    & Crisis/Recovery & 0.3567                        & 0.3440                        & {\color{red} 0.3980} & 0.3529 & 0.3235                        & 0.3415 & 0.3879                        & 0.2938 \\
\multirow{-3}{*}{2} & Full            & {\color{red} 0.4314} & 0.4042                        & 0.4222                        & 0.3686 & 0.3774                        & 0.3965 & 0.3762                        & 0.3588 \\
\midrule
                    & Pre-crisis      & 0.3094                        & 0.3080                        & {\color{red} 0.3154} & 0.2860 & 0.2699                        & 0.3082 & 0.3020                        & 0.2833 \\
                    & Crisis/Recovery & 0.2555                        & 0.2825                        & 0.2510                        & 0.2796 & 0.2919                        & 0.2633 & {\color{red} 0.3278} & 0.2687 \\
\multirow{-3}{*}{3} & Full            & 0.2875                        & 0.2976                        & 0.2892                        & 0.2834 & 0.2788                        & 0.2899 & {\color{red} 0.3166} & 0.2774 \\
\midrule
                    & Pre-crisis      & 0.2425                        & {\color{red} 0.2509} & 0.2446                        & 0.1919 & 0.2251                        & 0.1759 & 0.1856                        & 0.2425 \\
                    & Crisis/Recovery & 0.2470                        & 0.2519                        & 0.2549                        & 0.2592 & {\color{red} 0.2654} & 0.2230 & 0.2316                        & 0.1966 \\
\multirow{-3}{*}{4} & Full            & 0.2443                        & {\color{red} 0.2513} & 0.2488                        & 0.2193 & 0.2415                        & 0.1951 & 0.2043                        & 0.2238 \\
                   \bottomrule
\end{tabular}}
\begin{tablenotes}
 \item This table gives average predictive scores of different models for forecasting inflation rate. See table III for detailed explanation.   
\end{tablenotes}
\end{table}
Table \ref{table9} presents the average predictive scores of different models for predicting inflation rate over the whole evaluation period and two subperiods.
It can be observed that, the ``AR+$\hat{\bf{F}}_{M.t}$" model is hardly affected by the bandwidth setting, and its average predictive scores surpass those of the other models in the majority of cases. This indicates that the MFA factors provide valuable information for density forecasting of inflation rate.

\section{Conclusions}
This paper proposes a modal factor model to extract factors influencing the conditional mode of the distribution of the observables. An AMEM algorithm is developed to obtain the factor and loading estimators and  two model selection methods are introduced for selecting the number of factors. The asymptotic properties of the proposed estimators are established and numerical results demonstrate the nice finite sample performance of these estimators in both simulations and empirical applications to forecasting macroeconomic variables.

The current paper can be extended in several directions.
First, further investigation is needed on how to select the bandwidth in MFM, and the cross-validation method could be considered following \citet{CGTW2016}.
Second, the conditional cross-sectional independence assumption on the error terms may be further relaxed, as supported by the simulation results. Third, our results can be extended to cover the case where the conditional mode has a non-linear factor representation \citep{Wang2024,MT2023b}. Fourth, it would be interesting to further consider inference on the role of modal factors in factor-augmented models. Fifth, it is possible to extend the static MFM to dynamic MFM by allowing the factor loadings to be time-variant \citep{SW2017} or by including lagged factors \citep{FHLR2005}. These issues involve new technical challenges and deserve separate future efforts.

\section*{Acknowledgments}
The authors would like to thank seminar and conference participants at Beihang University, Nanjing Audit University and Tsinghua University for comments that help improve the paper. The authors thank the partial support from National Natural Science Foundation of China (Grant 72425009, 72073002), the Center for Statistical Science at Peking University, and Key Laboratory of Mathematical Economics and Quantitative Finance (Peking University), Ministry of Education.

\section*{Appendix}\label{appA}
\appendix
This Appendix contains the proofs of Theorems \ref{Thm1}, \ref{Thm2} and \ref{Thm3}, together with some auxiliary lemmas. The proofs of Theorems \ref{Thm4} and \ref{Thm5} are relegated to the Supplementary Material due to space consideration.

Throughout the Appendix, let $K_1,K_2,\cdots$ denote some positive constants that do not depend on $N,T$. The symbol $\lesssim$ means that the left side is bounded by a positive constant times the right side ($\gtrsim$ is defined similarly). For any sequence $\{a_{it}\}_{i=1,t=1}^{N,T}$ and $\varepsilon_{NT}\to 0$,  write $a_{it}=\bar{o}_p(\varepsilon_{NT})$ if $\max_{i\leq N,t\leq T}\|a_{it}\|=o_p(\varepsilon_{NT})$. $\bar{o}(\cdot),\bar{O}(\cdot), \bar{O}_p(\cdot)$  are defined in a similar fashion.
For any random variable $Y$, define the Orlicz norm  $\|Y\|_{{\psi}}$ as
\begin{equation*}
\|Y\|_{{\psi}}=\inf\{C>0:E{\psi}(|Y|/C)\leq 1\},
\end{equation*}
where ${\psi}$ is a non-decreasing, convex function with ${\psi}(0)=0$. In particular, when ${\psi}(X)=e^{x^p}-1$, the norm is written as $\|Y\|_{{\psi}_p}$. Additionally, $\|Y\|_{\infty}=\sup\left\{y: P(|Y|>y)>0\right\}$. For any square matrix $A$, let $\|A\|_{S}$ and $\rho_{j}(A)$ denote its spectral norm and $j$-th largest eigenvalue, respectively. $D(\cdot, \text g, \mathcal{G})$ denotes the packing number of space $\mathcal{G}$ endowed with semimetric $\text g$.

\section{Proof of Theorem \ref{Thm1}}
\begin{lemma}\label{lem1} For any ${\bm{\theta}}\in {\bm{\Theta}}^{r_0}$, let $c_{it}={\bm{\lambda}}_i^{\prime}{\bf{f}}_t, c_{it}^0={\bm{\lambda}}_{0i}^{\prime}{\bf{f}}_{0t}$, and $e_{it}^0=X_{it}-c_{it}^0$, $e_{it}=X_{it}-c_{it}$.
Under Assumptions \ref{Ass3}, \ref{Ass4}-\ref{Ass5}, for $i=1,\cdots,N; t=1,\cdots, T$, we have
\begin{enumerate}[(i)]
\item $E[K_h(e_{it}^0))]=g_{it}(0)+\bar{O}(h^2),E[K_h(e_{it}))]=g_{it}(c_{it}-c_{it}^0)+\bar{O}(h^2)$.
\item $E[K_h^{(1)}(e_{it}^0)]=\bar{O}(h^2),E[K_h^{(1)}(e_{it})]=-g^{(1)}_{it}(c_{it}-c_{it}^0)+\bar{O}(h^2)$.
\item $E[K_h^{(2)}(e_{it}^0)]=g^{(2)}_{it}(0)+\bar{O}(h^2),E[K_h^{(2)}(e_{it})]=g^{(2)}_{it}(c_{it}-c_{it}^0)+\bar{O}(h^2)$.
\item $E[K^{(1)}_h(e_{it}^0)^2]=\bar{O}(h^{-3}),E[K^{(1)}_h(e_{it})^2]=\bar{O}(h^{-3})$.
\end{enumerate}
\end{lemma}

\bigskip

\noindent \textbf{Proof of Lemma \ref{lem1}.}	
The proofs for the above results follow from standard derivations in the analysis of kernel density estimators, and thus are omitted here to save space. Similar results can be found in \citet{KS2012} and \citet{YL2014}.

\bigskip

\begin{lemma}\label{lem6}
Suppose $S_{it}$'s are zero-mean real-valued
processes, independent across $i$, with $\underset{1\leq i\leq N, 1\leq t\leq T}{\sup}\|S_{it}\|_{\infty}\leq b$.
Let $\mathbb{S}_{NT}=\sum_{i=1}^N\sum_{t=1}^TS_{it}$, then for each integer $q\in [1,\frac{T}{2}]$ and each $\epsilon>0$, we have
\begin{equation*}
P(|\mathbb{S}_{NT}|/(NT)\geq \epsilon)\leq 4exp(-\frac{\epsilon^2}{8v^2(\mathbb{S}_{NT})})+22(1+\frac{4b}{\epsilon})^{1/2}Nq\alpha([p]),
\end{equation*}
where
\begin{equation*}
v^2(\mathbb{S}_{NT})=\frac{4{\sigma}^2(\mathbb{S}_{NT})}{(NT)^2}+\frac{bp\epsilon}{NT},
\end{equation*}
with $p=T/(2q)$ and ${\sigma}^2(\mathbb{S}_{NT})=\sum_{j=0}^{2q-1}\sum_{i=1}^NE(([jp]+1-jp)S_{i,([jp]+1)}+S_{i,([jp]+2)}$ $
+\cdots+S_{i,([(j+1)p])}+((j+1)p-[(j+1)p])S_{i,([(j+1)p]+1)})^2.$
\end{lemma}

\bigskip

\noindent \textbf{Proof of Lemma \ref{lem6}.}	
The proof of Lemma \ref{lem6} proceeds in a way similar to that for Theorem 1.3 of \citet{Bosq2012}, and is therefore omitted here.


\bigskip

For any $\bm{\theta} \in \bm{\Theta}^{r_0}$, define $\mathbb{M}^*_{NT}({\bm{\theta}})=\mathbb{M}_{NT}({\bm{\theta}}_0)-\mathbb{M}_{NT}({\bm{\theta}})$, $
 \bar{\mathbb{M}}^*_{NT}({\bm{\theta}})=E[\mathbb{M}^*_{NT}({\bm{\theta}})]$, and $\mathbb{W}_{NT}({\bm{\theta}})=\mathbb{M}^*_{NT}({\bm{\theta}})-\bar{\mathbb{M}}^*_{NT}({\bm{\theta}})$.

 \bigskip

\begin{lemma}\label{lem2}
Under Assumptions \ref{Ass1}-\ref{Ass5}, it holds that $d(\hat{{\bm{\theta}}},{\bm{\theta}}_0)=o_p(1)$, as $N,T \to \infty$.
\end{lemma}

\bigskip

\noindent \textbf{Proof of Lemma \ref{lem2}.}	
Expanding $E[K_h(X_{it}-c_{it})]$ around $c_{it}^0$, by Lemma \ref{lem1} (ii), (iii), we have
\begin{equation} \label{Lem2-eq2}
E[K_h(X_{it}-c_{it}^0)-K_h(X_{it}-{c}_{it})]=({c}_{it}-c_{it}^0)\cdot O(h^2)+(-\frac{1}{2}g_{it}^{(2)}(c_{it}^*-c_{it}^0)+O(h^2))({c}_{it}-c_{it}^0)^2,
\end{equation}
where $c_{it}^*$ is between ${c}_{it}$ and $c_{it}^0$. Assumptions \ref{Ass3} (iii) and (iv) imply that
there exists $d_1>0$ such that for $|u|\leq d_1$
$-g_{it}^{(2)}(u)\geq \underline{g_1}/2>0$.
Then, if $|c_{it}-c_{it}^0|\leq d_1$, the second term on the right-hand side (RHS) of \eqref{Lem2-eq2} satisfies
\begin{equation}\label{Lem2-eq30}
(-\frac{1}{2}g_{it}^{(2)}(c_{it}^*-c_{it}^0)+O(h^2))({c}_{it}-c_{it}^0)^2\geq \frac{\underline{g_1}}{4}\cdot(c_{it}-c_{it}^0)^2(1+o(1))>0.
\end{equation}
By \eqref{Lem2-eq2} and \eqref{Lem2-eq30}, there exists $d>0$ such that for $dh^2<|c_{it}-c_{it}^0|\leq d_1$,
the RHS of \eqref{Lem2-eq2} will be dominated by the second term, and
\begin{equation}\label{Lem2-eq44}
E[K_h(X_{it}-c_{it}^0)-K_h(X_{it}-{c}_{it})]\geq \frac{\underline{g_1}}{8}\cdot(c_{it}-c_{it}^0)^2(1+o(1))>0.
\end{equation}
Additionally, for $|c_{it}-c_{it}^0|>d_1$, by Lemma \ref{lem1} (i) and Assumption \ref{Ass3} (ii), there exists $\underline{g}>0$ such that
\begin{equation}\label{Lem2-eq31}
E[K_h(X_{it}-c_{it}^0)-K_h(X_{it}-{c}_{it})]=g_{it}(0)-g_{it}(c_{it}-c^0_{it})+O(h^2)\geq \underline{g}/2>0.
\end{equation}
Since $c_{it}$ is bounded by Assumption \ref{Ass1}, then by \eqref{Lem2-eq44} and \eqref{Lem2-eq31}, for $|c_{it}-c_{it}^0|>dh^2$,
\begin{equation}\label{Lem2-eq32}
    E[K_h(X_{it}-c_{it}^0)-K_h(X_{it}-{c}_{it})]\gtrsim (c_{it}-c^0_{it})^2.
\end{equation}
Moreover, for $|c_{it}-c_{it}^0|<dh^2$, \eqref{Lem2-eq2} and Assumption \ref{Ass3} (iii) imply
\begin{equation}\label{Lem2-eq33}
    |E[K_h(X_{it}-c_{it}^0)-K_h(X_{it}-{c}_{it})]|\leq K_1h^4.
\end{equation}

Further, for any ${\bm{\theta}}\in {\bm{\Theta}}^{r_0}$, divide ${c}_{it}$ into groups $S_1$ and $S_2$, for which $S_1=\{{c}_{it}, |c_{it}-c_{it}^0|\geq dh^2\}$ and $S_2=\{{c}_{it}, |c_{it}-c_{it}^0|<dh^2\}$. Define $d^2_{S_1}({{\bm{\theta}}},{\bm{\theta}}_0)=\frac{1}{NT}\sum_{{c}_{it}\in S_1}({c}_{it}-c_{it}^0)^2$, $d^2_{S_2}({{\bm{\theta}}},{\bm{\theta}}_0)=\frac{1}{NT}\sum_{{c}_{it}\in S_2}({c}_{it}-c_{it}^0)^2$,
and
\begin{equation*}
\begin{split}
&\bar{\mathbb{M}}_{S_1}^*({{\bm{\theta}}})=\frac{1}{NT}\sum_{{c}_{it}\in S_1}E[K_h(X_{it}-c_{it}^0)-K_h(X_{it}-{c}_{it})],\\
&\bar{\mathbb{M}}_{S_2}^*({{\bm{\theta}}})=\frac{1}{NT}\sum_{{c}_{it}\in S_2}E[K_h(X_{it}-c_{it}^0)-K_h(X_{it}-{c}_{it})].
\end{split}
\end{equation*}
Then, 
$d^2_{S_2}({\bm{\theta}},{\bm{\theta}}_0)<d^2h^4$ by the definition of $S_2$, and $d^2_{S_1}({\bm{\theta}},{\bm{\theta}}_0)\lesssim \bar{\mathbb{M}}^*_{S_1}({\bm{\theta}})$ by \eqref{Lem2-eq32}. Additionally, by \eqref{Lem2-eq33}, it is easy to show that $|\bar{\mathbb{M}}^*_{S_2}({\bm{\theta}})|\leq K_1h^4$, and thus
\begin{equation}\label{Thm1-eq15}
  \underset{{\bm{\theta}}\in {\bm{\Theta}}^{r_0}}{\sup}|\bar{\mathbb{M}}^*_{S_2}({\bm{\theta}})|\leq K_1h^4.
\end{equation}

By the definition of $\hat{{\bm{\theta}}}$, we have $\mathbb{M}^*_{NT}(\hat{{\bm{\theta}}})=\mathbb{M}_{NT}({\bm{\theta}}_0)-\mathbb{M}_{NT}(\hat{{\bm{\theta}}})\leq 0$ or equivalently $\mathbb{W}_{NT}(\hat{{\bm{\theta}}})+\bar{\mathbb{M}}^*_{NT}(\hat{{\bm{\theta}}})=\mathbb{W}_{NT}(\hat{{\bm{\theta}}})+\bar{\mathbb{M}}^*_{S_1}(\hat{{\bm{\theta}}})+\bar{\mathbb{M}}^*_{S_2}(\hat{{\bm{\theta}}})\leq 0$. Since $\bar{\mathbb{M}}^*_{S_1}(\hat{{\bm{\theta}}})>0 $, then
\begin{equation*}\label{Lem2-eq5}
0\leq d^2_{S_1}(\hat{{\bm{\theta}}},{\bm{\theta}}_0)\lesssim \bar{\mathbb{M}}_{S_1}^*(\hat{{\bm{\theta}}})\leq -(\mathbb{W}_{NT}(\hat{{\bm{\theta}}})+\bar{\mathbb{M}}^*_{S_2}(\hat{{\bm{\theta}}}))\leq \underset{{\bm{\theta}}\in{\bm{\Theta}}^{r_0}}{\sup}|\bar{\mathbb{M}}_{S_2}^*({{\bm{\theta}}})|+\underset{{\bm{\theta}}\in{\bm{\Theta}}^{r_0}}{\sup}|\mathbb{W}_{NT}({\bm{\theta}})|.
\end{equation*}
Thus,
to prove $d^2(\hat{{\bm{\theta}}},{\bm{\theta}}_0)=o_p(1)$, we only need to show that $\underset{{\bm{\theta}}\in{\bm{\Theta}}^{r_0}}{\sup}|\mathbb{W}_{NT}({\bm{\theta}})|=o_p(1)$.

To this end, first choose $K_2$ large enough such that $\|{\bm{\lambda}}_i\|,\|{\bf{f}}_t\|\leq K_2$ for all $i,t$ and for any ${\bm{\theta}} \in {\bm{\Theta}}^{r_0}$. Let $B_r(K_2)$ be a Euclidean ball in $\mathbb{R}^{r_0}$ with radius $K_2$, and let ${\bm{\lambda}}_{(1)},\cdots,{\bm{\lambda}}_{(J)}$ and ${\bf{f}}_{(1)},\cdots, {\bf{f}}_{(J)}$  be two maximal sets of points in $B_r(K_2)$ such that $\|{\bm{\lambda}}_{(j)}-{\bm{\lambda}}_{(h)}\|>L^{-1}_{NT},\|{\bf{f}}_{(j)}-{\bf{f}}_{(h)}\|>L^{-1}_{NT}$ for any $j\neq h$.
Then $J$, the packing number of $B_r(K_2)$, is equal to $K_3L_{NT}^{r_0}$. For any ${\bm{\theta}}\in {\bm{\Theta}}^{r_0}$, define ${{\bm{\theta}}^*}=({{{\bm{\lambda}}}}^{*^{\prime}}_1,\cdots, {{{\bm{\lambda}}}}^{*^{\prime}}_N,{{\bf{f}}^*_1}^{\prime},\cdots,{{\bf{f}}^*_T}^{\prime})^{\prime}$, where ${{\bm{\lambda}}}^*_i=\left\{{\bm{\lambda}}_{(j)}: j\leq J, \|{\bm{\lambda}}_{(j)}-{\bm{\lambda}}_i\|\leq L^{-1}_{NT}\right\}, {\bf{f}}^*_t=\left\{{\bf{f}}_{(j)}: j\leq J, \|{\bf{f}}_{(j)}-{\bf{f}}_t\|\leq L^{-1}_{NT}\right\}$. Note that,
by Assumption \ref{Ass4} (v), $|K_h(X_{it}-{\bm{\lambda}}_i^{*^{\prime}}{\bf{f}}_t^*)-K_h(X_{it}-{\bm{\lambda}}_{i}^{\prime}{\bf{f}}_{t})|\leq c_1|{\bm{\lambda}}_i^{*^{\prime}}{\bf{f}}_t^*-{\bm{\lambda}}_{i}^{\prime}{\bf{f}}_{t}|/h^2\leq c_1(\|{\bm{\lambda}}_i\|\|{\bf{f}}_t-{\bf{f}}_t^*\|+\|{\bf{f}}_t^*\|\|{\bm{\lambda}}_i-{\bm{\lambda}}_i^*\|)/h^2 \leq 2c_1K_2/(L_{NT}h^2)$. Thus,
\begin{equation}\label{Thm1-eq7}
\underset{{\bm{\theta}}\in {\bm{\Theta}}^{r_0}}{\sup} |\mathbb{W}_{NT}({{\bm{\theta}}})-\mathbb{W}_{NT}({\bm{\theta}}^*)|\leq 4c_1K_2/(L_{NT}h^2).
\end{equation}

Second, let $c_{it}^*={\bm{\lambda}}^{*^\prime}_i{\bf{f}}^*_t$, $w_{it}=K_{h}(X_{it}-c_{it}^0)-K_{h}(X_{it}-c_{it}^*)$,  $S_{it}=w_{it}-E(w_{it})$, and $\mathbb{S}_{NT}=\sum_{i=1}^N\sum_{t=1}^TNThS_{it}$. Then, by Lemma \ref{lem6}, we have
\begin{align}\label{Lem2-eq12}
{\sigma}^2(\mathbb{S}_{NT})&\leq (NTh)^2\cdot [\sum_{t=1}^T\sum_{i=1}^NE(S^2_{it})+2\sum_{t=1}^{T-1}\sum_{l=t+1}^{t+p}\sum_{i=1}^N|E(S_{it}S_{il})|].
\end{align}
Note that $w_{it}=K_{h}^{(1)}(X_{it}-c_{it}^{**})(c_{it}^*-c_{it}^0)$, where $c_{it}^{**}$ is between $c_{it}^*$ and $c_{it}^0$. By Lemma \ref{lem1} (ii), (iv), we have $E(S^2_{it})=Var(w_{it})\leq K_4h^{-3}(c^*_{it}-c_{it}^0)^2$. By Assumption \ref{Ass6}, \ref{Ass4} (viii), and Lemma \ref{lem1} (ii), it is easy to show that $|E(S_{it}S_{il})|=|Cov(w_{it},w_{il})|\leq K_5/h^2\cdot |c^*_{it}-c_{it}^0|\cdot |c^*_{il}-c_{il}^0|$. Further, let $p=T^{1/\gamma}$, then $ph=o(1)$ by Assumption \ref{Ass5}. Thus, by \eqref{Lem2-eq12}, we have
\begin{equation}\label{Lem2-eq45}
{\sigma}^2(\mathbb{S}_{NT})\leq K_6(NT)^2h^{-1}\sum_{t=1}^T\sum_{i=1}^N{(c^*_{it}-c_{it}^0)^2}.
\end{equation}
Moreover, by Assumption \ref{Ass4} (iv), $|w_{it}|\leq 2c_0/h$ and $|S_{it}|\leq 4c_0/h$. Then, it follows from Lemma \ref{lem6} and \eqref{Lem2-eq45} that, for any $\epsilon>0$,
\begin{equation}\label{Lem2-eq16}
P(\frac{|\mathbb{S}_{NT}|}{NT}>\epsilon)\leq 4\exp{(-\frac{\epsilon^2}{8v_1^2(\mathbb{S}_{NT})})}+
11(1+\frac{4c_0NT}{\epsilon})^{1/2}T^{1-1/\gamma}N\alpha([T^{1/\gamma}]),
\end{equation}
where we have used the fact that
\begin{equation}\label{Lem2-eq42}
v^2(\mathbb{S}_{NT})\leq 4K_6NTh^{-1}d^2({\bm{\theta}}^*,{\bm{\theta}}_0)+4c_0T^{1/\gamma}\epsilon := v_1^2(\mathbb{S}_{NT}).
\end{equation}
Assumptions \ref{Ass2} (ii) and \ref{Ass5} imply the second term on the RHS of \eqref{Lem2-eq16} vanishes as $T\to \infty$. Hence, the RHS of \eqref{Lem2-eq16} will be dominated by the first term and there exists $K_7$ such that
\begin{equation}\label{Lem2-eq34}
\begin{split}
P(|\mathbb{S}_{NT}|/(NT)>\epsilon)\leq K_7\exp{(-\frac{\epsilon^2}{8v_1^2(\mathbb{S}_{NT})})}.
\end{split}
\end{equation}

Further, let $\mathbb{W}^*_{NT}({\bm{\theta}}^*)=NTh\mathbb{W}_{NT}({\bm{\theta}}^*)$, then $\mathbb{W}^*_{NT}({\bm{\theta}}^*)=\mathbb{S}_{NT}/(NT)$.  Since $\bm{{\theta}}^*$ can take at most $J^{N+T}\lesssim L_{NT}^{r_0(N+T)}$ different values, and $d({\bm{\theta}}^*,{\bm{\theta}}_0)$ is bounded,
then it follows from \eqref{Lem2-eq42}, \eqref{Lem2-eq34} and  Lemma 2.2.10 of \citet{VW1996} that
\begin{align*}
\|\underset{{\bm{\theta}}\in {\bm{\Theta}}^{r_0}}{\sup}|\mathbb{W}^*_{NT}({{\bm{\theta}^*}})|\|_{{\psi}_1}
\lesssim \sqrt{(N+T)}\sqrt{\log{(L_{NT})}}\cdot \sqrt{NTh^{-1}}+(N+T)\cdot \log{(L_{NT})}\cdot T^{1/\gamma}.
\end{align*}
Therefore, it follows that
\begin{align}\label{Lem2-eq35}
&E[\underset{{\bm{\theta}}\in {\bm{\Theta}}^{r_0}}{\sup}|\mathbb{W}_{NT}({{\bm{\theta}^*}})|]\leq \|\underset{{\bm{\theta}}\in {\bm{\Theta}}^{r_0}}{\sup}|\mathbb{W}_{NT}({{\bm{\theta}^*}})|\|_{{\psi}_1}\nonumber\\
\lesssim &\sqrt{(N+T)}\sqrt{\log{(L_{NT})}}/\sqrt{NTh^3}+(N+T)\cdot \log{(L_{NT})}\cdot T^{1/\gamma}/(NTh)\nonumber\\
\lesssim &\sqrt{\frac{\log{L_{NT}}}{L_{NT}h^3}}+\frac{\log{(L_{NT})}T^{1/\gamma}}{L_{NT}h}.
\end{align}

Finally, by Markov's inequality, for any $\delta>0$,
\begin{equation*}
    \begin{split}
P[\underset{{\bm{\theta}}\in{\bm{\Theta}}^{r_0}}{\sup}|\mathbb{W}_{NT}({\bm{\theta}})|> \delta]&\leq P[\underset{{\bm{\theta}}\in{\bm{\Theta}}^{r_0}}{\sup}|\mathbb{W}_{NT}({{\bm{\theta}^*}})|> \frac{\delta}{2}]+P[\underset{{\bm{\theta}}\in{\bm{\Theta}}^{r_0}}{\sup}|\mathbb{W}_{NT}({\bm{\theta}})-\mathbb{W}_{NT}({{\bm{\theta}^*}})|>\frac{\delta}{2}]\\
&\leq \frac{2}{\delta}\cdot E[\underset{{\bm{\theta}}\in{\bm{\Theta}}^{r_0}}{\sup}|\mathbb{W}_{NT}({{\bm{\theta}^*}})|]+P[\underset{{\bm{\theta}}\in{\bm{\Theta}}^{r_0}}{\sup}|\mathbb{W}_{NT}({\bm{\theta}})-\mathbb{W}_{NT}({{\bm{\theta}^*}})|>\frac{\delta}{2}].
\end{split}
\end{equation*}
It then follows from \eqref{Thm1-eq7}, \eqref{Lem2-eq35} and Assumption \ref{Ass5} that the above term is $o(1)$, and thus  $\underset{{\bm{\theta}}\in{\bm{\Theta}}^{r_0}}{\sup}|\mathbb{W}_{NT}({\bm{\theta}})|=o_p(1)$. This completes the proof.

\bigskip



\begin{lemma}\label{lem4}
Under Assumption \ref{Ass1}, for sufficiently small $\delta$ and any ${\bm{\theta}} \in {\bm{\Theta}}^{r_0}(\delta)$, we have
\begin{equation*}\label{L2}
\|{\bm{{\bm{\Lambda}}}}-{\bm{{\bm{\Lambda}}}}_0 {\bf{S}}\|/\sqrt{N}\lesssim \delta,\qquad \|{\bf{F}}-{\bf{F}}_0 {\bf{S}}\|/\sqrt{T}\lesssim \delta,
\end{equation*}
where $\bf{S}$=$\operatorname{sgn}({\bf{F}}^{\prime}{\bf{F}}_0/T)$, and ${\bm{\Theta}}^{r_0}(\delta)=\{{\bm{\theta}}\in {\bm{\Theta}}^{r_0}, d({\bm{\theta}},{\bm{\theta}}_0)\leq \delta\}, \delta>0$.
\end{lemma}

\bigskip

\noindent\textbf{Proof of Lemma \ref{lem4}.}	
The proof is similar to that of Lemma 2 of \citet{CDG2021}.


\bigskip

\begin{lemma}\label{lem3} Under Assumptions \ref{Ass1}-\ref{Ass5}, for sufficiently small $\delta$, we have
\begin{equation*}
E[\underset{{\bm{\theta}}\in {\bm{\Theta}}^{r_0}(\delta)}{\sup}|\mathbb{W}_{NT}({\bm{\theta}})|]\lesssim\frac{\delta}{\sqrt{L_{NT}h^3}}+\frac{\log{L_{NT}}}{L_{NT}h^2}.
\end{equation*}
\end{lemma}

\bigskip

\noindent \textbf{Proof of Lemma \ref{lem3}.}	
 Note first that $d({\bm{\theta}}^*,{\bm{\theta}}^0)\leq \sqrt{2}(d({\bm{\theta}}^*,{\bm{\theta}})+d({\bm{\theta}},{\bm{\theta}}^0))$. Further, by the definition of ${\bm{\theta}}^*$ in the proof of Lemma \ref{lem2}, it holds that $d({\bm{\theta}}^*,{\bm{\theta}})\leq 2K_1/L_{NT}$. Then if ${\bm{\theta}}\in {\bm{\Theta}}^{r_0}(\delta)$, we have ${\bm{\theta}}^*\in {\bm{\Theta}}^{r_0}(\delta^*)$ for $\delta^*=\sqrt{2}(\delta+2K_1/L_{NT})$. Hence, it follows that
\begin{align}\label{Lem3-eq15}
\underset{{\bm{\theta}}\in {\bm{\Theta}}^{r_0}(\delta)}{\sup}|\mathbb{W}_{NT}({\bm{\theta}})|&\leq \underset{{\bm{\theta}}\in {\bm{\Theta}}^{r_0}(\delta)}{\sup}|\mathbb{W}_{NT}({\bm{\theta}})-\mathbb{W}_{NT}({\bm{\theta}}^*)|+\underset{{\bm{\theta}}^*\in {\bm{\Theta}}^{r_0}(\delta^*)}{\sup}|\mathbb{W}_{NT}({\bm{\theta}}^*)|.
\end{align}

We now establish the bound for $E[\underset{{\bm{\theta}^*}\in {\bm{\Theta}}^{r_0}(\delta)}{\sup}|\mathbb{W}_{NT}({\bm{\theta}}^*)|]$.
First, let $a=4c_0T^{1/\gamma}$, $b({\bm{\theta}}^*)=4K_6NTh^{-1}d^2({\bm{\theta}}^*,{\bm{\theta}}_0)$, and $c({\bm{\theta}}^*)=b({\bm{\theta}}^*)/a$. By \eqref{Lem2-eq42}, \eqref{Lem2-eq34} and Lemma 2.2.10 of \citet{VW1996}, it holds that
\begin{align}\label{Lem3-eq5}
E[\underset{{\bm{\theta}}^*\in{\bm{\Theta}}^{r_0}(\delta)}{\sup}|\mathbb{W}^*_{NT}({\bm{\theta}}^*)|]\lesssim& \|\underset{{\bm{\theta}^*}\in{\bm{\Theta}}^{r_0}(\delta)}{\sup}|\mathbb{W}^*_{NT}({\bm{\theta}}^*)|\|_{{\psi}_1}\nonumber\\
\lesssim & \|\underset{{\bm{\theta}}^*\in{\bm{\Theta}}^{r_0}(\delta)}{\sup}|\mathbb{W}^*_{NT}({\bm{\theta}}^*)|\mathbf{1}\{\mathbb{W}^*_{NT}({\bm{\theta}}^*)> c({\bm{\theta}}^*)\}\|_{\psi_1}\nonumber\\
+& \|\underset{{\bm{\theta}}^*\in{\bm{\Theta}}^{r_0}(\delta)}{\sup}|\mathbb{W}^*_{NT}({\bm{\theta}}^*)|\mathbf{1}\{\mathbb{W}^*_{NT}({\bm{\theta}}^*)\leq  c({\bm{\theta}}^*)\}\|_{\psi_2}.
\end{align}
Next, by \eqref{Lem2-eq42}, \eqref{Lem2-eq34} and Lemma 2.2.1 of \citet{VW1996}, we have
\begin{equation}\label{Lem3-eq13}
\|\mathbb{W}^*_{NT}({\bm{\theta}}^*) \mathbf{1}\{\mathbb{W}^*_{NT}({\bm{\theta}}^*)> c({\bm{\theta}}^*)\}\|_{{\psi}_1}\lesssim a\lesssim T^{1/\gamma},
\end{equation}
and
\begin{equation}\label{Lem3-eq14}
\|\mathbb{W}^*_{NT}({\bm{\theta}}^*) \mathbf{1}\{\mathbb{W}^*_{NT}({\bm{\theta}}^*)\leq  c({\bm{\theta}}^*)\}\|_{{\psi}_2}\lesssim \sqrt{b({\bm{\theta}}^*)}\lesssim \sqrt{NTh^{-1}}d({\bm{\theta}}^*,{\bm{\theta}}_0).
\end{equation}
Since $\bm{\theta}^*\in {\bm{\Theta}^{r_0}}(\delta)$ can take at most $J\lesssim (L_{NT})^{r_0(N+T)}$ values, then it follows from \eqref{Lem3-eq13} and Lemma 2.2.2 of \citet{VW1996} that
\begin{equation}\label{Lem3-eq3}
\|\underset{{\bm{\theta}^*}\in{\bm{\Theta}}^{r_0}(\delta)}{\sup}|\mathbb{W}^*_{NT}({\bm{\theta}}^*)|\mathbf{1}\{\mathbb{W}^*_{NT}({\bm{\theta}}^*)> c({\bm{\theta}}^*)\}\|_{{\psi}_1}\lesssim (N+T) \log{L_{NT}}\cdot {T^{1/\gamma}}.
\end{equation}
Additionally, by \eqref{Lem3-eq14} and Theorem 2.2.4 of \citet{VW1996}, we have
\begin{equation}\label{Lem3-eq12}
\|\underset{{\bm{\theta}}^*\in{\bm{\Theta}}^{r_0}(\delta)}{\sup}|\mathbb{W}^*_{NT}({\bm{\theta}}^*)|/\sqrt{NTh^{-1}}\mathbf{1}\{\mathbb{W}^*_{NT}({\bm{\theta}}^*)\leq c({\bm{\theta}}^*)\}\|_{{\psi}_2}\lesssim \int_{0}^\delta\sqrt{\log{D(\epsilon,d, {\bm{\Theta}}^{r_0}(\delta))}}d\epsilon,
\end{equation}
which is $O(\sqrt{N+T}\delta)$ by Lemma 3 of \citet{CDG2021}.
Then, it follows from \eqref{Lem3-eq5}, \eqref{Lem3-eq3} and \eqref{Lem3-eq12} that
\begin{equation*}
E[\underset{{\bm{\theta}}^*\in{\bm{\Theta}}^{r_0}(\delta)}{\sup}|\mathbb{W}^*_{NT}({\bm{\theta}}^*)|]\lesssim \sqrt{N+T}\delta\cdot \sqrt{NTh^{-1}}+(N+T) \log{L_{NT}}\cdot {T^{1/\gamma}}.
\end{equation*}
Note that $\mathbb{W}^*_{NT}({\bm{\theta}}^*)=NTh\mathbb{W}_{NT}({\bm{\theta}}^*)$, then equivalently, we have
\begin{equation}\label{Lem3-eq16}
E[\underset{{\bm{\theta}}^*\in{\bm{\Theta}}^{r_0}(\delta)}{\sup}|\mathbb{W}_{NT}({\bm{\theta}}^*)|\}]\lesssim \frac{\delta}{\sqrt{L_{NT}h^3}}+\frac{(\log{L_{NT})\cdot T^{1/\gamma}}}{L_{NT}h}.
\end{equation}
Finally, combining \eqref{Thm1-eq7}, \eqref{Lem3-eq15} and \eqref{Lem3-eq16}, together with Assumption 6, we obtain
\begin{equation*}
\begin{split}
&E[\underset{{\bm{\theta}}\in {\bm{\Theta}}^{r_0}(\delta)}{\sup}|\mathbb{W}_{NT}({\bm{\theta}})|]\leq \frac{4c_1K_2}{L_{NT}h^2}+E[\underset{{\bm{\theta}}^*\in {\bm{\Theta}}^{r_0}(\delta^*)}{\sup}|\mathbb{W}_{NT}({\bm{\theta}}^*)|]\\
& \lesssim \frac{1}{L_{NT}h^2}+\frac{\delta+L_{NT}^{-1}}{\sqrt{L_{NT}h^3}}+\frac{(\log{L_{NT})\cdot T^{1/\gamma}}}{L_{NT}h}\lesssim \frac{\delta}{\sqrt{L_{NT}h^3}}+\frac{\log{L_{NT}}}{L_{NT}h^2}.
\end{split}
\end{equation*}

\bigskip

\noindent \textbf{Proof of Theorem \ref{Thm1}.}	
 The parameter space ${\bm{\Theta}}^{r_0}$ can be partitioned into shells $S_j=\{{\bm{\theta}}\in {\bm{\Theta}}^{r_0}:2^{j-1}<\delta_{NT}^{-1}\cdot d({\bm{\theta}},{\bm{\theta}}_0)\leq 2^{j}\}$ for $j\geq1$, and $S_0=\{{\bm{\theta}}\in {\bm{\Theta}}^{r_0}:0<\delta_{NT}^{-1}\cdot d({\bm{\theta}},{\bm{\theta}}_0)\leq 1\}$. Note that if $\hat{\bm{\theta}}\in S_j$, then the infimum of the mapping ${\bm{\theta}} \mapsto \mathbb{M}_{NT}^*({\bm{\theta}})=\mathbb{M}_{NT}({\bm{\theta}}_0)-\mathbb{M}_{NT}({\bm{\theta}})$ over $S_j$ is nonpositive by the definition of $\hat{{\bm{\theta}}}$. Then, for any given integer $V$ and each positive constant $\eta$, we have
\begin{align}\label{Thm1-eq5}
P[\delta_{NT}^{-1}\cdot d(\hat{{\bm{\theta}}},{\bm{\theta}}_0)&>2^V]
\leq P[2^V< \delta_{NT}^{-1}\cdot d(\hat{{\bm{\theta}}},{\bm{\theta}}_0)\leq \eta \delta_{NT}^{-1}]+P[d(\hat{{\bm{\theta}}},{\bm{\theta}}_0)>\eta]\nonumber\\
\leq & \sum_{j>V, 2^{j-1}\leq \eta \delta_{NT}^{-1}} P\big[\hat{\bm{\theta}}\in S_j]+P\big[d(\hat{{\bm{\theta}}},{\bm{\theta}}_0)>\eta\big]\nonumber\\
\leq & \sum_{j>V, 2^{j-1}\leq \eta \delta_{NT}^{-1}} P\big[\underset{{\bm{\theta}}\in S_j}{\inf}\mathbb{M}^*_{NT}({\bm{\theta}})<0\big]+P\big[d(\hat{{\bm{\theta}}},{\bm{\theta}}_0)>\eta\big].
\end{align}
By Lemma \ref{lem2}, the second term on RHS of \eqref{Thm1-eq5} is $o(1)$. We then consider the first term.

First, for any ${\bm{\theta}}\in S_j$, where $j\geq 1$,  we have $d^2({\bm{{\theta}}}, {\bm{\theta}}_0)\geq \delta_{NT}^22^{2j-2}$. Since $d^2_{S_2}({\bm{{\theta}}}, {\bm{\theta}}_0)<d^2h^4\leq \delta_{NT}^2d^2$, then for $j$ large enough, we have $d^2_{S_1}({\bm{{\theta}}}, {\bm{\theta}}_0) \geq \delta_{NT}^{2}2^{2j-3}$. Thus,
\begin{equation}\label{Thm1-eq14}
-\bar{\mathbb{M}}^*_{S_1}({\bm{\theta}})\lesssim -d^2_{S_1}({\bm{\theta}},{\bm{\theta}}_0)\leq -\delta_{NT}^22^{2j-3}.
\end{equation}
Since $\mathbb{M}^*_{NT}({\bm{\theta}})=\bar{\mathbb{M}}^*_{S_1}({\bm{\theta}})+\bar{\mathbb{M}}^*_{S_2}({\bm{\theta}})+\mathbb{W}_{NT}({\bm{\theta}})$, then $\underset{{\bm{\theta}}\in S_j}{\inf}\mathbb{M}^*_{NT}({\bm{\theta}})<0$ and \eqref{Thm1-eq14} imply
\begin{equation*}
\underset{{\bm{\theta}}\in S_j}{\inf}(\mathbb{W}_{NT}({\bm{\theta}})+\bar{\mathbb{M}}^*_{S_2}({\bm{\theta}}))\leq -\delta_{NT}^22^{2j-3}.
\end{equation*}
Hence, for $j$ large enough, we have
\begin{equation*}
P\big[\underset{{\bm{\theta}}\in S_j}{\inf}\mathbb{M}^*_{NT}({\bm{\theta}})<0\big]\leq
P\big[\underset{{\bm{\theta}}\in S_j}{\sup}|\mathbb{W}_{NT}({\bm{\theta}})+\bar{\mathbb{M}}^*_{S_2}({\bm{\theta}})|\geq \delta_{NT}^22^{2j-3}].
\end{equation*}
Further, by \eqref{Thm1-eq15}, Lemma \ref{lem3} and Markov's inequality, it holds that
\begin{align*}
P\big[\underset{{\bm{\theta}}\in {S}_j}{\sup}|\mathbb{W}_{NT}({\bm{\theta}})+\bar{\mathbb{M}}^*_{S_2}({\bm{\theta}})|\geq \delta_{NT}^{2}2^{2j-3}]&\lesssim \frac{1}{\delta_{NT}^22^{2j}}\cdot E[\underset{{\bm{\theta}}\in S_j}{\sup}|\mathbb{W}_{NT}({\bm{\theta}})+\bar{\mathbb{M}}^*_{S_2}({\bm{\theta}})|]\\
&\lesssim \frac{1}{\delta_{NT}^22^{2j}}\cdot(\frac{\delta_{NT}2^j}{\sqrt{L_{NT}h^3}}+\frac{\log{L_{NT}}}{L_{NT}h^2}+h^4)\lesssim 2^{-j}.
\end{align*}

Finally, combining above two inequalities, we obtain
\begin{equation*}
 \sum_{j>V, 2^{j-1}\leq \eta \delta_{NT}^{-1}} P\big[\underset{{\bm{\theta}}\in S_j}{\inf}\mathbb{M}^*_{NT}({\bm{\theta}})<0\big]\lesssim \sum_{j>V}2^{-j},
\end{equation*}
which converges to 0 as $V\to \infty$. Thus, by \eqref{Thm1-eq5}, we have $\delta_{NT}^{-1}d(\hat{{\bm{\theta}}},{\bm{\theta}}_0)=O_p(1)$ or $d(\hat{{\bm{\theta}}},{\bm{\theta}}_0)=O_p(\delta_{NT})$. This proves (c) of Theorem \ref{Thm1}. Further, the proofs of (a) and (b) then follow from Lemma \ref{lem4}. 

\bigskip

\section{Proof of Theorem \ref{Thm2}}

\noindent For ${\bm{\theta}}_a=({\bm{\lambda}}_{a1}^{\prime},\cdots,{\bm{\lambda}}_{aN}^{\prime},{\bf{f}}_{a1}^{\prime},\cdots,{\bf{f}}_{aT}^{\prime})^{\prime}\in {\bm{\Theta}}^{a}, {\bm{\theta}}_b=({\bm{\lambda}}_{b1}^{\prime},\cdots,{\bm{\lambda}}_{bN}^{\prime},{\bf{f}}_{b1}^{\prime},\cdots,{\bf{f}}_{bT}^{\prime})^{\prime}\in {\bm{\Theta}}^{b}$, let
\begin{equation*}
d({\bm{\theta}}_a,{\bm{\theta}}_b)=\sqrt{\sum_{i=1}^N\sum_{t=1}^T({\bm{\lambda}}_{ai}^{\prime}{\bf{f}}_{at}-{\bm{\lambda}}_{bi}^{\prime}{\bf{f}}_{bt})^2}.
\end{equation*}
Here ${\bm{\theta}}_a$ and ${\bm{\theta}}_b$ are allowed to belong to different spaces. 
For any $\delta>0$, define ${\bm{\Theta}}^r(\delta)=\{{\bm{\theta}}\in {\bm{\Theta}}^{r}: d({\bm{\theta}}^r,{\bm{\theta}}_0)\leq \delta\}$. For any ${\bm{\theta}}^r\in {\bm{\Theta}}^r$, write $\mathbb{M}^*_{NT}({\bm{\theta}}^r)=\mathbb{M}_{NT}({\bm{\theta}}_0)-\mathbb{M}_{NT}({\bm{\theta}}^r)$, $
 \bar{\mathbb{M}}^*_{NT}({\bm{\theta}}^r)=E[\mathbb{M}^*_{NT}({\bm{\theta}}^r)]$, and $\mathbb{W}_{NT}({\bm{\theta}})=\mathbb{M}^*_{NT}({\bm{\theta}}^r)-\bar{\mathbb{M}}^*_{NT}({\bm{\theta}}^r)$. For $r>r_0$, let ${\bf{F}}^{r,r_0}$ denote the first $r_0$ columns of ${\bf{F}}^{r}$, and ${\bf{F}}^{r,-r_0}$ denote the remaining $r-r_0$ columns. ${\bm{\Lambda}}^{r,r_0}$ and ${\bm{\Lambda}}^{r,-r_0}$ are defined similarly.

\bigskip

\begin{lemma}\label{lem5}
Let ${\bf{S}}=\operatorname{sgn}(({\bf{F}}^{r,r_0})^{\prime}{\bf{F}}_0)/T).$ Suppose that Assumption \ref{Ass1} holds and $r_0<r<\infty$, then for any ${\bm{\theta}}^r\in {\bm{\Theta}}^r(\delta)$ and sufficiently small $\delta$, we have
\begin{equation*}
	\|{\bf{F}}^{r,r_0}-{\bf{F}}_0{\bf{S}}\|/\sqrt{T}\lesssim \delta,\qquad \|{\bm{\Lambda}}^{r,r_0}-{\bm{\Lambda}}_0{\bf{S}}\|/\sqrt{N}\lesssim \delta,\qquad \|{\bm{\Lambda}}^{r,-r_0}\|/\sqrt{N}\lesssim \delta.
  \end{equation*}
\end{lemma}

\noindent \textbf{Proof of Lemma \ref{lem5}.}	
First, let $\tilde{\bf{U}}=[{\bf{U}}, {\bf{0}}_{r_0\times (r-r_0)}]$, where $ {\bf{U}}   \in \mathbb{R}^{r_0\times r_0} $ is a diagonal matrix whose diagonal elements are either 1 or -1. Since ${\bf{F}}_0^{\prime}{\bf{F}}_0={{\bf{F}}^r}^{\prime}{\bf{F}}^{r}=\mathbb{I}_{r_0}$, and $\|{\bm{\Lambda}}_0\|/\sqrt{N}\leq K_8$ by Assumption \ref{Ass1} (ii), then we have
\begin{align}\label{Lem5-eq9}
		\| {\bm{\Lambda}}^r-{\bm{\Lambda}}_0 \tilde{\bf{U}}\|/\sqrt{N}&
        =\| {\bm{\Lambda}}^r {{\bf{F}}^{r}}^{\prime}-{\bm{\Lambda}}_0{\bf{F}}_0^{\prime}+{\bm{\Lambda}}_0{\bf{F}}_0^{\prime}-{\bm{\Lambda}}_0 \tilde{\bf{U}} {{\bf{F}}^{r}}^{\prime}\|/\sqrt{NT}\nonumber\\
		& \leq \| {\bm{\Lambda}}^r {{\bf{F}}^{r}}^{\prime}-{\bm{\Lambda}}_0{\bf{F}}_0^{\prime}\|/\sqrt{NT} + \|{\bm{\Lambda}}_0{\bf{F}}_0^{\prime}-{\bm{\Lambda}}_0 {\bf{U}} {{\bf{F}}^{r,r_0}}^{\prime}\|/\sqrt{NT}\nonumber\\
		&\leq d({\bm{\theta}}^r,{\bm{\theta}}_{0})+K_{8} \|{\bf{F}}^{r,r_0}{\bf{U}}^{\prime}-{\bf{F}}_0\|/\sqrt{T}.
\end{align}

Second, let ${\bf{P}}_{\bf{A}}={\bf{A}}({\bf{A}}^{\prime}{\bf{A}})^{-1}{\bf{A}}^{\prime}$ and ${\bf{M}}_{\bf{A}}=\mathbb{I}-{\bf{P}}_{\bf{A}}$, then ${\bf{P}}_{{\bf{F}}_0}={\bf{F}}_0{\bf{F}}_0^{\prime}/T$, and
\begin{align}\label{Lem5-eq6}
\|{\bf{F}}^{r,r_0}-{\bf{F}}_0{\bf{U}}\|/\sqrt{T}&=\|{\bf{F}}^{r,r_0}-{\bf{F}}_0({\bf{F}}_0^{\prime}{\bf{F}}^{r,r_0}/T)+{\bf{F}}_0({\bf{F}}_0^{\prime}{\bf{F}}^{r,r_0}/T)-{\bf{F}}_0{\bf{U}}\|/\sqrt{T}\nonumber\\
&\leq \|{\bf{M}}_{{\bf{F}}_0}{\bf{F}}^{r,r_0}\|/\sqrt{T}+\|{\bf{F}}_0^{\prime}{\bf{F}}^{r,r_0}/T-{\bf{U}}\|.
\end{align}

Third,
\begin{align*}
\|({\bm{\Lambda}}^r {{\bf{F}}^{r}}^{\prime}- {\bm{\Lambda}}_0{\bf{F}}_0^{\prime}){\bf{M}}_{{\bf{F}}_0}\|&\leq \sqrt{\operatorname{rank}[({\bm{\Lambda}}^r {{\bf{F}}^{r}}^{\prime}-{\bm{\Lambda}}_0{\bf{F}}_0^{\prime}){\bf{M}}_{{\bf{F}}_0}]}\cdot \|{\bf{M}}_{{\bf{F}}_0}\|_S\cdot \|{\bm{\Lambda}}^r {{\bf{F}}^{r}}^{\prime}-{\bm{\Lambda}}_0 {\bf{F}}_0^{\prime}\|_S\nonumber\\
& \lesssim \|{\bm{\Lambda}}^r {{\bf{F}}^{r}}^{\prime}-{\bm{\Lambda}}_0 {\bf{F}}_0^{\prime}\|=\sqrt{NT}d({\bm{\theta}}^r,{\bm{\theta}}_0),
\end{align*}
and
\begin{equation*}
	\begin{split}
		&\|({\bm{\Lambda}}^r {{\bf{F}}^r}^{\prime}-{\bm{\Lambda}}_0 {\bf{F}}_0^{\prime}){\bf{M}}_{{\bf{F}}_0}\|=\|{\bm{\Lambda}}^r {{\bf{F}}^r}^{\prime}{\bf{M}}_{{\bf{F}}_0}\|=\sqrt{\operatorname{Tr}[({{\bm{\Lambda}}^r}^{\prime}{\bm{\Lambda}}^r)\cdot({{\bf{F}}}^{r^\prime}{\bf{M}}_{{\bf{F}}_0}{\bf{F}}^r)]}\\
		&\geq \sqrt{\rho_{\mathrm{min}}({{\bm{\Lambda}}^{r,r_0}}^{\prime}{\bm{\Lambda}}^{r,r_0})}\sqrt{\operatorname{tr}({{\bf{F}}^{r,r_0}}^{\prime}{\bf{M}}_{{\bf{F}}_0}{\bf{F}}^{r,r_0})}=\sqrt{\rho_{\mathrm{min}}({{\bm{\Lambda}}^{r,r_0}}^{\prime}{\bm{\Lambda}}^{r,r_0})}\cdot \|{\bf{M}}_{{\bf{F}}_0}{\bf{F}}^{r,r_0}\|,
	\end{split}
\end{equation*}
which imply that
\begin{equation}\label{lem2-eq4}
	\|{\bf{M}}_{{\bf{F}}_0}{\bf{F}}^{r,r_0}\|/\sqrt{T}\lesssim\sqrt{\frac{1}{\rho_{\mathrm{min}}({{\bm{\Lambda}}^{r,r_0}}^{\prime}{\bm{\Lambda}}^{r,r_0}/N)}}d({\bm{\theta}}^r,{\bm{\theta}}_{0}).
\end{equation}
We assert that there exists $\underline{\rho}>0$ such that
$\rho_{\mathrm{min}}({{\bm{\Lambda}}^{r,r_0}}^{\prime}{\bm{\Lambda}}^{r,r_0}/N)\geq \underline{\rho}$.
To see this,
if
$\rho_{\mathrm{min}}({{\bm{\Lambda}}^{r,r_0}}^{\prime}{\bm{\Lambda}}^{r,r_0}/N)<\delta/(2r-2r_0)$ for $\delta>0$ sufficiently small, then $\rho_{j}({{\bm{\Lambda}}^{r}}^{\prime}{\bm{\Lambda}}^{r}/N)<\delta/(2r-2r_0)$ for $j\geq r_0$, and  $\|{\bm{\Lambda}}^{r,-r_0}\|/\sqrt{N}<\delta/2$. Hence,
\begin{equation*}
\|{\bm{\Lambda}}^{r,r_0}{{\bf{F}}^{r,r_0}}^{\prime}-{\bm{\Lambda}}_0{\bf{F}}_0^{\prime}\|/\sqrt{NT}\leq \|{\bm{\Lambda}}^{r}{{\bf{F}}^{r}}^{\prime}-{\bm{\Lambda}}_0{\bf{F}}_0^{\prime}\|/\sqrt{NT}+\|{\bm{\Lambda}}^{r,-r_0}{{\bf{F}}^{r,-r_0}}^{\prime}\|/\sqrt{NT}=\delta.
\end{equation*}
That is, $\|{\bm{\Lambda}}^{r,r_0}{{\bf{F}}^{r,r_0}}^{\prime}-{\bm{\Lambda}}_0{\bf{F}}_0^{\prime}\|/\sqrt{NT}<\delta$ while $\rho_{\mathrm{min}}({{\bm{\Lambda}}^{r,r_0}}^{\prime}{\bm{\Lambda}}^{r,r_0}/N)$ can be sufficiently small, which contradicts the result in the proof of Lemma 2 of \citet{CDG2021}
that for $\|{\bm{\Lambda}}{{\bf{F}}}^{\prime}-{\bm{\Lambda}}_0{\bf{F}}_0^{\prime}\|/\sqrt{NT}<\delta$, $\rho_{\mathrm{min}}({{\bm{\Lambda}}}^{\prime}{\bm{\Lambda}}/N)$ is bounded below, where ${\bm{\Lambda}}\in \mathbb{R}^{N\times r_0}, {\bf{F}}\in \mathbb{R}^{T\times r_0}$. This validates our assertion. Hence, by \eqref{lem2-eq4}, we have
\begin{equation}\label{Lem5-eq7}
	\|{\bf{M}}_{{\bf{F}}_0}{\bf{F}}^{r,r_0}\|/\sqrt{T}\lesssim d({\bm{\theta}}^r,{\bm{\theta}}_{0}).
\end{equation}

Fourth, note that
\begin{align}\label{Lem5-eq3}
		\frac{1}{\sqrt{NT}}\|({\bm{\Lambda}}^r {{\bf{F}}^r}^{\prime}&-{\bm{\Lambda}}_0{\bf{F}}_0^{\prime}){\bf{P}}_{{\bf{F}}^{r,r_0}}\|
  =\frac{1}{\sqrt{N}}\|{\bm{\Lambda}}^{r,r_0} -{\bm{\Lambda}}_0({\bf{F}}_0^{\prime}{\bf{F}}^{r,r_0}/T)\|
  \nonumber\\
  &\leq\frac{1}{\sqrt{NT}}\|{\bm{\Lambda}}^r {{\bf{F}}^r}^{\prime}-{\bm{\Lambda}}_0{\bf{F}}_0^{\prime}\|\cdot\|{\bf{P}}_{{\bf{F}}^{r,r_0}}\|=\sqrt{r_0}d({\bm{\theta}}^r,{\bm{\theta}}_{0}).
\end{align}

Fifth, define ${\bf{R}}_T={\bf{F}}^{\prime}_0{\bf{F}}^{r,r_0}/T$, then ${\bf{R}}_T^{\prime}{\bf{R}}_T={{\bf{F}}^{r,r_0}}^{\prime}{\bf{P}}_{{\bf{F}}_0}{\bf{F}}^{r,r_0}/T$, and
\begin{equation} \label{lem2-eq1}
	\begin{split}
		\mathbb{I}_{r_0}={{\bf{F}}^{r,r_0}}^{\prime}{\bf{F}}^{r,r_0}/T
   ={\bf{R}}_T^{\prime}{\bf{R}}_T+{{\bf{F}}^{r,r_0}}^{\prime}{\bf{M}}_{{\bf{F}}_0}{\bf{F}}^{r,r_0}/T.\\
	\end{split}
\end{equation}
In addition,
\begin{equation*}
	\begin{split}
		{{\bm{\Lambda}}^{r,r_0}}^{\prime}{\bm{\Lambda}}^{r,r_0}/N=&{\bf{R}}_T^{\prime}({{\bm{\Lambda}}^{\prime}_0}{\bm{\Lambda}}_0/N){\bf{R}}_T+({{\bm{\Lambda}}^{r,r_0}}^{\prime}{\bm{\Lambda}}^{r,r_0}/N-{\bf{R}}_T^{\prime}({{\bm{\Lambda}}^{\prime}_0}{\bm{\Lambda}}_0/N){\bf{R}}_T)\\
		=&{\bf{R}}_T^{\prime}({{\bm{\Lambda}}^{\prime}_0}{\bm{\Lambda}}_0/N)({\bf{R}}_T^{\prime})^{-1}+{\bf{R}}_T^{\prime}({{\bm{\Lambda}}^{\prime}_0}{\bm{\Lambda}}_0/N)({\bf{R}}_T^{\prime})^{-1}({\bf{R}}_T^{\prime}{\bf{R}}_T-\mathbb{I}_{r_0})\\
		+&{{\bm{\Lambda}}^{r,r_0}}^{\prime}({\bm{\Lambda}}^{r,r_0}-{\bm{\Lambda}}_0 {\bf{R}}_T)/N+({\bm{\Lambda}}^{r,r_0}-{\bm{\Lambda}}_0 {\bf{R}}_T)^{\prime}{\bm{\Lambda}}_0{\bf{R}}_T/N.
	\end{split}
\end{equation*}
This, together with \eqref{lem2-eq1}, implies that
\begin{equation}\label{lem2-eq14}({{\bm{\Lambda}}^{r,r_0}}^{\prime}{\bm{\Lambda}}^{r,r_0}/N+{\bf{D}}_{NT}){\bf{R}}_T^{\prime}={\bf{R}}_T^{\prime}({{\bm{\Lambda}}^{\prime}_0}{\bm{\Lambda}}_0/N),
\end{equation}
where
${\bf{D}}_{NT}={\bf{R}}_T^{\prime}({{\bm{\Lambda}}^{\prime}_0}{\bm{\Lambda}}_0/N)({\bf{R}}_T^{\prime})^{-1}{{\bf{F}}^{r,r_0}}^{\prime}{\bf{M}}_{{\bf{F}}_0}{\bf{F}}^{r,r_0}/T-{{\bm{\Lambda}}^{r,r_0}}^{\prime}{\bf{R}}_N/N-{\bf{R}}_N^{\prime}{\bm{\Lambda}}_0 {\bf{R}}_T/N$, ${\bf{R}}_N={\bm{\Lambda}}^{r,r_0}-{\bm{\Lambda}}_0 {\bf{R}}_T$.
By \eqref{lem2-eq14} and the Bauer-Fike theorem \citep[Theorem 7.7.2]{GW2013}, for $d({\bm{\theta}}^r,{\bm{\theta}}_{0})$  sufficiently small, it holds that
\begin{equation}\label{Lem5-eq12}
|\rho_{j}[{{\bm{\Lambda}}^{r,r_0}}^{\prime}{\bm{\Lambda}}^{r,r_0}/N]-\rho_{j}[{\bm{\Lambda}}_0^{\prime}{\bm{\Lambda}}_0/N]|\leq \|{\bf{D}}_{NT}\|_S\leq \|{\bf{D}}_{NT}\|\lesssim d({\bm{\theta}}^r,{\bm{\theta}}_0),
\end{equation}
for $j=1,\cdots,r_0$, where the last inequality follows from \eqref{Lem5-eq7} and \eqref{Lem5-eq3}. Assumption \ref{Ass1} (ii)  and \eqref{Lem5-eq12} imply that for $N$ large enough and $d({\bm{\theta}}^r,{\bm{\theta}}_0)$ sufficiently small, the eigenvalues for ${{\bm{\Lambda}}^{r,r_0}}^{\prime}{\bm{\Lambda}}^{r,r_0}/N$ will be different. Hence,
by \eqref{lem2-eq14} and the perturbation theory for eigenvectors \citep[Section 6.12]{Franklin2012}, we have
\begin{equation*}
	\|{{\bf{R}}^{\prime}_{T}}{\bf{V}}_T{\bf{S}}-\mathbb{I}_{r_0}\|=\|{{\bf{R}}^{\prime}_{T}}{\bf{V}}_T-{\bf{S}}\|\lesssim \|{\bf{D}}_{NT}\| \lesssim d({\bm{\theta}}^r,{\bm{\theta}}_{0}),
\end{equation*}
where ${\bf{V}}_T=\operatorname{diag}(({\bf{R}}_{T,1}{\bf{R}}^{\prime}_{T,1})^{-1/2},\cdots,({\bf{R}}_{T,r_0}{\bf{R}}^{\prime}_{T,r_0})^{-1/2})$ and  ${\bf{R}}^{\prime}_{T,j}$ is the $j$-th column of ${\bf{R}}^{\prime}_{T}$.
Additionally, the triangular inequality implies that
\begin{equation*}
	\|{\bf{R}}_T-{\bf{S}}\|\leq \|{\bf{R}}_T^{\prime}{\bf{V}}_T-{\bf{S}}\|+\|{\bf{R}}_T^{\prime}{\bf{V}}_T-{\bf{R}}^{\prime}_T\|\leq \|{\bf{R}}_T^{\prime}{\bf{V}}_T-{{\bf{S}}}\|+\|{\bf{R}}_T\|\cdot\|{\bf{V}}_T-\mathbb{I}_{r_0}\|.
\end{equation*}
By \eqref{Lem5-eq7} and \eqref{lem2-eq1}, for $d({\bm{\theta}}^r,{\bm{\theta}}_{0})$ small enough, it holds that
\begin{equation*}\label{lem2-eq6}
\|{\bf{V}}_T-\mathbb{I}_{r_0}\|\lesssim\|{\bf{R}}_T{\bf{R}}_T^{\prime}-\mathbb{I}_{r_0}\|=
\|{\bf{R}}_T^{\prime}{\bf{R}}_T-\mathbb{I}_{r_0}\|=\|{\bf{M}}_{{\bf{F}}_0}{\bf{F}}^{r,r_0}\|^2/T\lesssim d^2({\bm{\theta}}^r,{\bm{\theta}}_{0}).
\end{equation*}
Then, it follows from the above three inequalities
that for $d({\bm{\theta}}^r,{\bm{\theta}}_{0})$ small enough,
\begin{equation}\label{Lem5-eq8}
	\|{\bf{F}}_0^{\prime}{\bf{F}}^{r,r_0}/T-{\bf{S}}\|=\|{\bf{R}}_T-{\bf{S}}\|\lesssim d({\bm{\theta}}^r,{\bm{\theta}}_{0}).
\end{equation}

Finally, setting ${\bf{U}}={\bf{S}}$, from \eqref{Lem5-eq6}, \eqref{Lem5-eq7} and \eqref{Lem5-eq8}, we obtain
\begin{equation}\label{Lem5-eq10}
	\begin{split}
\|{\bf{F}}^{r,r_0}-{\bf{F}}_0{\bf{S}}\|/\sqrt{T}\lesssim d({\bm{\theta}}^r,{\bm{\theta}}).
	\end{split}
\end{equation}
Then the desired result follows from \eqref{Lem5-eq9} and \eqref{Lem5-eq10}.

\bigskip

\begin{lemma}\label{lem8}
Suppose that Assumption \ref{Ass1} holds and $r_0<r<\infty$, then for sufficiently small $\delta$, we have
\begin{equation*}
E[\underset{{\bm{\theta}}^r\in {\bm{\Theta}}^{r}(\delta)}{\sup}|\mathbb{W}_{NT}({\bm{\theta}})|\}]\lesssim\frac{\delta}{\sqrt{L_{NT}h^3}}+\frac{\log{L_{NT}}}{L_{NT}h^2}.
\end{equation*}
\end{lemma}

\noindent \textbf{Proof of Lemma \ref{lem8}.}	
The proof is similar to the proof of Lemma \ref{lem3}, and it is thus omitted.

\bigskip

\noindent \textbf{Proof of Theorem \ref{Thm2}.}	
First, following the proof of Lemma \ref{lem2}, we can show that $d(\hat{{\bm{\theta}}}^r,{\bm{\theta}}_{0})=o_p(1)$ when $r>r_0$. Second, like in the proof of Theorem \ref{Thm1}, it follows from Lemma \ref{lem8} that
\begin{equation}\label{Thm2-eq5}
d(\hat{{\bm{\theta}}}^r,{\bm{\theta}}_{0})=O_p(\delta_{NT}), \qquad \text{for } r>r_0.
\end{equation}
Third, similar to \eqref{Lem5-eq12}, we can show that
\begin{equation}\label{Thm2-eq2}
	 |\hat{{\sigma}}_{N,j}^{\bar{r}}-{\sigma}_{Nj}|=|\rho_{j}[\hat{\bm{\Lambda}}^{{\bar{r},r_0}^{\prime}}\hat{\bm{\Lambda}}^{\bar{r},r_0}/N]-\rho_{j}[{\bm{\Lambda}}_0^{\prime}{\bm{\Lambda}}_0/N]|\lesssim d(\hat{{\bm{\theta}}}^{\bar{r}},{\bm{\theta}}_0), \quad \text{for
 } j=1,\cdots, r_0.
\end{equation}
Thus, by \eqref{Thm2-eq5}, \eqref{Thm2-eq2} and Assumption \ref{Ass1} (ii),
\begin{equation}\label{Thm2-eq4}
\hat{{\sigma}}_{N,r_0}^{\bar{r}}={\sigma}_{Nr_0}+o_p(1)\to \sigma_{r_0}>0,
\end{equation}
Fourth, by Lemma \ref{lem5} and \eqref{Thm2-eq5}, it holds that
\begin{equation}\label{Thm2-eq3}
\hat{{\sigma}}_{N,r_0+1} ^{\bar{r}}\leq \sum_{j=r_0+1}^{\bar{r}}\hat{{\sigma}}_{N,j} ^{\bar{r}}=\|\hat{{\bm{\Lambda}}}^{\bar{r},-r_0}\|^2/N\lesssim d(\hat{{\bm{\theta}}}^{\bar{r}},{\bm{\theta}}_{0})^2=O_p(\delta_{NT}^{2}).
\end{equation}
Then, it follows from \eqref{Thm2-eq4} and \eqref{Thm2-eq3} that
\begin{align*}
		P[\hat{r}_{\mathrm{rank}}\neq r_0]&=P[\hat{r}_{\mathrm{rank}}< r_0]+ P[\hat{r}_{\mathrm{rank}}> r_0]\nonumber\\
		&\leq P[\hat{{\sigma}}_{N,r_0}^{\bar{r}}\leq P_{NT}]+ P[\hat{{\sigma}}_{N,r_0+1}^{\bar{r}}>P_{NT}]=o(1).
\end{align*}
Thus, $P[\hat{r}_{\mathrm{rank}}=r_0]\to 1$.

\section{Proof of Theorem \ref{Thm3}}
Following the proof of \citet{BN2002}, to prove Theorem \ref{Thm3}, we just need to show that, for $l>r_0$,
\begin{equation}\label{Thm3-eq8}
    \mathbb{M}_{NT}(\hat{{\bm{\theta}}}^{r_0})-\mathbb{M}_{NT}(\hat{{\bm{\theta}}}^{l})=O_p(\delta_{NT}^2),
\end{equation}
and for $l<r_0$, there exists $C>0$, such that
\begin{equation}\label{Thm3-eq7}
    \mathbb{M}_{NT}(\hat{{\bm{\theta}}}^{r_0})-\mathbb{M}_{NT}(\hat{{\bm{\theta}}}^{l})>C+o_p(1).
\end{equation}

We first consider ${\mathbb{M}}_{NT}(\hat{{\bm{\theta}}}^{r_0})$. Note that for any $\bm{\theta}\in {\bm{\Theta}}^r$, where $0<r\leq \bar{r}$, $\mathbb{M}_{NT}({{\bm{\theta}}}^r)=\mathbb{W}_{NT}({{\bm{\theta}}}^r)+\bar{\mathbb{M}}^*_{NT}({{\bm{\theta}}}^r)$.
Similar to \eqref{Thm1-eq5}, for any given integer $V$,
we have
\begin{align}\label{thm3-eq5}
&P[\delta_{NT}^{-2}|\mathbb{W}_{NT}(\hat{\bm{\theta}}^{r_0})|>2^V]\nonumber\\
\leq &P[\delta_{NT}^{-2}
|\mathbb{W}_{NT}(\hat{\bm{\theta}}^{r_0})|>2^V, \delta_{NT}^{-1}
\cdot d(\hat{\bm{\theta}}^{r_0},\bm{\theta}_0)\leq 2^{V/2}]+P[\delta_{NT}^{-1}\cdot d(\hat{\bm{\theta}}^{r_0},\bm{\theta}_0)>2^{V/2}]\nonumber\\
\leq&\sum_{j\leq V/2}
P[\underset{\bm{\theta}^{r_0}\in S_j}{\sup}|\mathbb{W}_{NT}(\bm{\theta}^{r_0})|>\delta_{NT}^22^{V}]+P[d(\hat{\bm{\theta}}^{r_0},\bm{\theta}_0)>\delta_{NT}2^{V/2}].
 \end{align}
 By Theorem \ref{Thm1}, the second term on the RHS of \eqref{thm3-eq5} converges to $0$ as $V\to \infty$. As for the first term, by Markov's inequality and Lemma \ref{lem3}, it holds that
\begin{equation*}
\begin{split}
P[\underset{\bm{\theta}^{r_0}\in {S}_j}{\sup}|\mathbb{W}_{NT}(\bm{\theta}^{r_0})|>\delta_{NT}^22^V]\lesssim \frac{1}{2^{V}\delta_{NT}^2}\cdot (\frac{2^{j}\delta_{NT}}{\sqrt{L_{NT}h^3}}+\frac{\log{L_{NT}}}{L_{NT}h^2})\lesssim2^{j-V}.
\end{split}
\end{equation*}
Then
\begin{equation*}
\sum_{j\leq V/2}P[\underset{\bm{\theta}^{r_0}\in S_j}{\sup}|\mathbb{W}_{NT}(\bm{\theta}^{r_0})|>\delta_{NT}^22^{V}]\lesssim \sum_{j\leq V/2}2^{j-V}<\frac{V}{2}\cdot 2^{-V/2},
\end{equation*}
which converges to $0$ as $V\to \infty $. Therefore, 
$\mathbb{W}_{NT}(\hat{\bm{\theta}}^{r_0})=O_p(\delta_{NT}^2)$.
In addition,  by \eqref{Lem2-eq2},
\begin{equation*}
    |E[K_h(X_{it}-{\bm{\lambda}}^{\prime}_{0i}{\bf{f}}_{0t})-K_h(X_{it}-{\bm{\lambda}}^{r_0^{\prime}}_{i}{\bf{f}}^{r_0}_{t})]|\lesssim ({\bm{\lambda}}^{\prime}_{0i}{\bf{f}}_{0t}-{\bm{\lambda}}^{r_0^{\prime}}_{i}{\bf{f}}^{r_0}_{t})^2+h^4.
\end{equation*}
Then
\begin{equation*}
|\bar{\mathbb{M}}^*_{NT}(\hat{{\bm{\theta}}}^{r_0})|\lesssim d^2(\hat{{\bm{\theta}}}^{r_0},{\bm{\theta}}_0)+h^4=O_p(\delta_{NT}^2).
\end{equation*}
Thus,
\begin{equation}\label{Thm3-eq9}
   {\mathbb{M}}_{NT}(\hat{{\bm{\theta}}}^{r_0})=\bar{\mathbb{M}}^*_{NT}(\hat{{\bm{\theta}}}^{r_0})+\mathbb{W}_{NT}(\hat{\bm{\theta}}^{r_0})=O_p(\delta_{NT}^2).
\end{equation}
We then consider Case 1: $l>r_0$ and Case 2: $l<r_0$, respectively.

Case 1: $l>r_0$. Following the proof of the case where $l=r_0$, by Lemma \ref{lem8} and \eqref{Thm2-eq5}, we can show that  $
\mathbb{W}_{NT}(\hat{\bm{\theta}}^{l})=O_p(\delta_{NT}^2)$ and $\bar{\mathbb{M}}^*_{NT}(\hat{{\bm{\theta}}}^{l})=O_p(\delta_{NT}^2)$, thus ${\mathbb{M}}_{NT}(\hat{{\bm{\theta}}}^{l})=O_p(\delta_{NT}^2)$. Combining this and \eqref{Thm3-eq9}, we have \eqref{Thm3-eq8} established.

Case 2: $l<r_0$. By \eqref{Thm3-eq9}, to prove \eqref{Thm3-eq7}, we only need to show that $\mathbb{M}_{NT}(\hat{{\bm{\theta}}}^{l})\geq C+o_p(1)$ for some $C>0$. For this purpose, first, following the proof of Lemma \ref{lem2}, we can obtain $\underset{{\bm{\theta}}\in {\bm{\Theta}}^l}{\sup}|\mathbb{W}_{NT}({\bm{\theta}^l})|=o_p(1)$.
Further, for any ${\bm{\theta}}^l\in {\bm{\Theta}}^l$, divide  $c_{it}^l={\bm{\lambda}}_i^{l^\prime}{\bf{f}}^l_t$ into groups $S_1, S_2$, and define $\bar{\mathbb{M}}^*_{S_1}(\bm{\theta}^l),\bar{\mathbb{M}}^*_{S_2}(\bm{\theta}^l)$, $d^2_{S_1}({{\bm{\theta}}}^l,{\bm{\theta}}_0),d^2_{S_2}({{\bm{\theta}}}^l,{\bm{\theta}}_0)$ similarly as in the proof of Lemma \ref{lem2}. Then, $d^2_{S_1}({{\bm{\theta}}}^l,{\bm{\theta}}_0)\lesssim \bar{\mathbb{M}}^*_{S_1}(\bm{\theta}^l), d^2_{S_2}({{\bm{\theta}}}^l,{\bm{\theta}}_0)<d^2h^4$ and $\underset{{\bm{\theta}}\in {\bm{\Theta}}^l}{\sup}|\bar{\mathbb{M}}^*_{S_2}(\bm{\theta}^l)|\leq K_1h^4$. Note that $\mathbb{M}_{NT}(\hat{{\bm{\theta}}}^{l})=\mathbb{W}_{NT}(\hat{\bm{\theta}}^l)+\bar{\mathbb{M}}^*_{S_1}(\hat{\bm{\theta}}^l)+\bar{\mathbb{M}}^*_{S_2}(\hat{\bm{\theta}}^l)$. Since $|\mathbb{W}_{NT}(\hat{\bm{\theta}}^l)|=o_p(1), |\bar{\mathbb{M}}^*_{S_2}(\hat{\bm{\theta}}^l)|=o_p(1)$, then it remains to show that $d^2_{S_1}(\hat{{\bm{\theta}}}^l,{\bm{\theta}}_0)\geq C+o_p(1)$.

To prove the above claim, similar to \eqref{Lem5-eq7}, we can show that
\begin{equation}\label{thm4-eq51}
d^2(\hat{{\bm{\theta}}}^l,{\bm{\theta}}_0)\gtrsim\|{\bf{M}}_{\hat{\bf{F}}^l}{\bf{F}}_0\|^2/T.
\end{equation}
Note that
\begin{equation}\label{thm4-eq52}
\|{\bf{M}}_{\hat{\bf{F}}^l}{\bf{F}}_0\|^2/T=\operatorname{tr}(\mathbb{I}_{r_0}-{\bf{F}}_0^{\prime}\hat{\bf{F}}^l\hat{\bf{F}}^{l^{\prime}}{\bf{F}}_0/T^2)\geq \rho_{\mathrm{max}}(\mathbb{I}_{r_0}-{\bf{F}}_0^{\prime}\hat{\bf{F}}^l\hat{\bf{F}}^{l^{\prime}}{\bf{F}}_0/T^2).
\end{equation}
Moreover, by  Lemma A.5 of \citet{AH2013}, it holds that
\begin{equation} \label{thm4-eq53}
	\rho_{\mathrm{max}}(\mathbb{I}_{r_0}-{\bf{F}}_0^{\prime}\hat{\bf{F}}^l\hat{\bf{F}}^{l^{\prime}}{\bf{F}}_0/T^2)+\rho_{\mathrm{min}}({\bf{F}}_0^{\prime}\hat{\bf{F}}^l\hat{\bf{F}}^{l^{\prime}}{\bf{F}}_0/T^2)\geq \rho_{\mathrm{min}}(\mathbb{I}_{r_0}).
\end{equation}
Since $ {\bf{F}}_0^{\prime}\hat{\bf{F}}^l\hat{\bf{F}}^{l^{\prime}}{\bf{F}}_0 $ is an $r_0\times r_0$ symmetric matrix with rank less than or equal to $l<r_0$, then $\rho_{\mathrm{min}}({\bf{F}}_0^{\prime}\hat{\bf{F}}^l\hat{\bf{F}}^{l^{\prime}}{\bf{F}}_0/T^2)=0$. Therefore, it follows from \eqref{thm4-eq51}-\eqref{thm4-eq53} that there exists $C>0$, such that $d^2(\hat{{\bm{\theta}}}^l,{\bm{\theta}}_0)\geq C+o_p(1)$. Since $d^2_{S_2}(\hat{{\bm{\theta}}}^l,{\bm{\theta}}_0)=o_p(1)$, then $d^2_{S_1}(\hat{{\bm{\theta}}}^l,{\bm{\theta}}_0)\geq C+o_p(1)$.
This completes the proof.

\renewcommand{\theequation}{S.\arabic{equation}}
\renewcommand{\thefigure}{S.\arabic{figure}}
\renewcommand{\thetable}{S.\arabic{table}}
\renewcommand{\thelemma}{S.\arabic{lemma}}
\renewcommand{\thesection}{S.\arabic{section}}

\def\calplongarrow{\stackrel{{\mathcal{P}}}{\longrightarrow}}
\def\darrow{\stackrel{\mathcal{D}}{\longrightarrow}}

\newpage
\section*{Supplementary Material}

\setcounter{section}{0}

This supplementary document contains two parts. Appendix \ref{S-Sec1} provides the proofs for Theorems \ref{Thm4} and \ref{Thm5}, together with some useful lemmas.
Appendix \ref{S-Sec3} presents the additional simulation results for error specifications S2-S3 and asymptotic distributions.
\section{Proofs of Theorems \ref{Thm4} and \ref{Thm5}}\label{S-Sec1}
\noindent Let $K_h^{(j)}(u)=(\partial/\partial u)^{j}K_h(u)$ for $j=1,2,3$. For fixed ${\bm{\lambda}}_{i}\in \mathcal{A}$, ${\bf{f}}_t\in\mathcal{F}$, define
\begin{equation*}
\begin{split}
&\bar{K}_h^{(j)}(X_{it}-{\bm{\lambda}}_{i}^{\prime}{\bf{f}}_t)=E[K_h^{(j)}(X_{it}-{\bm{\lambda}}_{i}^{\prime}{\bf{f}}_t)]\\
&\tilde{K}_h^{(j)}(X_{it}-{\bm{\lambda}}_{i}^{\prime}{\bf{f}}_t)=K_h^{(j)}(X_{it}-{\bm{\lambda}}_{i}^{\prime}{\bf{f}}_t)-\bar{K}_h^{(j)}(X_{it}-{\bm{\lambda}}_{i}^{\prime}{\bf{f}}_t), \qquad \text{for } j=1,2,3.
\end{split}
\end{equation*}
When the functions are evaluated at the true parameters, we suppress $K_h(X_{it}-{\bm{\lambda}}_{0i}^{\prime}{\bf{f}}_{0t})$ as $K_{h,it}$, $\bar{K}_h^{(j)}(X_{it}-{\bm{\lambda}}_{0i}^{\prime}{\bf{f}}_{0t})$ as $\bar{K}_{h,it}^{(j)}$, and $\tilde{K}_h^{(j)}(X_{it}-{\bm{\lambda}}_{0i}^{\prime}{\bf{f}}_{0t})$ as $\tilde{K}_{h,it}^{(j)}$ for $j=1,2,3$, to simplify the notations. For any matrix $A$ with elements $A_{kl}$, let $\|A\|_{\max}=\max_{k,l}|A_{kl}|$.
Let $D_1,D_2,\cdots$ denote some positive constants that do not depend on $N$ and $T$.
\begin{lemma}\label{LemC-1}
Under Assumptions \ref{Ass3}, \ref{Ass4}-\ref{Ass8}, for any ${\bm{\theta}}\in {\bm{\Theta}}^{r_0}$, let $c_{it}={\bm{\lambda}}_i^{\prime}{\bf{f}}_t, c_{it}^0={\bm{\lambda}}_{0i}^{\prime}{\bf{f}}_{0t}$, then
\begin{enumerate}[(i)]
\item $\bar{K}_{h,it}^{(1)}=\bar{O}(h^5)$.
\item $\bar{K}_h^{(1)}(X_{it}-c_{it})=-g_{it}^{(1)}(c_{it}-c_{it}^0)-g_{it}^{(3)}(c_{it}-c_{it}^0)h^2/2-g_{it}^{(5)}(c_{it}-c_{it}^0)h^4/8+\bar{O}(h^5)$.
\item $\bar{K}_{h,it}^{(3)}=\bar{O}(h^2),\bar{K}_h^{(3)}(X_{it}-c_{it})=-g_{it}^{(3)}(c_{it}-c_{it}^0)+\bar{O}(h^2)$.
\item $E[(K^{(1)}_{h,it})^2]=g_{it}(0)Lh^{-3}+\bar{O}(h^{-2}),E[K^{(2)}_h(X_{it}-c_{it})^2]=\bar{O}(h^{-5})$.
\end{enumerate}
\end{lemma}

\bigskip

\noindent \textbf{Proof of Lemma \ref{LemC-1}.}	
The proofs follow the standard steps when computing the means of kernel density estimators, therefore they are omitted here to save space. 
\bigskip

\begin{lemma}\label{LemC-2}
Under Assumptions \ref{Ass1}-\ref{Ass9},  $d(\hat{\bm{\theta}}, {\bm{\theta}}_0)=O_p(1/\sqrt{L_{NT}h^3})$, as $N,T \to \infty$.
\end{lemma}

\bigskip

\noindent \textbf{Proof of Lemma \ref{LemC-2}.}	
Following the proof of Theorem \ref{Thm1}, by Lemma \ref{LemC-1} (i), we can show that $d(\hat{\bm{\theta}}, {\bm{\theta}}_0)=O_p(1/\sqrt{L_{NT}h^3}+h^5)$. Since $L_{NT}h^{13} \to 0$ by Assumption \ref{Ass9}, then
$d(\hat{\bm{\theta}}, {\bm{\theta}}_0)=O_p(1/\sqrt{L_{NT}h^3})$, and thus $\|\hat{\bf{F}}-{\bf{F}}_0\hat{\bf{S}}\|/\sqrt{T}=O_p(1/\sqrt{L_{NT}h^3})$, $ \|\hat{\bm{\Lambda}}-{\bf{\Lambda}}_0\hat{\bf{S}}\|/\sqrt{N}=O_p(1/\sqrt{L_{NT}h^3})$,
Without loss of generality, we assume $\hat{\bf{S}}=\mathbb{I}_{r_0}$
to simplify the notations.





\bigskip

\begin{lemma}\label{LemC-13}
Under Assumptions \ref{Ass1}-\ref{Ass9}, we have
\begin{align*}\label{LemC13-eq1}
\underset{{\bm{\lambda}} \in \mathcal{A}}{\sup}|\frac{1}{T}\sum_{t=1}^T\tilde{K}_h(X_{it}-{\bm{\lambda}}^{\prime}{\bf{f}}_{t})|=o_p(1),\quad
\underset{{\bm{\lambda}} \in \mathcal{A}}{\sup}|\frac{1}{T}\sum_{t=1}^T\tilde{K}^{(2)}_h(X_{it}-{\bm{\lambda}}^{\prime}{\bf{f}}_{t}){\bf{f}}_{0t}{\bf{f}}_{0t}^{\prime}|=o_p(1).
\end{align*}
\end{lemma}

\bigskip

\noindent \textbf{Proof of Lemma \ref{LemC-13}.}	
To save space, we only give the proof for the second result, as the proof for the first result is similar. Define
\begin{equation*}
   \mathbb{M}^{2}_{i,T}({\bm{\lambda}},{\bf{F}})=\frac{1}{T}\sum_{t=1}^T[K^{(2)}_{h}(X_{it}-{\bm{\lambda}}^{\prime}{\bf{f}}_t)]{\bf{f}}_{0t}{\bf{f}}_{0t}^{\prime},\; \bar{\mathbb{M}}^{2}_{i,T}({\bm{\lambda}},{\bf{F}})=\frac{1}{T}\sum_{t=1}^TE[K_{h}^{(2)}(X_{it}-{\bm{\lambda}}^{\prime}{\bf{f}}_t)]{\bf{f}}_{0t}{\bf{f}}_{0t}^{\prime},
\end{equation*}
then $\mathbb{M}^{2}_{i,T}({\bm{\lambda}},{\bf{F}})-\bar{\mathbb{M}}^{2}_{i,T}({\bm{\lambda}},{\bf{F}})=\frac{1}{T}\sum_{t=1}^T\tilde{K}^{(2)}_h(X_{it}-{\bm{\lambda}}^{\prime}{\bf{f}}_{t}){\bf{f}}_{0t}{\bf{f}}_{0t}^{\prime}$. Following 
the proof of Lemma \ref{lem2}, for any ${\bm{\lambda}}\in\mathcal{A}$, let ${\bm{\lambda}}^*=\{{\bm{\lambda}}_{(j)},j\leq J,\|{\bm{\lambda}}-{\bm{\lambda}}_{(j)}\|\leq 1/L_{NT}\}$. Fix any ${\bm{\eta}}\in \mathbb{R}^{r_0},{\bm{\eta}}\neq 0$, and
define
\begin{equation*}
\begin{split}
 &A_{1T}({\bm{\lambda}})={\bm{\eta}}^{\prime}(\mathbb{M}^{2}_{i,T}({\bm{\lambda}},{\bf{F}}_0)-\mathbb{M}^{2}_{i,T}({\bm{\lambda}}^*,{\bf{F}}_0)){\bm{\eta}},\\
 &A_{2T}({\bm{\lambda}})={\bm{\eta}}^{\prime}(\mathbb{M}^{2}_{i,T}({\bm{\lambda}}^*,{\bf{F}}_0)-\bar{\mathbb{M}}^{2}_{i,T}({\bm{\lambda}}^*,{\bf{F}}_0)){\bm{\eta}},\\
 &A_{3T}({\bm{\lambda}})={\bm{\eta}}^{\prime}(\bar{\mathbb{M}}^{2}_{i,t}({\bm{\lambda}}^*,{\bf{F}}_0)-\bar{\mathbb{M}}^{2}_{i,t}({\bm{\lambda}},{\bf{F}}_0)){\bm{\eta}}.\\
\end{split}
\end{equation*}
We analyze each of the above terms in order. First, for any ${\bm{\lambda}}\in\mathcal{A}$, by Assumption \ref{Ass7} (i),
\begin{equation*}
|A_{1T}({\bm{\lambda}})|
\leq c_3K_2^3\cdot\|{\bm{\eta}}\|^2/h^4\cdot \|{\bm{\lambda}}-{\bm{\lambda}}^*\|\leq c_3K_2^3\|{\bm{\eta}}\|^2/(L_{NT}h^4).
\end{equation*}
Hence,
$\underset{{\bm{{\bm{\lambda}}}}\in\mathcal{A}}{\sup}|A_{1T}({\bm{\lambda}})|\leq c_3K_2^3\|{\bm{\eta}}\|^2/(L_{NT}h^4)=o_p(1).$
Since $A_{3T}({\bm{\lambda}})=-E[A_{1T}({\bm{\lambda}})]$, then
\begin{equation*}
\underset{{\bm{{\bm{\lambda}}}}\in\mathcal{A}}{\sup}|A_{3T}({\bm{\lambda}})|=\underset{{\bm{{\bm{\lambda}}}}\in\mathcal{A}}{\sup}|E[A_{1T}({\bm{\lambda}})]|\leq E[\underset{{\bm{{\bm{\lambda}}}}\in\mathcal{A}}{\sup}|A_{1T}({\bm{\lambda}})|]=o_p(1).
\end{equation*}

Second,
write $A_{2T}({\bm{{\bm{\lambda}}}})=\mathbb{Z}_T/T=\sum_{t=1}^TZ_{t}/T$, where $w_{t}=K^{(2)}_h(X_{it}-{\bm{\lambda}}^{*^\prime}{\bf{f}}_{0t})({\bm{\eta}}^{\prime}{\bf{f}}_{0t})^2=K_h^{(2)}(e_{it}^0-({\bm{\lambda}}^*-{\bm{\lambda}}_{0i})^{\prime}{\bf{f}}_{0t})({\bm{\eta}}^{\prime}{\bf{f}}_{0t})^2$ and
$Z_{t}=w_{t}-E(w_{t})$. By Lemma \ref{lem6}, we have
\begin{equation}\label{LemC3-eq3}
\sigma^2(\mathbb{Z}_T)\leq \sum_{t=1}^TVar(w_{t})+2\sum_{t=1}^{T-1}\sum_{l=t+1}^{t+p}|Cov[w_{t},w_{l}]| := I+II.
\end{equation}
Since $|w_{t}|\leq c_2K^2_2\|{\bm{\eta}}\|^2/h^3$ by Assumption \ref{Ass4} (vi), then we can show that
$I\leq \sum_{t=1}^TD_1/h^6=TD_1/h^6.$
In addition, for $II$, by Davydov's Lemma and Assumption \ref{Ass2} (ii), we have
\begin{equation*}
\begin{split}
II&\leq 2\sum_{t=1}^{T-1}\sum_{l=t+1}^{t+p}\alpha^{1/2}(l-t)E[|w_{t}|^4]^{1/4}E[|w_{l}|^4]^{1/4}\leq 
D_2T/h^6.
\end{split}
\end{equation*}
Therefore, $\sigma^2(\mathbb{Z}_T)\leq (D_1+D_2)T/h^6$. Then by Lemma \ref{lem6}, for any $\epsilon>0$,
\begin{equation*}
\begin{split}
P(|A_{2T}({\bm{\lambda}})|\geq \epsilon)&=P(|\mathbb{Z}_T|/T\geq \epsilon)\leq \exp{(-\frac{\epsilon^2}{8v_1^2(\mathbb{Z}_T)})}+q(1+\frac{8c_2K^2_2\|{\bm{\eta}}\|^2}{h^3\epsilon})\alpha([p]),
\end{split}
\end{equation*}
where
\begin{equation*}
v_1^2(\mathbb{Z}_T)=\frac{4(D_1+D_2)}{Th^6}+\frac{2c_2K^2_2\|{\bm{\eta}}\|^2p}{Th^3}.
\end{equation*}
Let $p=T^{1/2}, q=T^{1/2}/2$. Since ${\bm{\lambda}}^*$ can take at most $J\lesssim L_{NT}^{r_0}$ values, then for any $\epsilon>0$,
\begin{equation*}
\begin{split}
&P(\underset{{\bm{\lambda}}\in\mathcal{A}}{\sup}|A_{2T}({\bm{\lambda}})|\geq \epsilon)\lesssim L_{NT}^{r_0}P(|A_{2T}({\bm{\lambda}})|\geq \epsilon)\\
\lesssim &\exp{(-Th^6(a(\epsilon)-\frac{r_0\log{L_{NT}}}{Th^6}))}+T^{1/2}L_{NT}^{r_0}(1+\frac{8c_2K^2_2\|{\bm{\eta}}\|^2}{h^3\epsilon})\cdot \rho^{T^{1/2}}=o_p(1),
\end{split}
\end{equation*}
where $a(\epsilon)=\epsilon^2/(32(D_1+D_2)+16c_2K_2^2\|{\bm{\eta}}\|^2ph^3)$. Hence $\underset{{\bm{\lambda}}\in\mathcal{A}}{\sup}|A_{2T}({\bm{\lambda}})|=o_p(1)$.

Combining above results, we obtain
$\underset{{\bm{\lambda}}\in\mathcal{A}}{\sup}|{\bm{\eta}}^{\prime}(\mathbb{M}^{(2)}_{i,t}({\bm{\lambda}},{\bf{F}}_0)-\bar{\mathbb{M}}_{i,t}({\bm{\lambda}},{\bf{F}}_0)){\bm{\eta}}|=o_p(1).$
Since ${\bm{\eta}}\neq 0$ and was set arbitrarily, then
$\underset{{\bm{\lambda}} \in \mathcal{A}}{\sup}|\mathbb{M}^{(2)}_{i,t}({\bm{\lambda}},{\bf{F}}_0)-\bar{\mathbb{M}}^{(2)}_{i,t}({\bm{\lambda}},{\bf{F}}_0)|=o_p(1).$

\bigskip

\begin{lemma}\label{LemC-3} Under Assumptions \ref{Ass1}-\ref{Ass9}, we have
$\|\hat{\bm{\lambda}}_i-{\bm{\lambda}}_{0i}\|=o_p(1)$, for each $i$ .
\end{lemma}

\bigskip

\noindent \textbf{Proof of Lemma \ref{LemC-3}.}	
Let $\mathbb{M}^*_{i,T}({\bm{\lambda}},{\bf{F}})=\frac{1}{T}\sum_{t=1}^T[K_h(X_{it}-{\bm{\lambda}}_{0i}^{\prime}{\bf{f}}_{0t})-K_{h}(X_{it}-{\bm{\lambda}}^{\prime}{\bf{f}}_t)]$ and $\bar{\mathbb{M}}^*_{i,T}({\bm{\lambda}},{\bf{F}})=E[\mathbb{M}^*_{i,T}({\bm{\lambda}},{\bf{F}})]$. Note that
$\hat{{\bm{\lambda}}}_i=\underset{{\bm{\lambda}}\in \mathcal
{A}}{\arg\max}  \mathbb{M}^*_{i,T}({\bm{\lambda}},\hat{{\bf{F}}}).$
First, we show that
\begin{equation}\label{LemC3-eq1}
    \underset{{\bm{\lambda}}\in \mathcal{A}}{\sup}|\mathbb{M}^*_{i,T}({\bm{\lambda}},\hat{{\bf{F}}})-\bar{\mathbb{M}}^*_{i,T}({\bm{\lambda}},{\bf{F}}_0)|=o_p(1).
\end{equation}
Note that
\begin{equation*}
\begin{split}
\underset{{\bm{\lambda}}\in \mathcal{A}}{\sup}|\mathbb{M}^*_{i,T}({\bm{\lambda}},\hat{{\bf{F}}})&-\bar{\mathbb{M}}^*_{i,T}({\bm{\lambda}},{\bf{F}}_0)|
=\underset{{\bm{\lambda}}\in \mathcal{A}}{\sup}|-(\mathbb{M}_{i,T}({\bm{\lambda}},\hat{{\bf{F}}})-{\mathbb{M}}_{i,T}({\bm{\lambda}},{\bf{F}}_0))\\
&-({\mathbb{M}}_{i,T}({\bm{\lambda}},{\bf{F}}_0)-\bar{\mathbb{M}}_{i,T}({\bm{\lambda}},{\bf{F}}_0))+({\mathbb{M}}_{i,T}({\bm{\lambda}}_{0i},{\bf{F}}_0)-\bar{\mathbb{M}}_{i,T}({\bm{\lambda}}_{0i},{\bf{F}}_0))|\\
&\leq \underset{{\bm{\lambda}}\in \mathcal{A}}{\sup}|\mathbb{M}_{i,T}({\bm{\lambda}},\hat{{\bf{F}}})-{\mathbb{M}}_{i,T}({\bm{\lambda}},{\bf{F}}_0)|+2\underset{{\bm{\lambda}}\in \mathcal{A}}{\sup}|{\mathbb{M}}_{i,T}({\bm{\lambda}},{\bf{F}}_0)-\bar{\mathbb{M}}_{i,T}({\bm{\lambda}},{\bf{F}}_0)|.
\end{split}
\end{equation*}
The second term is $o_p(1)$ by Lemma \ref{LemC-13} (i). In addition, by Assumption \ref{Ass4} (v)  and Lemma \ref{LemC-2}, it is easy to show that the first term is also $o_p(1)$. Thus, we have \eqref{LemC3-eq1} established.

Second, for any $\epsilon>0$, let $B_i(\epsilon)=\{{\bm{\lambda}}\in \mathcal{A}, \|{\bm{\lambda}}-{\bm{\lambda}}_{0i}\|>\epsilon\}$. For any ${\bm{\lambda}}\in B_i^C(\epsilon)$, by Lemma \ref{lem1} (i) and Assumption \ref{Ass3} (ii), we have
\begin{equation*}
\bar{\mathbb{M}}^*_{i,T}({\bm{\lambda}},{\bf{F}}_0)-\bar{\mathbb{M}}^*_{i,T}({\bm{\lambda}}_{0i},{\bf{F}}_0)=\frac{1}{T}\sum_{t=1}^T[g_{it}({\bm{\lambda}}^{\prime}{\bf{f}}_{0t}-{\bm{\lambda}}_{0i}^{\prime}{\bf{f}}_{0t})-g_{it}(0)]+{O}(h^2)<0.
\end{equation*}
Hence, $\underset{{\bm{\lambda}}\in B_i^C(\epsilon)}{\sup}\bar{\mathbb{M}}^*_{i,T}({\bm{\lambda}},{\bf{F}}_0)<\bar{\mathbb{M}}^*_{i,T}({\bm{\lambda}}_{0i},{\bf{F}}_0)$. Given this and \eqref{LemC3-eq1}, the consistency of $\hat{{\bm{\lambda}}}$ follows from a standard consistency argument for $M$-estimators \citep[Theorem 2.1]{NM1994}. 

\bigskip

To derive the asymptotic distribution of $\hat{\bm{\lambda}}_i$,
we need to obtain the stochastic expansion of $\hat{\bf{f}}_t$. Define
\begin{equation*}
\begin{split}
{\mathcal{S}}^*({\bm{\theta}})=\left[\underbrace{\cdots,\frac{1}{\sqrt{NT}}\sum_{t=1}^T\bar{K}_h^{(1)}(X_{it}-{\bm{\lambda}}_{i}^{\prime}{\bf{f}}_t){\bf{f}}_t^{\prime},\cdots}_{1\times Nr_0},\underbrace{\cdots, \frac{1}{\sqrt{NT}}\sum_{i=1}^N\bar{K}_h^{(1)}(X_{it}-{\bm{\lambda}}_{i}^{\prime}{\bf{f}}_t){\bm{\lambda}}_{i}^{\prime},\cdots}_{1\times Tr_0}\right]^{\prime},
\end{split}
\end{equation*}
and $\mathcal{H}({\bm{\theta}})=\partial {\mathcal{S}}^*({\bm{\theta}})/\partial {\bm{\theta}}^{\prime}$, $\mathcal{H}=\mathcal{H}({\bm{\theta}}_0)$. Expanding $\mathcal{S}^*(\hat{{\bm{\theta}}})$ around $\mathcal{S}^*({\bm{\theta}}_0)$,
\begin{equation}\label{lem2-2-eq1}
    \mathcal{S}^*(\hat{{\bm{\theta}}})=\mathcal{S}^*({\bm{\theta}}_0)+\mathcal{H}\cdot(\hat{{\bm{\theta}}}-{\bm{\theta}}_0)+\frac{1}{2}\mathcal{R}(\hat{{\bm{\theta}}}),
\end{equation}
where
$\mathcal{R}(\hat{{\bm{\theta}}})=\left(\sum_{j=1}^M\partial \mathcal{H}(\hat{\bm{\theta}}^*)/\partial {\theta}_j \cdot(\hat{{\theta}}_j-{\theta}_{0j})\right)(\hat{{\bm{\theta}}}-{\bm{\theta}}_0)$,
and $\theta_{0j},{\theta}_j$ denote the $j$-th element of ${\bm{\theta}}_0, {\bm{\theta}}$, respectively.  $\hat{\bm{\theta}}^*$ lies between $\hat{{\bm{\theta}}}$ and ${\bm{\theta}}_0$. Define
\begin{equation*}
\begin{split}
\mathcal{H}_d=\begin{pmatrix}
\mathcal{H}_d^A & 0\\
0 & \mathcal{H}_d^F
\end{pmatrix}, \; \mathcal{H}_d^A=\frac{\sqrt{T}}{\sqrt{N}}\text{diag}[{\bm{\Phi}}_{T,1},\cdots,{\bm{\Phi}}_{T,N} ], \; \mathcal{H}_d^F=\frac{\sqrt{N}}{\sqrt{T}}\text{diag}[{\bm{\Psi}}_{N,1},\cdots,{\bm{\Psi}}_{N,T} ],
\end{split}
\end{equation*}
where ${\bm{\Phi}}_{T,i}=\frac{1}{T}\sum_{t=1}^T\bar{K}_{h,it}^{(2)}{\bf{f}}_{0t}{\bf{f}}_{0t}^{\prime}, {\bm{\Psi}}_{N,t}=\frac{1}{N}\sum_{i=1}^N\bar{K}_{h,it}^{(2)}{\bm{\lambda}}_{0i}{\bm{\lambda}}_{0i}^{\prime}.$

\bigskip

\begin{lemma}\label{LemC-5}
Under Assumptions \ref{Ass1}-\ref{Ass8}, there exists a matrix $\tilde{\mathcal{H}}$, such that $\tilde{\mathcal{H}}$ is invertible, and $\|\tilde{\mathcal{H}}-\mathcal{H}\|_{\max}=O(1/T^2), \|\tilde{\mathcal{H}}^{-1}-\mathcal{H}_d^{-1}\|_{\max}=O(1/T)$.
\end{lemma}

\bigskip

\noindent \textbf{Proof of Lemma \ref{LemC-5}.}	
To simplify notations, we only consider $r_0=2$, but the proof can be generalized to the case
where
$r_0>2$ easily. Write ${\bm{\lambda}}_{0i}=(\lambda_{0i,1},\lambda_{0i,2})^{\prime}$, ${\bf{f}}_{0t}=(f_{0t,1},f_{0t,2})^{\prime}.$ First, define
\begin{equation*}
\begin{split}
&{\bm{\omega}}_{1}^{\prime}=[\underbrace{\left(\lambda_{01,1}, 0\right) / \sqrt{N}, \ldots,\left(\lambda_{0 N, 1}, 0\right) / \sqrt{N}}_{{\bm{\omega}}_{1 \Lambda}^{\prime}}, \underbrace{\left(-f_{01,1}, 0\right) / \sqrt{T}, \ldots,\left(-f_{0 T, 1}, 0\right) / \sqrt{T}}_{{\bm{\omega}}_{1 F}^{\prime}}], \\
&{\bm{\omega}}_{2}^{\prime}=[\underbrace{\left(0, \lambda_{01,2}\right) / \sqrt{N}, \ldots,\left(0, \lambda_{0 N, 2}\right) / \sqrt{N}}_{{\bm{\omega}}_{2 \Lambda}^{\prime}}, \underbrace{\left(0,-f_{01,2}\right) / \sqrt{T}, \ldots,\left(0,-f_{0 T, 2}\right) / \sqrt{T}}_{{\bm{\omega}}_{2 F}^{\prime}}], \\
&{\bm{\omega}}_{3}^{\prime}=[\underbrace{\left(\lambda_{01,2}, 0\right) / \sqrt{N}, \ldots,\left(\lambda_{0 N, 2}, 0\right) / \sqrt{N}}_{{\bm{\omega}}_{3 \Lambda}^{\prime}}, \underbrace{\left(0,-f_{01,1}\right) / \sqrt{T}, \ldots,\left(0,-f_{0 T, 1}\right) / \sqrt{T}}_{{\bm{\omega}}_{3 F}^{\prime}}], \\
&{\bm{\omega}}_{4}^{\prime}=[\underbrace{\left(0,\lambda_{01,1}\right) / \sqrt{N}, \ldots,\left(0,\lambda_{0 N, 1}\right) / \sqrt{N}}_{{\bm{\omega}}_{4 \Lambda}^{\prime}}, \underbrace{\left(-f_{01,2},0\right) / \sqrt{T}, \ldots,\left(-f_{0 T, 2},0\right) / \sqrt{T}}_{{\bm{\omega}}_{4 F}^{\prime}}],
\end{split}
\end{equation*}
and ${\bm{\omega}}=[{\bm{\omega}}_1,{\bm{\omega}}_2,{\bm{\omega}}_3,{\bm{\omega}}_4].$ It is easy to check that ${\bm{\omega}}_p^{\prime}{\bm{\omega}}_q=0,p\neq q$. Moreover, we have
\begin{equation}\label{LemC5-eq3}
{\bm{\omega}} {\bm{\omega}}^{\prime}
=\left(\begin{array}{cc}
\sum_{k=1}^{4} {\bm{\omega}}_{k \Lambda} {\bm{\omega}}_{k \Lambda}^{\prime}, & -(N T)^{-1 / 2}\left\{{\bf{f}}_{0t} {\bm{\lambda}}_{0i}^{\prime}\right\} \\
-(N T)^{-1 / 2}\left\{{\bm{\lambda}}_{0i} {\bf{f}}_{0t}^{\prime}\right\}, &\sum_{k=1}^{4} {\bm{\omega}}_{k F} {\bm{\omega}}_{k F}^{\prime}
\end{array}\right),
\end{equation}
where $\left\{{\bf{f}}_{0 t} {\bm{\lambda}}_{0 i}^{\prime}\right\}$ denotes a  $2 N \times 2 T$  matrix whose  $\{i, t\}$-th block is  ${\bf{f}}_{0t} {\bm{\lambda}}_{0i}^{\prime}$, and $\left\{{\bm{\lambda}}_{0i} {\bf{f}}_{0t}^{\prime}\right\}$ denotes a  $2 T \times 2 N$  matrix whose  $\{t,i\}$-th block is  ${\bm{\lambda}}_{0i} {\bf{f}}_{0t}^{\prime}$.

Second, we can write
\begin{align}\label{LemC5-eq5}
&\mathcal{H}=
\left(\begin{array}{cc}
-(NT)^{-1/2}\operatorname{diag}\left[\left\{\sum_{t=1}^T\bar{K}_{h,it}^{(2)}{\bf{f}}_{0t}{\bf{f}}_{0t}^{\prime}\right\}\right] \; &  -(NT)^{-1/2}\left\{ \bar{K}_{h,it}^{(2)}{\bf{f}}_{0t}{\bm{\lambda}}_{0i}^{\prime}-\bar{K}_{h,it}^{(1)}\cdot \mathbb{I}_2\right\}\\
-(NT)^{-1/2}\left\{ \bar{K}_{h,it}^{(2)}{\bm{\lambda}}_{0i}{\bf{f}}_{0t}^{\prime}-\bar{K}_{h,it}^{(1)}\cdot \mathbb{I}_2\right\} \; & -(NT)^{-1/2}\operatorname{diag}\left[\left\{\sum_{i=1}^N\bar{K}_{h,it}^{(2)}{\bm{\lambda}}_{0i}{\bm{\lambda}}_{0i}^{\prime}\right\}\right]
\end{array}\right)\nonumber\\
&= b\underbrace{\left(\begin{array}{cc}
(NT)^{-1/2}\operatorname{diag}\left[\left\{\sum_{t=1}^T{\bf{f}}_{0t}{\bf{f}}_{0t}^{\prime}\right\}\right] &  (NT)^{-1/2}\left\{ {\bf{f}}_{0t}{\bm{\lambda}}_{0i}^{\prime}\right\}\\
(NT)^{-1/2}\left\{ {\bm{\lambda}}_{0i}{\bf{f}}_{0t}^{\prime}\right\} & (NT)^{-1/2}\operatorname{diag}\left[\left\{\sum_{i=1}^N{\bm{\lambda}}_{0i}{\bm{\lambda}}_{0i}^{\prime}\right\}\right]
\end{array}\right)}_{I}\nonumber\\
&+\underbrace{\left(\begin{array}{cc}
(NT)^{-1/2}\operatorname{diag}\left[\left\{\sum_{t=1}^Tb_{it}{\bf{f}}_{0t}{\bf{f}}_{0t}^{\prime}\right\}\right] &  (NT)^{-1/2}\left\{ b_{it}{\bf{f}}_{0t}{\bm{\lambda}}_{0i}^{\prime}+\bar{K}_{h,it}^{(1)}\cdot \mathbb{I}_2\right\}\\
(NT)^{-1/2}\left\{ b_{it}{\bm{\lambda}}_{0i}{\bf{f}}_{0t}^{\prime}+\bar{K}_{h,it}^{(1)}\cdot \mathbb{I}_2\right\}& (NT)^{-1/2}\operatorname{diag}\left[\left\{\sum_{i=1}^Nb_{it}{\bm{\lambda}}_{0i}{\bm{\lambda}}_{0i}^{\prime}\right\}\right]
\end{array}\right) }_{II},
\end{align}
where $b_{it}=-\bar{K}_{h,it}^{(2)}-b, 0<b\leq g_1/2$. For I, by \eqref{LemC5-eq3}, we can write
\begin{equation}\label{LemC5-eq2}
I+b{\bm{\omega}} {\bm{\omega}}^{\prime}=b\left(\begin{array}{cc}
\sqrt{\frac{T}{N}} \cdot \mathbb{I}_{2 N}+\sum_{k=1}^{4} {\bm{\omega}}_{k \Lambda} {\bm{\omega}}_{k \Lambda}^{\prime} & \mathbf{0}_{2 N \times 2 T} \\
\mathbf{0}_{2 T \times 2 N} & \sqrt{\frac{N}{T}} \cdot \mathbb{I}_{T} \otimes \operatorname{diag}\left(\sigma_{N 1}, \sigma_{N 2}\right)+\sum_{k=1}^{4} {\bm{\omega}}_{k F} {\bm{\omega}}_{k F}^{\prime}
\end{array}\right).
\end{equation}
By Assumption \ref{Ass1}, there exists a constant  $c$ such that $I+b\cdot{\bm{\omega}} {\bm{\omega}}^{\prime}\geq c \cdot \mathbb{I}_{2(N+T)}$.
For $II$, by Lemma \ref{LemC-1} (i), we have $\bar{K}_{h,it}^{(1)}=O(h^5)$, then
\begin{equation}\label{LemC5-eq7}
II=\frac{1}{\sqrt{NT}}\sum_{i=1}^N\sum_{t=1}^T(-\bar{K}^{(2)}_{h,it}-b){\bm{\mu}}_{it}{\bm{\mu}}_{it}^{\prime}+O(h^5),
\end{equation}
where ${\bm{\mu}}_{it}=[\underbrace{\mathbf{0}_{1\times2},\cdots,{\bf{f}}_{0t}^{\prime}\cdots,\mathbf{0}_{1\times 2}}_{2T},\underbrace{{\mathbf{0}}_{1\times2},\cdots,{\bm{\lambda}}_{0i}^{\prime},\cdots,{\mathbf{0}}_{1\times2}}_{2N}]^{\prime}$. Lemma \ref{lem1} (iii) and Assumption \ref{Ass3} (iv) imply that for $N,T$ large enough, $-\bar{K}_{h,it}^{(2)}>\underline{g_1}/2\geq b$ for all $i,t$, thus $II\geq 0$. Then, it follows from \eqref{LemC5-eq5}-\eqref{LemC5-eq7} that
\begin{equation}\label{LemC5-eq11}
\mathcal{H}_b=\mathcal{H}+{b}\cdot {\bm{\omega}}{\bm{\omega}}^{\prime}\geq {c}\cdot \mathbb{I}_{2(N+T)},\qquad
 \mathcal{H}_b^{-1}\leq c^{-1}\cdot \mathbb{I}_{2(N+T)}.
\end{equation}

Write $\mathcal{H}_b=\mathcal{H}_d+\mathcal{C}$, where
\begin{equation*}
\mathcal{C}=\left(\begin{array}{cc}
     b\cdot \sum_{k=1}^{4} {\bm{\omega}}_{k \Lambda} {\bm{\omega}}_{k \Lambda}^{\prime} & (NT)^{-1/2}\left\{ b_{it}{\bf{f}}_{0t}{\bm{\lambda}}_{0i}^{\prime}+\bar{K}_{h,it}^{(1)}\cdot \mathbb{I}_2\right\}\\
    (NT)^{-1/2}\left\{ b_{it}{\bm{\lambda}}_{0i}{\bf{f}}_{0t}^{\prime}+\bar{K}_{h,it}^{(1)}\cdot \mathbb{I}_2\right\} &  b\cdot \sum_{k=1}^{4} {\bm{\omega}}_{k F} {\bm{\omega}}_{k F}^{\prime}
\end{array}\right).
\end{equation*}
Note that $\mathcal{H}_b^{-1}-\mathcal{H}_d^{-1}=-\mathcal{H}_d^{-1}\mathcal{C}\mathcal{H}_d^{-1}+\mathcal{H}_d^{-1}\mathcal{C}\mathcal{H}_b^{-1}\mathcal{C}\mathcal{H}_d^{-1}$.
By \eqref{LemC5-eq11}, $\mathcal{H}_d^{-1}\mathcal{C}\mathcal{H}_b^{-1}\mathcal{C}\mathcal{H}_d^{-1}\leq {c}^{-1}\mathcal{H}_d^{-1}\mathcal{C}^2\mathcal{H}_d^{-1}$, thus the $j$-th diagonal element of $\mathcal{H}_d^{-1}\mathcal{C}\mathcal{H}_b^{-1}\mathcal{C}\mathcal{H}_d^{-1}$ is smaller than that of ${c}^{-1}\mathcal{H}_d^{-1}\mathcal{C}^2\mathcal{H}_d^{-1}$.
Since the entry with the largest absolute value of a positive semideﬁnite matrix is always on the diagonal, then 
\begin{equation*}
    \|\mathcal{H}_b^{-1}-\mathcal{H}_d^{-1}\|_{\max}\leq \|\mathcal{H}_d^{-1}\mathcal{C}\mathcal{H}_d^{-1}\|_{\max}+{c}^{-1}\|\mathcal{H}_d^{-1}\mathcal{C}^2\mathcal{H}_d^{-1}\|_{\max}.
\end{equation*}
Since $\mathcal{H}_d^{-1}$ is a block-diagonal matrix whose elements are all $O(1)$ 
and both $\|\mathcal{C}\|_{\max}$ and $\|\mathcal{C}^2\|_{\max}$ can be shown to be $O(1/T)$, then $\|\mathcal{H}_b^{-1}-\mathcal{H}_d^{-1}\|_{\max}=O(1/T)$.

Third, \eqref{LemC5-eq11} implies
$\operatorname{rank}(\mathcal{H})=2(N+T)-s,$ for $0\leq s\leq 4$, then there exists an orthogonal matrix $\mathcal{P}$, and a diagonal matrix whit nonzero elements $\mathcal{Q}$, such that
\begin{equation*}
\mathcal{H}=\mathcal{P}\left(\begin{array}{cc}
    \mathcal{Q} & \mathbf{0}_{[2(N+T)-s]\times s} \\
    \mathbf{0}_{[2(N+T)-s]\times s} & \mathbf{0}_{s\times s}
\end{array}\right)\mathcal{P}^{\prime},
\end{equation*}
Let $\mathcal{P}_s$ denote the last $s$ columns of $\mathcal{P}$, and let $ \tilde{\mathcal{H}}=
\mathcal{H}_b-b\cdot \tilde{{\bm{\omega}}}\tilde{{\bm{\omega}}}^{\prime}$, $\tilde{{\bm{\omega}}}=[{\bm{\omega}}, \mathcal{P}_s/T]$, then $\tilde{H}$ is invertible.
By Woodbury matrix identity \citep[Theorem 18.2.8]{Harville1997}, we can write
\begin{equation}\label{LemC6-eq7}
\tilde{\mathcal{H}}^{-1}=\mathcal{H}_b^{-1}+b\cdot\mathcal{H}_b^{-1}\tilde{{\bm{\omega}}}(\mathbb{I}_{4+s}-b\cdot\tilde{{\bm{\omega}}}^{\prime}\mathcal{H}_b^{-1}\tilde{{\bm{\omega}}})^{-1}\tilde{{\bm{\omega}}}^{\prime}\mathcal{H}_b^{-1}.
\end{equation}
Since $\|\mathcal{H}_b^{-1}-\mathcal{H}_d^{-1}\|_{\max}=O(1/T)$, then
there exists $\overline{c}$ such that $0<\mathcal{H}_b^{-1}\leq \overline{c}\cdot \mathbb{I}_{2\times(N+T)}$ for all $\mathcal{H}_b$. Then, we can choose $b=b_1$ such that $0<b_1\cdot\tilde{{\bm{\omega}}}^{\prime}\mathcal{H}_{b_1}^{-1}\tilde{{\bm{\omega}}}\leq 1/2\cdot \mathbb{I}_{4+s}$, and thus
\begin{equation}\label{LemC6-eq8}
b_1\cdot\|\mathcal{H}_{b_1}^{-1}\tilde{{\bm{\omega}}}(\mathbb{I}_{4+s}-b_1\cdot\tilde{{\bm{\omega}}}^{\prime}\mathcal{H}_{b_1}^{-1}\tilde{{\bm{\omega}}})^{-1}\tilde{{\bm{\omega}}}^{\prime}\mathcal{H}_{b_1}^{-1}\|_{\max}\leq 2b_1\cdot\|\mathcal{H}_{b_1}^{-1}\tilde{{\bm{\omega}}}\tilde{{\bm{\omega}}}^{\prime}\mathcal{H}_{b_1}^{-1}\|_{\max}.
\end{equation}
It is easy to show that $\|\mathcal{H}_{b_1}^{-1}\tilde{{\bm{\omega}}}\tilde{{\bm{\omega}}}^{\prime}\mathcal{H}_{b_1}^{-1}\|_{\max}=O(1/T)$, then by \eqref{LemC6-eq7}-\eqref{LemC6-eq8}, $\|\tilde{\mathcal{H}}^{-1}-\mathcal{H}_{b_1}^{-1}\|_{\max}=O(1/T)$. Thus, $\|\tilde{\mathcal{H}}^{-1}-\mathcal{H}_{d}^{-1}\|_{\max}=O(1/T)$, $\|\mathcal{H}-\tilde{\mathcal{H}}\|=O(1/T^2)$.

\bigskip

\begin{lemma}\label{LemC-12}
Under Assumptions \ref{Ass2}-\ref{Ass9}, we have
$$E[(\frac{1}{T}\sum_{t=1}^T\tilde{K}_{h,it}^{(1)})^2]=\bar{O}((Th^3)^{-1}),\;\;
E[(\frac{1}{T}\sum_{t=1}^T\tilde{K}_{h,it}^{(1)}\tilde{K}_{h,jt}^{(2)})^2]=\bar{O}((Th^8)^{-1}), j\neq i.$$
\end{lemma}

\bigskip

\noindent \textbf{Proof of Lemma \ref{LemC-12}.}	
We only prove the first result. By Lemma \ref{lem1} (ii) (iv), $Var(K_{h,it}^{(1)})=O(h^{-3})$,
then it suffices to show that
$\sum_{t=1}^T\sum_{s=1,s\neq t}^TCov[K_{h,it}^{(1)},K_{h,is}^{(1)}]=O(Th^{-3}).$
Decompose $\sum_{s=1,s\neq t}^TCov[K_{h,it}^{(1)},K_{h,is}^{(1)}]$ as
\begin{equation*}
\begin{split}
&\sum_{s=1}^{t-\pi_T}Cov[K_{h,it}^{(1)},K_{h,is}^{(1)}]+\sum_{s=t-\pi_T+1,s\neq t}^{t+\pi_T}Cov[K_{h,it}^{(1)}\cdot K_{h,is}^{(1)}]+\sum_{s=\pi_T+t+1}^{T}Cov[K_{h,it}^{(1)},K_{h,is}^{(1)}]\\
:= &J_{21}+J_{22}+J_{23},
\end{split}
\end{equation*}
where $\pi_T\to \infty, \pi_Th\to 0$. For $J_{22}$, by Lemma \ref{LemC-1} (i), Assumptions \ref{Ass6} and \ref{Ass4} (viii),
it is easy to show that 
$Cov[K_{h,it}^{(1)},K_{h,is}^{(1)}]=O(h^{-2})$, then
$|J_{22}|=O(\pi_T h^{-2})=o(h^{-3})$.
For $J_{21}$ and $J_{23}$, by Davydov's Lemma, it holds that
\begin{equation*}
|Cov[K_{h,it}^{(1)},K_{h,is}^{(1)}]|\leq 8\alpha(|t-s|)^{1-2/b}E[|K_{h,it}^{(1)}|^b]^{1/b}\cdot E[|K_{h,is}^{(1)}|^b]^{1/b},
\end{equation*}
where $b>2$. Assumption \ref{Ass3} (i), \ref{Ass4} (v) (vii) imply $E[|K_{h,it}^{(1)}|^b]=O(h^{1-2b})$, then
\begin{equation*}
|J_{21}|\lesssim \sum_{l=\pi_T}[\alpha(l)]^{1-2/b}h^{2/b-4}\lesssim (\rho^{\pi_T}h^{-1})^{1-2/b}\sum_{l=1}[\alpha(l)]^{1-2/b}\cdot h^{-3}.
\end{equation*}
Let $\pi_T=T^{1/\gamma}$, then $\pi_Th\to 0, \rho^{\pi_T}h^{-1}\to 0$ by Assumption \ref{Ass5}. Moreover, by Assumption \ref{Ass2} (ii), $\sum_{l=1}[\alpha(l)]^{1-2/b}$ is bounded, then, $J_{21}=o(h^{-3})$. Likewise, $J_{23}$ is also $o(h^{-3})$, then $\sum_{s=1,s\neq t}^TCov[K_{h,it}^{(1)},K_{h,is}^{(1)}]=o(h^{-3})$. Thus, $E[(\frac{1}{T}\sum_{t=1}^T\tilde{K}_{h,it}^{(1)})^2]=\bar{O}((Th^{3})^{-1})$.

\bigskip

We then consider the stochastic expansion of $\hat{\bf{f}}_t$. By Lemma \ref{LemC-5} and \eqref{lem2-2-eq1}, we can write
\begin{equation}\label{LemC6-eq1}
\hat{{\bm{\theta}}}-{\bm{\theta}}_0=\mathcal{H}_d^{-1}\mathcal{S}^*(\hat{{\bm{\theta}}})+\mathcal{D}\mathcal{S}^*(\hat{{\bm{\theta}}})-\tilde{\mathcal{H}}^{-1}\mathcal{S}^*({\bm{\theta}}_0)-\mathcal{U}\mathcal{V}(\hat{\bm{\theta}})-0.5\tilde{\mathcal{H}}^{-1}\mathcal{R}(\hat{{\bm{\theta}}}),
\end{equation}
where $\mathcal{D}=\tilde{\mathcal{H}}^{-1}-\mathcal{H}_d^{-1}, \mathcal{U}=\tilde{\mathcal{H}}^{-1}\cdot (\mathcal{H}-\tilde{\mathcal{H}}), \mathcal{V}(\hat{\bm{\theta}})=\hat{\bm{\theta}}-{\bm{
\theta}}_0$. 
Second, define
\begin{equation*}
\begin{split}
\mathcal{S}({\bm{\theta}})=\left[\cdots,\frac{1}{\sqrt{NT}}\sum_{t=1}^TK_{h}^{(1)}(X_{it}-{\bm{\lambda}}_{i}^{\prime}{\bf{f}}_t){\bf{f}}_t^{\prime},\cdots,\frac{1}{\sqrt{NT}}\sum_{i=1}^NK_{h}^{(1)}(X_{it}-{\bm{\lambda}}_{i}^{\prime}{\bf{f}}_t){\bm{\lambda}}_{i}^{\prime},\cdots\right]^{\prime},
\end{split}
\end{equation*}
and $\tilde{\mathcal{S}}({\bm{\theta}})=\mathcal{S}({\bm{\theta}})-\mathcal{S}^*({\bm{\theta}})$. 
By first-order condition,
$\mathcal{S}(\hat{{\bm{\theta}}})=0$, then by \eqref{LemC6-eq1},
\begin{equation}\label{LemC6-eq6}
\hat{{\bm{\theta}}}-{\bm{\theta}}_0
=-\mathcal{H}_d^{-1}\tilde{\mathcal{S}}(\hat{{\bm{\theta}}})-\mathcal{D}\tilde{\mathcal{S}}(\hat{{\bm{\theta}}})-\mathcal{U}\cdot \mathcal{V}(\hat{\bm{\theta}})-\tilde{\mathcal{H}}^{-1}\mathcal{S}^*({\bm{\theta}}_0)-0.5\tilde{\mathcal{H}}^{-1}\mathcal{R}(\hat{{\bm{\theta}}}).
\end{equation}
Next, let $\mathcal{R}(\hat{{\bm{\theta}}})_j,\mathcal{V}(\hat{{\bm{\theta}}})_j$ denote the vector containing the $(j-1)r_0+1$-th to the $jr_0$-th elements of $\mathcal{R}(\hat{{\bm{\theta}}}), \mathcal{V}(\hat{{\bm{\theta}}})$ for $j=1,\cdots,N+T$. Lemma \ref{lem1} (ii) (iii) and Lemma \ref{LemC-2} imply
\begin{equation}\label{LemC6-eq3}
\mathcal{R}(\hat{{\bm{\theta}}})_i=\bar{O}_p(1)\|\hat{\bm{\lambda}}_i-{\bm{\lambda}}_{0i}\|^2+\bar{O}_p((L_{NT}h^3)^{-1})\|\hat{\bm{\lambda}}_i-{\bm{\lambda}}_{0i}\|+\bar{O}_p((L_{NT}h^3)^{-2}),
\end{equation}
for $i=1,\cdots,N$, and
\begin{equation}\label{LemC6-eq4}
\mathcal{R}(\hat{{\bm{\theta}}})_{N+t}=\bar{O}_p(1)\|\hat{\bf{f}}_t-{\bf{f}}_{0t}\|^2+\bar{O}_p((L_{NT}h^3)^{-1})\|\hat{\bf{f}}_t-{\bf{f}}_{0t}\|+\bar{O}_p((L_{NT}h^3)^{-2})
\end{equation}
for $t=1,\cdots,T$.
Moreover, write $\mathcal{D}_{j,s}, \mathcal{U}_{j,s}$ as the $r_0\times r_0$ matrix containing the $(j-1)r_0+1$ to $jr_0$ rows and $(s-1)r_0+1$ to $sr_0$ columns of $\mathcal{D}$ and $\mathcal{U}$. Write $[\tilde{\mathcal{H}}^{-1}\mathcal{S}^*({\bm{\theta}}_0)]_{j}$ as the vector collecting the $(j-1)r_0+1$ to $jr_0$ rows of $\tilde{\mathcal{H}}^{-1}\mathcal{S}^*({\bm{\theta}}_0)$. Lemma \ref{lem1} (iii),
Lemma \ref{LemC-1} (i) and Lemma \ref{LemC-5} imply 
$\|\tilde{\mathcal{H}}^{-1}\mathcal{S}^*({\bm{\theta}}_0)\|_{\max}=\bar{O}(h^5)$.
Then by \eqref{LemC6-eq6}, we can write
\begin{align}\label{LemC6-eq5}
&\hat{\bf{f}}_t-{\bf{f}}_{0t}=-({\bm{\Psi}}_{N,t})^{-1}\frac{1}{N}\sum_{j=1}^N\tilde{K}_h^{(1)}(X_{jt}-\hat{\bm{\lambda}}_j^{\prime}\hat{\bf{f}}_t)\hat{\bm{\lambda}}_j-\frac{1}{\sqrt{NT}}\sum_{j=1}^N\sum_{s=1}^T\mathcal{D}_{N+t,j}\tilde{K}_h^{(1)}(X_{js}-\hat{\bm{\lambda}}_j^{\prime}\hat{\bf{f}}_s)\hat{\bf{f}}_s\nonumber\\
&-\frac{1}{\sqrt{NT}}\sum_{j=1}^N\sum_{s=1}^T\mathcal{D}_{N+t,N+s}\tilde{K}_h^{(1)}(X_{js}-\hat{\bm{\lambda}}_j^{\prime}\hat{\bf{f}}_s)\hat{\bm{\lambda}}_j-\sum_{j=1}^{N+T}\mathcal{U}_{N+t,j}\mathcal{V}(\hat{{\bm{\theta}}})_j\nonumber\\
&-0.5({\bm{\Psi}}_{N,t})^{-1}\sqrt{\frac{T}{N}}\mathcal{R}(\hat{{\bm{\theta}}})_{N+t}-0.5\sum_{j=1}^{N+T}\mathcal{D}_{N+t,j}\mathcal{R}(\hat{{\bm{\theta}}})_j+\bar{O}(h^5).
\end{align}


\bigskip

\begin{lemma}\label{LemC-7}
Let $b_1,\cdots,b_T$ be a sequence of uniformly bounded constants. Under Assumptions \ref{Ass1}-\ref{Ass9}, we have
\begin{equation*}
\frac{1}{T}\sum_{t=1}^Tb_t(\hat{\bf{f}}_t-{\bf{f}}_{0t})=O_p(h^5).
\end{equation*}
\end{lemma}

\bigskip

\noindent \textbf{Proof of Lemma \ref{LemC-7}.}	
Define $d_j=\sqrt{NT}\cdot T^{-1}\sum_{t=1}^Tb_t\mathcal{D}_{N+t,j}$ and $ u_j=\sqrt{NT}\cdot T^{-1}\sum_{t=1}^Tb_t\mathcal{U}_{N+t,j}$ for $j=1,\cdots, N+T$. Lemma \ref{LemC-5} and Assumption \ref{Ass9} imply that $\max_{1\leq j\leq N+T}\|d_j\|$ is bounded, and $\max_{1\leq j\leq N+T}\|u_j\|$ is $O(1/T)$.
By \eqref{LemC6-eq5}, we can write
\begin{align}\label{LemC7-eq1}
&\frac{1}{T}\sum_{t=1}^Tb_t(\hat{\bf{f}}_t-{\bf{f}}_{0t})=-\frac{1}{NT}\sum_{j=1}^N\sum_{t=1}^Tb_t({\bm{\Psi}}_{N,t})^{-1}\tilde{K}_{h,jt}^{(1)}{\bm{\lambda}}_{0j}-\frac{1}{NT}\sum_{j=1}^N\sum_{s=1}^Td_j\tilde{K}_{h,js}^{(1)}{\bf{f}}_{0s}\nonumber\\
-&\frac{1}{NT}\sum_{j=1}^N\sum_{s=1}^Td_{N+s}\tilde{K}_{h,js}^{(1)}{\bm{\lambda}}_{0j}-\frac{1}{NT}\sum_{j=1}^N\sum_{t=1}^Tb_t({\bm{\Psi}}_{N,t})^{-1}\{\tilde{K}_h^{(1)}(X_{jt}-\hat{\bm{\lambda}}_j^{\prime}\hat{\bf{f}}_t)\hat{\bm{\lambda}}_j-\tilde{K}_{h,jt}^{(1)}{\bm{\lambda}}_{0j}\}\nonumber\\
-&\frac{1}{NT}\sum_{j=1}^N\sum_{s=1}^Td_j\{\tilde{K}_{h}^{(1)}(X_{js}-\hat{\bm{\lambda}}_j^{\prime}\hat{\bf{f}}_s)\hat{\bf{f}}_s-\tilde{K}_{h,js}^{(1)}{\bf{f}}_{0s}\}\nonumber\\
-&\frac{1}{NT}\sum_{j=1}^N\sum_{s=1}^Td_{N+s}\{\tilde{K}_{h}^{(1)}(X_{js}-\hat{\bm{\lambda}}_j^{\prime}\hat{\bf{f}}_s)\hat{\bm{\lambda}}_j-\tilde{K}_{h,js}^{(1)}{\bm{\lambda}}_{0j}\}-\frac{1}{\sqrt{NT}}\sum_{j=1}^{N+T}u_j\mathcal{V}(\hat{{\bm{\theta}}})_j\nonumber\\
-&
\frac{1}{2\sqrt{NT}}\sum_{t=1}^Tb_t({\bm{\Psi}}_{N,t})^{-1}\mathcal{R}(\hat{{\bm{\theta}}})_{N+t}-\frac{1}{2\sqrt{NT}}\sum_{j=1}^{N+T}d_j\mathcal{R}(\hat{{\bm{\theta}}})_j+\bar{O}(h^5).
\end{align}

First, by Lemma \ref{LemC-12} (i), the first three terms on the RHS of \eqref{LemC7-eq1} are $O_p(1/\sqrt{NTh^3})$. Next, it follows from Lemma \ref{LemC-2} and Lemma \ref{LemC-5} that the $7$th term is $O_p((L_{NT}h)^{-3/2})$. Further, by \eqref{LemC6-eq3} and \eqref{LemC6-eq4}, the $8$th-$9$th terms are $O_p((L_{NT}h^3)^{-1})$.

We then consider the remaining three terms. For any ${\bm{\theta}}\in {\bm{\Theta}}^{r_0}$,
define
\begin{equation*}
\begin{split}
&\Delta_{NT}({\bm{\theta}},{\bm{\theta}}_0)=\frac{1}{NT}\sum_{j=1}^N\sum_{s=1}^Td_j\{\tilde{K}^{(1)}_{h}(X_{js}-{\bm{\lambda}}_j^{\prime}{\bf{f}}_s){\bf{f}}_s-\tilde{K}_{h,js}^{(1)}{\bf{f}}_{0s}\}\\
=&\frac{1}{NT}\sum_{j=1}^N\sum_{s=1}^Td_j\tilde{K}^{(1)}_{h}(X_{js}-{\bm{\lambda}}_{j}^{\prime}{\bf{f}}_{s})({\bf{f}}_{s}-{\bf{f}}_{0s})+\frac{1}{NT}\sum_{j=1}^N\sum_{s=1}^Td_j[\tilde{K}^{(1)}_{h}(X_{js}-{\bm{\lambda}}_{j}^{\prime}{\bf{f}}_{s})-\tilde{K}^{(1)}_{h,js}]{\bf{f}}_{0s}\\
:=&\Delta_{1,NT}({\bm{\theta}},{\bm{\theta}}_0)+\Delta_{2,NT}({\bm{\theta}},{\bm{\theta}}_0).
\end{split}
\end{equation*}
Thus,
$\|\Delta_{NT}({\bm{\theta}},{\bm{\theta}}_0)\|_{\psi_1}\leq \|\Delta_{1,NT}({\bm{\theta}},{\bm{\theta}}_0)\|_{\psi_1}+\|\Delta_{2,NT}({\bm{\theta}},{\bm{\theta}}_0)\|_{\psi_1}.$
Similar to Lemma \ref{lem2}, we can show that, for $d({\bm{\theta}},{\bm{\theta}}_0)$ sufficiently small,
\begin{equation*}
\begin{split}
\|\Delta_{1,NT}({\bm{\theta}},{\bm{\theta}}_0)\|_{\psi_1}\lesssim \frac{\|{\bf{F}}-{\bf{F}}_0\|}{T\sqrt{Nh^3}}+\frac{T^{1/\gamma}}{NTh^2}
\lesssim \frac{d({\bm{\theta}},{\bm{\theta}}_0)}{\sqrt{NTh^3}}+\frac{T^{1/\gamma}}{NTh^2},   \end{split}
\end{equation*}
and
\begin{equation*}
\|\Delta_{2,NT}({\bm{\theta}},{\bm{\theta}}_0)\}\|_{\psi_1}\lesssim \frac{d({\bm{\theta}},{\bm{\theta}}_0)}{\sqrt{NTh^5}}+\frac{T^{1/\gamma}}{NTh^2}.
\end{equation*}
Then, similar to Lemma \ref{lem3}, we can show that, for sufficiently small $\delta>0$,
\begin{equation}\label{LemC7-eq2}
E[\underset{{\bm{\theta}}\in {\bm{\Theta}}^{r_0}(\delta)}{\sup}\Delta_{NT}({\bm{\theta}})]\lesssim \frac{\delta}{\sqrt{L_{NT}h^5}}+\frac{\log{L_{NT}}}{L_{NT}h^3}.
\end{equation}
Then, by \eqref{LemC7-eq2} and Lemma \ref{LemC-2}, $\Delta_{NT}(\hat{{\bm{\theta}}})=O_p((L_{NT}h^{4})^{-1})$, and the fifth term on the RHS of \eqref{LemC7-eq1} is $O_p((L_{NT}h^{4})^{-1})$. Similar results can be obtained for the fourth and sixth terms. Combining all the results, by Assumption \ref{Ass9}, the desired result follows.

\bigskip

\begin{lemma}\label{LemC-8}
Under Assumptions \ref{Ass1}-\ref{Ass9}, for each $i$, we have
\begin{equation*}
\begin{split}
&\frac{1}{T}\sum_{t=1}^T\tilde{K}_{h,it}^{(1)}(X_{it}-
\hat{\bm{\lambda}}_i^{\prime}\hat{\bf{f}}^*_t)(\hat{\bf{f}}_t-{\bf{f}}_{0t})=O_p((L_{NT}h^6)^{-1})+o_p(\|\hat{\bm{\lambda}}_i-{\bm{\lambda}}_{0i}\|),\\
&\frac{1}{T}\sum_{t=1}^T\tilde{K}_{h,it}^{(2)}(X_{it}-
\hat{\bm{\lambda}}_i^{\prime}\hat{\bf{f}}^*_t){\bf{f}}_{0t}(\hat{\bf{f}}_t-{\bf{f}}_{0t})^{\prime}=O_p((L_{NT}h^7)^{-1})+o_p(\|\hat{\bm{\lambda}}_i-{\bm{\lambda}}_{0i}\|),
\end{split}
\end{equation*}
where $\hat{\bf{f}}_t^*$ lies between $\hat{\bf{f}}_t$ and ${\bf{f}}_{0t}$.
\end{lemma}

\bigskip

\noindent \textbf{Proof of Lemma \ref{LemC-8}.}	
We only prove the second result.
First, the second result can be written as
\begin{align}\label{LemC8-eq4}
\frac{1}{T}\sum_{t=1}^T\tilde{K}_{h,it}^{(2)}{\bf{f}}_{0t}(\hat{\bf{f}}_t-{\bf{f}}_{0t})^{\prime}+\frac{1}{T}\sum_{t=1}^T\tilde{K}_{h,it}^{(3)}(X_{it}-c_{it}^*){\bf{f}}_{0t}(\hat{\bf{f}}_t-{\bf{f}}_{0t})^{\prime}(\hat{\bm{\lambda}}_i^{\prime}\hat{\bf{f}}^*_t-{\bm{\lambda}}_{0i}^{\prime}{\bf{f}}_{0t}),
\end{align}
where $c_{it}^*$ is between $\hat{\bm{\lambda}}_i^{\prime}\hat{\bf{f}}^*_t$ and ${\bm{\lambda}}_{0i}^{\prime}{\bf{f}}_{0t}$. By Lemma \ref{LemC-2}, the second term on the RHS of \eqref{LemC8-eq4} is bounded by
\begin{equation*}
\begin{split}
&\frac{1}{T}\sum_{t=1}^T|\tilde{K}_{h,it}^{(3)}(X_{it}-c_{it}^*)|\cdot \|{\bf{f}}_{0t}\|\cdot \|\hat{\bf{f}}_t-{\bf{f}}_{0t}\|\cdot (\|\hat{\bf{f}}^*_t-{\bf{f}}_{0t}\|\|\hat{\bm{\lambda}}_i\|+\|\hat{\bm{\lambda}}_i-{\bm{\lambda}}_{0i}\|\|{\bf{f}}_{0t}\|)\\
\lesssim & O_p((L_{NT}h^7)^{-1})+O_p(\|\hat{\bm{\lambda}}_i-{\bm{\lambda}}_{0i}\|\cdot 1/\sqrt{L_{NT}h^{11}})=O_p((L_{NT}h^7)^{-1})+o_p(\|\hat{\bm{\lambda}}_i-{\bm{\lambda}}_{0i}\|).
\end{split}
\end{equation*}

We then consider the first term on the RHS of \eqref{LemC8-eq4}. From \eqref{LemC6-eq5}, we have
\begin{align}\label{LemC8-eq1}
&\frac{1}{T}\sum_{t=1}^T\tilde{K}_{h,it}^{(2)}{\bf{f}}_{0t}(\hat{\bf{f}}_t-{\bf{f}}_{0t})^{\prime}\nonumber=-\frac{1}{NT}\sum_{t=1}^T\sum_{j=1}^N\tilde{K}_{h,it}^{(2)}\tilde{K}_{h,jt}^{(1)}{\bf{f}}_{0t}{\bm{\lambda}}_{0j}({\bm{\Psi}}_{N,t})^{-1}\nonumber\\
-&\frac{1}{NT}\sum_{t=1}^T\sum_{j=1}^N\tilde{K}_{h,it}^{(2)}{\bf{f}}_{0t}\cdot \{\tilde{K}_{h}^{(1)}(X_{jt}-\hat{\bm{\lambda}}_j^{\prime}\hat{\bf{f}}_t)\hat{\bm{\lambda}}_j^{\prime}-\tilde{K}_{h,jt}^{(1)}{\bm{\lambda}}^{\prime}_{0j}\}({\bm{\Psi}}_{N,t})^{-1}\nonumber\\
+&\frac{1}{\sqrt{NT}}\sum_{j=1}^N\sum_{s=1}^T(\frac{1}{T}\sum_{t=1}^T\tilde{K}_{h,it}^{(2)}{\bf{f}}_{0t}\hat{\bf{f}}_s^{\prime}\mathcal{D}_{N+t,j}^{\prime})\bar{K}_h^{(1)}(X_{js}-\hat{\bm{\lambda}}_j^{\prime}\hat{\bf{f}}_s)\nonumber\\
+&\frac{1}{\sqrt{NT}}\sum_{j=1}^N\sum_{s=1}^T(\frac{1}{T}\sum_{t=1}^T\tilde{K}_{h,it}^{(2)}{\bf{f}}_{0t}\hat{\bm{\lambda}}_j^{\prime}
\mathcal{D}^{\prime}_{N+t,N+s})\bar{K}_h^{(1)}(X_{js}-\hat{\bm{\lambda}}_j^{\prime}\hat{\bf{f}}_s)\nonumber\\
-&\frac{1}{T}\sum_{t=1}^T\sum_{j=1}^{N+T}\tilde{K}_{h,it}^{(2)}{\bf{f}}_{0t}\mathcal{V}(\hat{{\bm{\theta}}})^{\prime}_j\mathcal{U}_{N+t,j}^{\prime}-\frac{1}{2\sqrt{NT}}\sum_{t=1}^T\tilde{K}_{h,it}^{(2)}{\bf{f}}_{0t}\mathcal{R}(\hat{{\bm{\theta}}})^{\prime}_{N+t}({\bm{\Psi}}_{N,t})^{-1}\nonumber\\
-&\frac{1}{2T}\sum_{t=1}^T\sum_{j=1}^{N}\tilde{K}_{h,it}^{(2)}{\bf{f}}_{0t}\mathcal{R}(\hat{{\bm{\theta}}})^{\prime}_j\mathcal{D}_{N+t,j}^{\prime}-\frac{1}{2T}\sum_{t=1}^T\sum_{s=1}^T\tilde{K}_{h,it}^{(2)}{\bf{f}}_{0t}\mathcal{R}(\hat{{\bm{\theta}}})^{\prime}_{N+s}\mathcal{D}_{N+t,N+s}^{\prime}+\bar{O}(h^2).
\end{align}

First, the first term on the RHS of \eqref{LemC8-eq1}  can be written as
\begin{equation*}
\frac{1}{NT}\sum_{t=1}^T\tilde{K}_{h,it}^{(2)}\tilde{K}_{h,it}^{(1)}{\bf{f}}_{0t}{\bm{\lambda}}_{0j}({\bm{\Psi}}_{N,t})^{-1}+\frac{1}{NT}\sum_{t=1}^T\sum_{j=1, j\neq i}^N\tilde{K}_{h,it}^{(2)}\tilde{K}_{h,jt}^{(1)}{\bf{f}}_{0t}{\bm{\lambda}}_{0j}({\bm{\Psi}}_{N,t})^{-1}.
\end{equation*}
The first term of above equation is $O_p(1/(Nh^5))$ by Assumption \ref{Ass4} (v) (vi), while the second term is $O_p(1/\sqrt{NTh^8})$ by Lemma \ref{LemC-12} (ii).
Thus, the first term of \eqref{LemC8-eq1} is $O_p(1/(Nh^5)$.

Second, the second term on the RHS of \eqref{LemC8-eq1} can be written as
\begin{equation*}
\begin{split}
\frac{1}{NT}\sum_{t=1}^T&\tilde{K}_{h,it}^{(2)}{\bf{f}}_{0t}\cdot\{\tilde{K}_{h}^{(1)}(X_{it}-\hat{\bm{\lambda}}_i^{\prime}\hat{\bf{f}}_t)\hat{\bm{\lambda}}_j^{\prime}-\tilde{K}_{h,it}^{(1)}{\bm{\lambda}}^{\prime}_{0i}\}({\bm{\Psi}}_{N,t})^{-1}\\
+&\frac{1}{NT}\sum_{t=1}^T\sum_{j=1, j\neq i}^N\tilde{K}_{h,it}^{(2)}{\bf{f}}_{0t}\cdot \{\tilde{K}_{h}^{(1)}(X_{jt}-\hat{\bm{\lambda}}_j^{\prime}\hat{\bf{f}}_t)\hat{\bm{\lambda}}_j^{\prime}-\tilde{K}_{h,jt}^{(1)}{\bm{\lambda}}^{\prime}_{0j}\}({\bm{\Psi}}_{N,t})^{-1}.
\end{split}
\end{equation*}
The first term is $O_p((Nh^5)^{-1})$, while similar to the proof of Lemma \ref{LemC-7}, the second term can be shown to be $O_p((L_{NT}h^7)^{-1})$. Thus, the second term on RHS of \eqref{LemC8-eq1} is $O_p((L_{NT}h^7)^{-1})$.

Next, for the third term on the RHS of \eqref{LemC8-eq1}, its $(p,q)$ element is given by
\begin{equation*}
\frac{1}{NT}\sum_{j=1}^N\sum_{s=1}^T\chi_{i,j}\cdot \bar{K}_h^{(1)}(X_{js}-\hat{\bm{\lambda}}_j^{\prime}\hat{\bf{f}}_s)\hat{\bf{f}}_s,
\end{equation*}
where $\chi_{i,j}=T^{-1}\sum_{t=1}^T(\sqrt{NT}\mathcal{D}_{N+t,j,q}) {f}_{0t,p}\cdot\tilde{K}_{h,it}^{(2)}$, and ${f}_{0t,p}$ is the $p$-th element of ${\bf{f}}_{0t}$, $\mathcal{D}_{N+t,j,q}$ is the $q$-th row of $D_{N+t,j}$.
Then, the third term is bounded by
\begin{equation*}
\sqrt{\frac{1}{N}\sum_{j=1}^N\|\chi_{i,j}\|^2}\cdot\sqrt{\frac{1}{NT}\sum_{j=1}^N\sum_{s=1}^T[\bar{K}_h^{(1)}(X_{js}-\hat{\bm{\lambda}}_j^{\prime}\hat{\bf{f}}_s)]^2\|\hat{\bf{f}}_s\|^2}.
\end{equation*}
Since $|\sqrt{NT}{f}_{0t,p}\mathcal{D}_{N+t,j,q}|$ is uniformly bounded by Lemma \ref{LemC-5}, then by Lemma \ref{LemC-1} (v), we have $\|\chi_{i,j}\|=O_p(1/\sqrt{Th^5})$.  Moreover, by Lemma \ref{LemC-1} (i) (ii), it holds that
\begin{equation*}
[\bar{K}_h^{(1)}(X_{js}-\hat{\bm{\lambda}}_j^{\prime}\hat{\bf{f}}_s)]^2\lesssim ({\bm{\lambda}}_{0j}^{\prime}{\bf{f}}_{0s}-\hat{\bm{\lambda}}_j^{\prime}\hat{\bf{f}}_s)^2+\bar{O}(h^{10}).
\end{equation*}
Then, it follows from Lemma \ref{LemC-2} and Assumption \ref{Ass9} that
\begin{equation*}
\sqrt{\frac{1}{NT}\sum_{j=1}^N\sum_{s=1}^T[\bar{K}_h^{(1)}(X_{js}-\hat{\bm{\lambda}}_j^{\prime}\hat{\bf{f}}_s)]^2\|\hat{\bf{f}}_s\|^2}\lesssim O(h^5)+d(\hat{{\bm{\theta}}},{\bm{\theta}}_0)=O_p(1/\sqrt{L_{NT}h^3}).
\end{equation*}
Therefore, the third term on the RHS of \eqref{LemC8-eq1} is $O_p((L_{NT}h^{4})^{-1})$. The $4$th term can be shown to be $O_p((L_{NT}h^{4})^{-1})$ in the same way. Additionally, since $\|\mathcal{U}\|_{\max}=O(1/T^2)$, then by Lemma \ref{LemC-2}, the $5$th  term is $O_p((L_{NT}^{3/2}h^{9/2})^{-1})$. Moreover, it follows from \eqref{LemC6-eq3} that the $6$th term is $O_p((L_{NT}h^6)^{-1})$. The $(p,q)$ element of the $7$th term can be written as $(2\sqrt{NT})^{-1}\sum_{j=1}^N\chi_{i,j}\mathcal{R}(\hat{{\bm{\theta}}})_{j}$, then, by \eqref{LemC6-eq3} and Lemma \ref{LemC-2}, it is bounded by
\begin{equation*}
\begin{split}
&\frac{\sqrt{N}}{2\sqrt{T}}\sqrt{\frac{1}{N}\sum_{j=1}^N\|\chi_{i,j}\|^2}\cdot\sqrt{\frac{1}{N}\sum_{j=1}^N\|\mathcal{R}(\hat{{\bm{\theta}}}_j)\|^2}
=O_p((L_{NT}h^4)^{-1}).
\end{split}
\end{equation*}
The same bound for the $8$th term can be obtained using the same argument.

Combining all the above results, by Assumption \ref{Ass9}, we have Lemma \ref{LemC-8} established.

\bigskip

\begin{lemma}\label{LemC-11}
Under Assumptions \ref{Ass1}-\ref{Ass10}, we have
\begin{equation}\label{LemC11-eq0}
\frac{\sqrt{Th^3}}{T}\sum_{t=1}^T\tilde{K}_{h,it}^{(1)}{\bf{f}}_{0t}\overset{d}{\rightarrow} \mathcal{N}(0, {\bm{\Sigma}}_i),
\end{equation}
where ${\bm{\Sigma}}_i=\lim_{T\to\infty}\frac{1}{T}\sum_{t=1}^TLg_{it}(0){\bf{f}}_{0t}{\bf{f}}_{0t}^{\prime}, L=\int |K^{(1)}(u)|^2du$.
\end{lemma}

\bigskip

\noindent \textbf{Proof of Lemma \ref{LemC-11}.}	
Let ${\bf{Z}}_{i,t}=\sqrt{h^{3}}\tilde{K}_{h,it}^{(1)}{\bf{f}}_{0t},{\bf{W}}_T=\sum_{t=1}^T{\bf{Z}}_{i,t}$, then it suffices to show
$\frac{1}{\sqrt{T}}{\bf{W}}_T \overset{d}{\rightarrow} \mathcal{N}(0,{\bm{\Sigma}}_i).$
Partition $\{1, \cdots, T\}$ into $2k+1$ subsets with large blocks of size $u=u_T$ and small blocks of size $v=v_T$. Let
$k=k_T=\lfloor\frac{T}{u_T+v_T}\rfloor,$
where $\lfloor\cdot\rfloor$ is the floor function.
Define
\begin{equation}\label{LemC11-eq6}
{\bm{\eta}}_j=\sum_{t=j(u+v)+1}^{j(u+v)+u-1}{\bf{Z}}_{i,t}, \quad 
{\bm{\xi}}_j=\sum_{t=j(u+v)+u}^{(j+1)(u+v)-1}{\bf{Z}}_{i,t}, 
\quad
{\bm{\zeta}}_j=\sum_{k(u+v)}^{T}{\bf{Z}}_{i,t},
\end{equation}
for $0\leq j\leq k-1$. Then
${\bf{W}}_T=\sum_{j=0}^{k-1}{\bm{\eta}}_j+\sum_{j=0}^{k-1}{\bm{\xi}}_j+{\bm{\zeta}}_j:={\bf{W}}_{T1}+{\bf{W}}_{T2}+{\bf{W}}_{T3}.$ Write ${\bm{\eta}}_j=(\eta_{1,j},\cdots, \eta_{r_0,j})^{\prime}, {\bf{W}}_{T1}=(w_{1,T1},\cdots, w_{r_0,T1})^{\prime}$.
According to \citet{Masry1996}, to prove the asymptotic normality of ${\bf{W}}_T$,
it suffices to show that, as $T\to \infty$, for every $\varepsilon>0$,
\begin{align}
&\frac{1}{T}E[{\bf{W}}_{T2}{\bf{W}}_{T2}^{\prime}]\to 0, \quad \frac{1}{T}E[{\bf{W}}_{T3}{\bf{W}}_{T3}^{\prime}]\to 0, \label{LemC11-eq1}\\
&|E[\exp{(itw_{s,T1})}]-\prod_{j=0}^{k-1}E[\exp{(it\eta_{s,j})}]|\to 0,\qquad \text{for } s=1,\cdots, r_0, \label{LemC11-eq2}\\
&\frac{1}{T}\sum_{j=0}^{k-1}E[{\bm{\eta}}_j{\bm{\eta}}_j^{\prime}]\to {\bm{\Sigma}}_i, \quad
\frac{1}{T}\sum_{j=0}^{k-1}E[{\bm{\eta}}_j{\bm{\eta}}_j^{\prime} 1\{\|{\bm{\eta}}_j\|>\varepsilon \|{\bm{\Sigma}}_i^{1/2}\|\sqrt{T}\}]\to 0. \label{LemC11-eq3}
\end{align}

First, we choose $u_T$. By Assumption \ref{Ass2} (ii), there exist integers $q_T\to \infty$ such that
\begin{equation}\label{LemC11-eq5}
q_Tv_T=o((Th)^{1/2}), \quad q_T(T/h)^{1/2}\alpha(v_T)\to 0, \qquad \text{as } T\to \infty.
\end{equation}
Define 
$u_T=[(Th)^{1/2}/q_T]$. By \eqref{LemC11-eq5}, we can show that,
as $T\to \infty$
\begin{equation}\label{LemC11-eq7}
v_T/u_T\to 0,\quad u_T/T\to 0,\quad u_T/(Th)^{1/2}\to 0,\quad \frac{T}{u_T}\alpha(v_T)\to 0.
\end{equation}

Second, we establish \eqref{LemC11-eq1}. Note that
\begin{equation*}
E[{\bf{W}}_{T2}{\bf{W}}_{T2}^{\prime}]=Var[\sum_{j=0}^{k-1}{\bm{\xi}}_j]=\sum_{j=0}^{k-1}Var[{\bm{\xi}}_j]+\sum_{q,j=0,q\neq j}^{k-1}Cov({\bm{\xi}}_q,{\bm{\xi}}_j):={\bf{F}}_1+{\bf{F}}_2.
\end{equation*}
By Lemma \ref{LemC-1} (iv),  
we have $Var({\bf{Z}}_{i,t})=Lg_{it}(0){\bf{f}}_{0t}{\bf{f}}_{0t}^{\prime}+\bar{O}(h)$. Further, following the proof of Lemma \ref{LemC-12}, we can obtain $\sum_{s=1,s\neq t}^TCov({\bf{Z}}_{i,t},{\bf{Z}}_{i,s})=o(1)$. Hence,
\begin{align}\label{LemC11-eq11}
Var[{\bm{\xi}}_j]
&=\sum_{t=j(u+v)+u}^{(j+1)(u+v)-1}(Lg_{it}(0){\bf{f}}_{0t}{\bf{f}}_{0t}^{\prime}+o_p(1))+\sum_{t=j(u+v)+u, t\neq s}^{(j+1)(u+v)-1}\{\sum_{s=j(u+v)+u}^{(j+1)(u+v)-1}Cov({\bf{Z}}_{i,t}, {\bf{Z}}_{i,s})\}\nonumber\\
&=v_T\cdot(\frac{1}{v_T}\sum_{t=j(u+v)+u}^{(j+1)(u+v)-1}Lg_{it}(0){\bf{f}}_{0t}{\bf{f}}_{0t}^{\prime})+o(v_T)=v_T{\bf{A}}_{j}+o(v_T),
\end{align}
Then
${\bf{F}}_1=\sum_{j=0}^{k-1}Var[{\bm{\xi}}_j]=v_T\sum_{j=0}^{k-1}{\bf{A}}_{j}+k_To(v_T)=o(T)$.
For ${\bf{F}}_2$, let $r_j=j(u+v)+u, j\geq 0$. Then if $q\neq j, 0\leq l_1,l_2\leq v-1$, we have $|r_q-r_j+l_1-l_2|\geq u$. Hence,
\begin{align*}\label{LemC11-eq10}
{\bf{F}}_2=\sum_{q,j=0,q\neq j}^{k-1}\sum_{l_1,l_2=0}^{v-1}Cov({\bf{Z}}_{i,r_q+l_1},{\bf{Z}}_{i,r_j+l_2})=2\sum_{l_1=1}^{T-u}\sum_{l_2=l_1+u}^{T}Cov({\bf{Z}}_{i,l_1},{\bf{Z}}_{i,l_2})=o(T).
\end{align*}
Thus,
$\frac{1}{T}E[{\bf{W}}_{T1}{\bf{W}}_{T1}^{\prime}]\to 0$, as $T\to\infty$. Similarly, $\frac{1}{T}E[{\bf{W}}_{T2}{\bf{W}}_{T2}^{\prime}]\to 0,$as $T\to \infty$.

Third,  to obtain \eqref{LemC11-eq2}, Lemma 1.1 of \citet{VR1959} implies
\begin{equation*}
|E[\exp{(itw_{s,T1})}]-\prod_{j=0}^{k-1}E[\exp{(it\eta_{s,j})}]|\leq 16k_T\alpha(v_T-1)\lesssim \frac{T}{u_T}\alpha(v_T)\to 0,
\end{equation*}
 for $s=1,\cdots,r_0$. Finally, to establish \eqref{LemC11-eq3}, similar to \eqref{LemC11-eq11}, we can write
\begin{equation*}
\frac{1}{T}\sum_{j=0}^{k-1}Var[{\bm{\eta}}_j]
=\frac{1}{T}\sum_{t=1}^{T}Lg_{it}(0){\bf{f}}_{0t}{\bf{f}}_{0t}^{\prime}-\frac{1}{T}\sum_{j=T-v}^{T}Lg_{it}(0){\bf{f}}_{0t}{\bf{f}}_{0t}^{\prime}+o(\frac{u_T}{u_T+v_T})\to {\bm{\Sigma}}_i.
\end{equation*}
In addition, 
since $\|{\bf{f}}_{0t}\|\leq K_2,|K^{(1)}(\cdot)|\leq c_1$,
then
$\|{\bf{Z}}_{i,t}\|\leq 2c_1K_2h^{-1/2}.$
This, together with \eqref{LemC11-eq6} and \eqref{LemC11-eq7}, implies that
$\underset{0\leq k\leq j-1}{\max}\|{\bm{\eta}}_j\|/\sqrt{T}\lesssim\frac{u_T}{(Th)^{1/2}}\to 0.$
Hence, when $T$ is large, the set $\{\|{\bm{\eta}}_j\|\geq \varepsilon\|{\bm{\Sigma}}_i^{1/2}\|\sqrt{T}\}$ becomes an empty and thus \eqref{LemC11-eq3} hold.

Combining all the above results, we establish  \eqref{LemC11-eq1}-\eqref{LemC11-eq3}, then we obtain \eqref{LemC11-eq0}.

\bigskip

Define $\mathbb{M}^{2}_{i,T}({\bm{\lambda}},{\bf{F}}), {\bar{\mathbb{M}}}^{2}_{i,T}({\bm{\lambda}},{\bf{F}})$ in the same manner as in Lemma \ref{LemC-13}, and
\begin{equation*}
\mathbb{M}^{1}_{i,T}({\bm{\lambda}},{\bf{F}})=\frac{1}{Th}\sum_{t=1}^T[K^{(1)}(\frac{X_{it}-{\bm{\lambda}}^{\prime}{\bf{f}}_t}{h})]^2{\bf{f}}_{0t}{\bf{f}}_{0t}^{\prime},
\bar{\mathbb{M}}^{1}_{i,T}({\bm{\lambda}},{\bf{F}})=\frac{1}{Th}\sum_{t=1}^TE[K^{(1)}(\frac{X_{it}-{\bm{\lambda}}^{\prime}{\bf{f}}_t}{h})]^2{\bf{f}}_{0t}{\bf{f}}_{0t}^{\prime}.
\end{equation*}

\medskip

\begin{lemma}\label{LemC-14}
Under Assumptions \ref{Ass1}-\ref{Ass9}, we have
\begin{equation*}
\mathbb{M}^{1}_{i,T}(\hat{\bm{\lambda}}_i, \hat{{\bf{F}}})=\frac{1}{T}\sum_{t=1}^TLg_{it}(0){\bf{f}}_{0t}{\bf{f}}_{0t}^{\prime}+o_p(1),\; \; 
\mathbb{M}^{2}_{i,T}(\hat{{\bm{\lambda}}}_i,\hat{{\bf{F}}})=\frac{1}{T}\sum_{t=1}^Tg^{(2)}_{it}(0){\bf{f}}_{0t}{\bf{f}}_{0t}^{\prime}+o_p(1).
\end{equation*}
\end{lemma}

\bigskip

\noindent \textbf{Proof of Lemma \ref{LemC-14}.}	
To save place, we only prove the second result.
First, we show that
\begin{equation}\label{LemC14-eq5}
\underset{{\bm{\lambda}}\in \mathcal{A}}{\sup}|{\mathbb{M}}^{2}_{i,T}({{\bm{\lambda}}},\hat{{\bf{F}}})-M_{i,2}({\bm{\lambda}},{\bf{F}}_0)|=o_p(1),
\end{equation}
where $M_{i,2}=\frac{1}{T}\sum_{t=1}^Tg^{(2)}_{it}(({\bm{\lambda}}-{\bm{\lambda}}_{0i})^{\prime}{\bf{f}}_{0t}){\bf{f}}_{0t}{\bf{f}}_{0t}^{\prime}$. Note that
\begin{align}\label{LemC14-eq1}
\underset{{\bm{\lambda}}\in \mathcal{A}}{\sup}|{\mathbb{M}}^{2}_{i,T}&({{\bm{\lambda}}},\hat{{\bf{F}}})-M_{i,2}({\bm{\lambda}},{\bf{{\bf{F}}}}_0)|\leq
+\underset{{\bm{\lambda}}\in \mathcal{A}}{\sup}|{\mathbb{M}}^{2}_{i,T}({{\bm{\lambda}}},\hat{{\bf{F}}})-{\mathbb{M}}^{2}_{i,T}({{\bm{\lambda}}},{{\bf{F}}}_0)|\nonumber\\
&\underset{{\bm{\lambda}}\in \mathcal{A}}{\sup}|{\mathbb{M}}^{2}_{i,T}({{\bm{\lambda}}},{{\bf{F}}}_0)-\bar{\mathbb{M}}^{2}_{i,T}({{{\bm{\lambda}}}},{{\bf{F}}}_0)|
+\underset{{\bm{\lambda}}\in \mathcal{A}}{\sup}|\bar{\mathbb{M}}^{2}_{i,T}({{\bm{\lambda}}},{\bf{F}}_0)-M_{i,2}({\bm{\lambda}},{\bf{F}}_0)|.
\end{align}
By Lemma \ref{LemC-13}, the second term on the RHS of \eqref{LemC14-eq1} is $o_p(1)$. For the first term, by Lemma \ref{LemC-2}, Assumptions \ref{Ass7} (i) and \ref{Ass9}, for any ${\bm{\lambda}}\in \mathcal{A}$, it holds that
\begin{equation*}
|{\mathbb{M}}^{2}_{i,T}({{\bm{\lambda}}},\hat{{\bf{F}}})-{\mathbb{M}}^{2}_{i,T}({{\bm{\lambda}}},{{\bf{F}}}_0)|\lesssim h^{-4}\cdot \|\hat{\bf{F}}-{\bf{F}}_0|/\sqrt{T}=O(1/\sqrt{L_{NT}h^{11}})=o_p(1).
\end{equation*}
In addition, Lemma \ref{lem1} (iii) and Assumption \ref{Ass1} (i) imply that for any ${\bm{\lambda}}\in \mathcal{A}$,
\begin{equation*}
\bar{\mathbb{M}}^{2}_{i,T}({{\bm{\lambda}}},{\bf{F}}_0)= \frac{1}{T}\sum_{t=1}^T(g^{(2)}_{it}(({\bm{\lambda}}-{\bm{\lambda}}_{0i})^{\prime}{\bf{f}}_{0t})+\bar{O}(h^2)){\bf{f}}_{0t}{\bf{f}}_{0t}^{\prime}=M_{i,2}({\bm{\lambda}},{\bf{F}}_0)+o_p(1).
\end{equation*}
Therefore, the last two terms on the RHS of \eqref{LemC14-eq1} are also $o_p(1)$, and thus we have \eqref{LemC14-eq5} established. Moreover, Lemma \ref{LemC-3} implies that $\hat{{\bm{\lambda}}}_i\overset{p}{\to}{\bm{\lambda}}_{0i}$, then it follows that $\mathbb{M}^{2}_{i,T}(\hat{{\bm{\lambda}}}_i,\hat{{\bf{F}}})$ converges in probability to $\frac{1}{T}\sum_{t=1}^Tg^{(2)}_{it}(0){\bf{f}}_{0t}{\bf{f}}_{0t}^{\prime}$ as desired.

\bigskip

\noindent \textbf{Proof of Theorem \ref{Thm4}.}	
For any ${\bm{\lambda}}_{i}\in \mathcal{A},{\bf{f}}_t\in \mathcal{F}$, expanding $K_h^{(1)}(X_{it}-{\bm{\lambda}}_{i}{\bf{f}}_t){\bf{f}}_t$ gives
\begin{align*}\label{LemC4-eq1}
&K_h^{(1)}(X_{it}-{\bm{\lambda}}_{i}^{\prime}{\bf{f}}_t){\bf{f}}_t\nonumber\\
=&K_h^{(1)}(X_{it}-{\bm{\lambda}}_{0i}^{\prime}{\bf{f}}_t){\bf{f}}_t-K_h^{(2)}(X_{it}-{\bm{\lambda}}_{0i}^{\prime}{\bf{f}}_t){\bf{f}}_t{\bf{f}}_t^{\prime}({\bm{\lambda}}_{i}-{\bm{\lambda}}_{0i})+\frac{1}{2}K_h^{(3)}(X_{it}-{{\bm{\lambda}}_{i}^*}^{\prime}{\bf{f}}_t){\bf{f}}_t[({\bm{\lambda}}_{i}-{\bm{\lambda}}_{0i})^{\prime}{\bf{f}}_t]^2\nonumber\\
=&K_{h,it}^{(1)}{\bf{f}}_{0t}+K_h^{(1)}(X_{it}-{\bm{\lambda}}_{0i}^{\prime}{\bf{f}}_t)({\bf{f}}_t-{\bf{f}}_{0t})-K_h^{(2)}(X_{it}-{\bm{\lambda}}_{0i}^{\prime}{\bf{f}}_t^{*}){\bf{f}}_{0t}{\bm{\lambda}}_{0i}^{\prime}({\bf{f}}_t-{\bf{f}}_{0t})-K_{h,it}^{(2)}{\bf{f}}_t{\bf{f}}_t^{\prime}({\bm{\lambda}}_{i}-{\bm{\lambda}}_{0i})\nonumber\\
+&K_{h}^{(3)}(X_{it}-{\bm{\lambda}}_{0i}^{\prime}{\bf{f}}_t^{*}){\bf{f}}_t{\bf{f}}_t^{\prime}\cdot ({\bm{\lambda}}_{i}-{\bm{\lambda}}_{0i}){\bm{\lambda}}_{0i}^{\prime}({\bf{f}}_t-{\bf{f}}_{0t})+\frac{1}{2}K_{h}^{(3)}(X_{it}-{{\bm{\lambda}}_{i}^*}^{{\prime}}{\bf{f}}_t){\bf{f}}_t[({\bm{\lambda}}_{i}-{\bm{\lambda}}_{0i})^{\prime}{\bf{f}}_t]^2,
\end{align*}
where ${\bm{\lambda}}_{i}^*$ is between ${\bm{\lambda}}_{i}$ and ${\bm{\lambda}}_{0i}$, ${\bf{f}}_t^{*}$ is between ${\bf{f}}_t$ and ${\bf{f}}_{0t}$. Taking expectations and setting ${\bm{\lambda}}_{i}=\hat{\bm{\lambda}}_i, {\bf{f}}_t=\hat{\bf{f}}_t$, then by Lemma \ref{lem1} (iii), Lemma \ref{LemC-1} (iii), Lemma \ref{LemC-2} and Lemma \ref{LemC-3},
\begin{equation*}
    \begin{split}
&\frac{1}{T}\sum_{t=1}^T\bar{K}_h^{(1)}(X_{it}-\hat{\bm{\lambda}}_i^{\prime}\hat{\bf{f}}_t)\hat{\bf{f}}_t=\frac{1}{T}\sum_{t=1}^T\bar{K}_{h,it}^{(1)}{\bf{f}}_{0t}-\frac{1}{T}\sum_{t=1}^T\bar{K}_h^{(1)}(X_{it}-{\bm{\lambda}}_{0i}^{\prime}\hat{\bf{f}}_t)(\hat{\bf{f}}_t-{\bf{f}}_{0t})\\
-(&\frac{1}{T}\sum_{t=1}^T\bar{K}_{h,it}^{(2)}\hat{\bf{f}}_t\hat{\bf{f}}_t^{\prime})(\hat{\bm{\lambda}}_i-{\bm{\lambda}}_{0i})-\frac{1}{T}\sum_{t=1}^T\bar{K}_{h}^{(2)}(X_{it}-{\bm{\lambda}}_{0i}^{\prime}\hat{\bf{f}}^*_t){\bf{f}}_{0t}(\hat{\bf{f}}_t-{\bf{f}}_{0t})^{\prime}{\bm{\lambda}}_{0i}+
o_p(\|\hat{\bm{\lambda}}_i-{\bm{\lambda}}_{0i}\|).
    \end{split}
\end{equation*}
where 
$\hat{\bf{f}}^*_t$ lies between  $\hat{\bf{f}}_t$ and ${\bf{f}}_{0t}$. Lemma \ref{lem1} (iii) and Lemma \ref{LemC-2} imply that
\begin{equation*}
\frac{1}{T}\sum_{t=1}^T\bar{K}^{(2)}_{h,it}\hat{\bf{f}}_t\hat{\bf{f}}_t^{\prime}=\frac{1}{T}\sum_{t=1}^T\bar{K}^{(2)}_{h,it}{\bf{f}}_{0t}{\bf{f}}_{0t}^{\prime}+o_p(1)={\bm{\Phi}}_{T,i}+o_p(1).
\end{equation*}
Then, it follows from  Lemma \ref{LemC-1} (i) (ii), Lemma \ref{lem1} (iii) and Lemma \ref{LemC-7} that
\begin{equation*}\label{Thm4-eq4}
{\bm{\Phi}}_{T,i}(\hat{\bm{\lambda}}_i-{\bm{\lambda}}_{0i})=\frac{1}{T}\sum_{t=1}^T\bar{K}_h^{(1)}(X_{it}-\hat{\bm{\lambda}}_i^{\prime}\hat{\bf{f}}_t)\hat{\bf{f}}_t+O_p(h^5)
+o_p(\|\hat{\bm{\lambda}}_{i}-{\bm{\lambda}}_{0i}\|).
\end{equation*}

Next, 
since $\frac{1}{T}\sum_{t=1}^T{K}_h^{(1)}(X_{it}-\hat{\bm{\lambda}}_i^{\prime}\hat{\bf{f}}_t)\hat{\bf{f}}_t=0$, then we can write
\begin{align}\label{Thm4-eq7}
 &-\frac{1}{T}\sum_{t=1}^T\bar{K}_h^{(1)}(X_{it}-\hat{\bm{\lambda}}_i^{\prime}\hat{\bf{f}}_t)\hat{\bf{f}}_t=\frac{1}{T}\sum_{t=1}^T\tilde{K}_h^{(1)}(X_{it}-\hat{\bm{\lambda}}_i^{\prime}\hat{\bf{f}}_t)\hat{\bf{f}}_t\nonumber\\
=&\frac{1}{T}\sum_{t=1}^T\tilde{K}_{h,it}^{(1)}{\bf{f}}_{0t}+\frac{1}{T}\sum_{t=1}^T[\tilde{K}_{h}^{(1)}(X_{it}-\hat{\bm{\lambda}}_i^{\prime}\hat{\bf{f}}_t)\hat{\bf{f}}_{t}-\tilde{K}_{h}^{(1)}(X_{it}-\hat{\bm{\lambda}}_i^{\prime}{\bf{f}}_{0t}){\bf{f}}_{0t}]\nonumber\\
&+\frac{1}{T}\sum_{t=1}^T[\tilde{K}_{h}^{(1)}(X_{it}-\hat{\bm{\lambda}}_i^{\prime}{\bf{f}}_{0t})-\tilde{K}_{h,it}^{(1)}]{\bf{f}}_{0t}.
\end{align}
The second term on the RHS of \eqref{Thm4-eq7} can be written as
\begin{equation*}
\begin{split}
&\frac{1}{T}\sum_{t=1}^T[\tilde{K}_{h}^{(1)}(X_{it}-\hat{\bm{\lambda}}_i^{\prime}\hat{\bf{f}}_t)\hat{\bf{f}}_{t}-\tilde{K}_{h}^{(1)}(X_{it}-\hat{\bm{\lambda}}_i^{\prime}{\bf{f}}_{0t}){\bf{f}}_{0t}]\\
=&\frac{1}{T}\sum_{t=1}^T\tilde{K}_{h}^{(1)}(X_{it}-\hat{\bm{\lambda}}_i^{\prime}\hat{\bf{f}}^*_t)(\hat{\bf{f}}_t-{\bf{f}}_{0t})-\frac{1}{T}\sum_{t=1}^T\tilde{K}_{h}^{(2)}(X_{it}-\hat{\bm{\lambda}}_i^{\prime}\hat{\bf{f}}^*_t)\hat{\bf{f}}^*_t\hat{\bm{\lambda}}_i^{\prime}(\hat{\bf{f}}_t-{\bf{f}}_{0t})\\
=&\frac{1}{T}\sum_{t=1}^T\tilde{K}_{h}^{(1)}(X_{it}-\hat{\bm{\lambda}}_i^{\prime}\hat{\bf{f}}^*_t)(\hat{\bf{f}}_t-{\bf{f}}_{0t})-\frac{1}{T}\sum_{t=1}^T\tilde{K}_{h}^{(2)}(X_{it}-\hat{\bm{\lambda}}_i^{\prime}\hat{\bf{f}}^*_t){\bf{f}}_{0t}(\hat{\bf{f}}_t-{\bf{f}}_{0t})^{\prime}(\hat{\bm{\lambda}}_i-{\bm{\lambda}}_{0i})\\
-&\frac{1}{T}\sum_{t=1}^T\tilde{K}_{h}^{(2)}(X_{it}-\hat{\bm{\lambda}}_i^{\prime}\hat{\bf{f}}^*_t){\bf{f}}_{0t}(\hat{\bf{f}}_t-{\bf{f}}_{0t})^{\prime}{\bm{\lambda}}_{0i}-\frac{1}{T}\sum_{t=1}^T\tilde{K}_{h}^{(2)}(X_{it}-\hat{\bm{\lambda}}_i^{\prime}\hat{\bf{f}}^*_t)(\hat{\bf{f}}^*_{t}-{\bf{f}}_{0t})(\hat{\bf{f}}_t-{\bf{f}}_{0t})^{\prime}\hat{\bm{\lambda}}_i,
\end{split}
\end{equation*}
where $\hat{\bf{f}}^*_t$ is between ${\bf{f}}_{0t}$ and $\hat{\bf{f}}_t$. By Lemma \ref{LemC-8}, the first term on the RHS of above equation is $O_p((L_{NT}h^6)^{-1})$, the second term is $o_p(\|\hat{\bm{\lambda}}_i-{\bm{\lambda}}_{0i}\|)$ and the third one is $O_p((L_{NT}h^7)^{-1})$. Moreover, by Lemma \ref{LemC-2} and Assumption \ref{Ass4} (vi), the fourth term is $O_p((L_{NT}h^6)^{-1})$.


Regarding the third term on the RHS of \eqref{Thm4-eq7}, we have
\begin{equation*}
\begin{split}
\frac{1}{T}\sum_{t=1}^T[\tilde{K}_{h}^{(1)}(X_{it}-\hat{\bm{\lambda}}_i^{\prime}{\bf{f}}_{0t})-\tilde{K}_{h,it}^{(1)}]{\bf{f}}_{0t}=\frac{1}{T}\sum_{t=1}^T\tilde{K}_{h}^{(2)}(X_{it}-\hat{\bm{\lambda}}_i^{*^{\prime}}{\bf{f}}_{0t}){\bf{f}}_{0t}{\bf{f}}_{0t}^{\prime}(\hat{\bm{\lambda}}_i-{\bm{\lambda}}_{0i}),
\end{split}
\end{equation*}
where $\hat{\bm{\lambda}}_i^{*}$ is between $\hat{\bm{\lambda}}_i$ and ${\bm{\lambda}}_{0i}$. By Lemma \ref{LemC-13}, the above equation is $o_p(\|(\hat{\bm{\lambda}}_i-{\bm{\lambda}}_{0i})\|)$.

Finally, combining all the above results, we obtain
\begin{equation}\label{Thm4-eq1}
{\bm{\Phi}}_{T,i}(\hat{\bm{\lambda}}_i-{\bm{\lambda}}_{0i})=\frac{1}{T}\sum_{t=1}^T\tilde{K}_{h,it}^{(1)}{\bf{f}}_{0t}+o_p(\|\hat{\bm{\lambda}}_i-{\bm{\lambda}}_{0i}\|)+O_p((L_{NT}h^7)^{-1})+O_p(h^5),
\end{equation}
By Lemma \ref{lem1} (iii) and Assumption \ref{Ass10}, we can show that
\begin{equation}\label{Thm4-eq2}
{\bm{\Phi}}_{T,i}=\frac{1}{T}\sum_{t=1}^T\bar{K}_{h,it}^{(2)}{\bf{f}}_{0t}{\bf{f}}_{0t}^{\prime}=\frac{1}{T}\sum_{t=1}^Tg_{it}^{(2)}(0){\bf{f}}_{0t}{\bf{f}}_{0t}^{\prime}+O(h^2)\rightarrow {\bm{\Phi}}_i<0.
\end{equation}
In addition, it follows from Lemma \ref{LemC-11} that
$\frac{\sqrt{Th^3}}{T}\sum_{t=1}^T\tilde{K}_{h,it}^{(1)}{\bf{f}}_{0t}\overset{d}{\rightarrow} \mathcal{N}(0,
{\bm{\Sigma}}_i).$
Combining this with \eqref{Thm4-eq1}-\eqref{Thm4-eq2} and Assumption \ref{Ass9}, the asymptotic distribution of $\hat{\bm{\lambda}}_i$ can be established. The proof for the asymptotic distribution of $\hat{\bf{f}}_t$ is similar and thus omitted here.

\bigskip


\noindent \textbf{Proof of Lemma Theorem \ref{Thm5}.}	
It suffices to show that $\|\hat{{\bm{\Phi}}}_i-{\bm{\Phi}}_i\|=o_p(1)$ and $\|\hat{{\bm{\Sigma}}}_i-{\bm{\Sigma}}_i\|=o_p(1)$.
Let $\hat{e}_{it}=X_{it}-\hat{\bm{\lambda}}_i^{\prime}\hat{\bf{f}}_t$. First, we consider $\hat{{\bm{\Phi}}}_i-{\bm{\Phi}}_i$. Write
\begin{equation}\label{Thm5-eq1}
\hat{{\bm{\Phi}}}_i
= \frac{1}{T}\sum_{t=1}^TK_{h}^{(2)}(\hat{e}_{it}){\bf{f}}_{0t}{\bf{f}}_{0t}^{\prime}+\frac{1}{T}\sum_{t=1}^TK_h^{(2)}(\hat{e}_{it})(\hat{\bf{f}}_t\hat{\bf{f}}_t^{\prime}-{\bf{f}}_{0t}{\bf{f}}_{0t}^{\prime}).
\end{equation}
For the first term on the RHS of \eqref{Thm5-eq1}, Lemma \ref{LemC-14} implies
\begin{equation}
\frac{1}{T}\sum_{t=1}^TK_{h}^{(2)}(\hat{e}_{it}){\bf{f}}_{0t}{\bf{f}}_{0t}^{\prime}=\frac{1}{T}\sum_{t=1}^Tg_{it}^{(2)}(0){\bf{f}}_{0t}{\bf{f}}_{0t}^{\prime}+o_p(1).
\end{equation}
As for the second term,
by Lemma \ref{LemC-2}, Assumptions \ref{Ass4} (vi) and \ref{Ass9}, it is bounded by
\begin{equation}\label{Thm5-eq3}
\sqrt{\frac{1}{T}\sum_{t=1}^T[K_{h}^{(2)}(\hat{e}_{it})]^2}\cdot\sqrt{\frac{1}{T}\sum_{t=1}^T\|\hat{\bf{f}}_t\hat{\bf{f}}_t^{\prime}-{\bf{f}}_{0t}{\bf{f}}_{0t}^{\prime}\|^2}=O_p(h^{-3}\cdot1/\sqrt{L_{NT}h^3})=o_p(1).
\end{equation}
From\eqref{Thm5-eq1}-\eqref{Thm5-eq3}, we obtain
$\hat{{\bm{\Phi}}}_i=\frac{1}{T}\sum_{t=1}^Tg_{it}^{(2)}(0){\bf{f}}_{0t}{\bf{f}}_{0t}^{\prime}+o_p(1).$
Then, it follows from the definition of ${{\bm{\Phi}}}_i$ 
that $\hat{{\bm{\Phi}}}_i-{\bm{\Phi}}_i=o_p(1)$. Similarly, we can show that $\hat{{\bm{\Sigma}}}_i-{\bm{\Sigma}}_i=o_p(1)$.


Therefore, $\hat{{\bm{\Phi}}}_i\overset{p}{\to}{\bm{\Phi}}_i, \hat{{\bm{\Sigma}}}_i\overset{p}{\to}{\bm{\Sigma}}_i$, and thus $\hat{{\bm{\Phi}}}_i^{-1}\hat{{\bm{\Sigma}}}_i\hat{{\bm{\Phi}}}_i^{-1}\overset{p}{\to}{{\bm{\Phi}}}_i^{-1}{{\bm{\Sigma}}}_i{{\bm{\Phi}}}_i^{-1}$. The proof for $\hat{{\bm{\Psi}}}_t^{-1}\hat{{\bm{\Omega}}}_t\hat{{\bm{\Psi}}}_t^{-1}\overset{p}{\to}{{\bm{\Psi}}}_t^{-1}{{\bm{\Omega}}}_t{{\bm{\Psi}}}_t^{-1}$ is similar, and thus omitted here.

\newpage
\section{Additional Simulation results}\label{S-Sec3}
\vspace{5em}

\renewcommand\arraystretch{0.95}
\begin{table*}[ht]
\centering
\caption{Factor estimation accuracy results for S2.}
\resizebox{1\linewidth}{!}{
\begin{tabular}{c l c c c c c c c c c c }
\toprule
&   & \multicolumn{3}{c}{$T=60$}& \multicolumn{3}{c}{$T=100$} & \multicolumn{3}{c}{$T=200$}\\
                \cmidrule(r){3-5}  \cmidrule(r){6-8} \cmidrule(r){9-11}
                  & &  \multicolumn{3}{c}{$N$} & \multicolumn{3}{c}{$N$} &\multicolumn{3}{c}{$N$}\\
                \cmidrule(r){3-5} \cmidrule(r){6-8} \cmidrule(r){9-11}
              &  & $60$    & $100$   & $200$   & $60$    & $100$   & $200$    & $60$    & $100$   & $200$  \\
                    \midrule
\multirow{5}{*}{(D1)} & $\operatorname{tr}(\hat{\bf{F}}_M) \; c=3$ & 0.963 & 0.980 & 0.990 & 0.966 & 0.981 & 0.991 	& 0.968 & 0.981 & 0.990 \\
                   & $\operatorname{tr}(\hat{\bf{F}}_M) \; c=5$ & 0.965 &	0.980 &	0.991 &	0.966 	& 0.981 &	0.991 &	0.968 &	0.982 	&0.991 \\
                   & $\operatorname{tr}(\hat{\bf{F}}_M) \; c=7$ & 0.963 &	0.979 & 0.990 &	0.964 & 0.981 &	0.990 &	0.966 &	0.980 &	0.990\\
                   & $\operatorname{tr}(\hat{\bf{F}}_P)$ & 0.902& 0.937 &	0.982 &	0.920 &	0.959& 	0.964 &	0.942 &	0.965 &	0.965 \\
                   & $\operatorname{tr}(\hat{\bf{F}}^{0.5}_Q)$ & 0.958 &	0.977 &	0.988 &	0.959 &	0.978 &	0.989 &	0.962 &	0.978 &	0.989  \\
                   \midrule
\multirow{5}{*}{(D2)} & $\operatorname{tr}(\hat{\bf{F}}_M) \; c=3$ & 0.935 & 0.966 & 0.982 & 0.931 & 0.967 & 0.984 & 0.937 & 0.969 & 0.983 \\
                   & $\operatorname{tr}(\hat{\bf{F}}_M) \; c=5$ & 0.944 & 0.969 & 0.983 & 0.949 & 0.972 & 0.986 & 0.955 & 0.974 & 0.987 \\
                   & $\operatorname{tr}(\hat{\bf{F}}_M) \; c=7$ & 0.942 & 0.968 & 0.982 & 0.947 & 0.972 & 0.985 & 0.954 & 0.973 & 0.987 \\
                   & $\operatorname{tr}(\hat{\bf{F}}_P)$ & 0.870 & 0.918 & 0.968 & 0.899 & 0.942 & 0.953 & 0.934 & 0.954 & 0.961 \\
                   & $\operatorname{tr}(\hat{\bf{F}}^{0.5}_Q)$ & 0.922 & 0.962 & 0.979 & 0.935 & 0.965 & 0.982 & 0.943 & 0.966 & 0.983 \\
                   \midrule
\multirow{5}{*}{(D3)} & $\operatorname{tr}(\hat{\bf{F}}_M) \; c=3$ & 0.882 & 0.938 & 0.966 & 0.901 & 0.944 & 0.973 & 0.924 & 0.951 & 0.973 \\
                   & $\operatorname{tr}(\hat{\bf{F}}_M) \; c=5$ & 0.907 & 0.949 & 0.970 & 0.923 & 0.957 & 0.978 & 0.938 & 0.962 & 0.980 \\
                   & $\operatorname{tr}(\hat{\bf{F}}_M) \; c=7$ & 0.906 & 0.948 & 0.969 & 0.923 & 0.958 & 0.977 & 0.938 & 0.963 & 0.981 \\
                   & $\operatorname{tr}(\hat{\bf{F}}_P)$ & 0.805 & 0.869 & 0.937 & 0.850 & 0.917 & 0.936 & 0.911 & 0.935 & 0.953 \\
                   & $\operatorname{tr}(\hat{\bf{F}}^{0.5}_Q)$ & 0.878 & 0.936 & 0.964 & 0.896 & 0.949 & 0.973 & 0.922 & 0.953 & 0.976 \\
                    \bottomrule
\end{tabular}}
\begin{tablenotes}
\item Simulation results over 500 repetitions. The DGP considered in this table:  $X_{it}=\sum_{j=1}^3{\lambda}_{ji}f_{jt}+ e_{it}$, ${\lambda}_{ji},f_{jt}\sim i.i.d \; \mathcal{N}(0,1),e_{it}=\rho e_{it-1}+v_{it}+\sum_{j\neq 0,j=-J}^{J}\beta v_{i-jt}, v_{it}\sim i.i.d\; t_3$.
$(D1):\rho=0.2,\beta=0; (D2):\rho=0,\beta=0.2,J=3; (D3):\rho=0.2,\beta=0.2,J=3.$
\end{tablenotes}
\end{table*}

\begin{table*}[ht]
\centering
\caption{Factor estimation accuracy results for S3.}
\resizebox{1\linewidth}{!}{
\begin{tabular}{c l c c c c c c c c c c }
\toprule
&   & \multicolumn{3}{c}{$T=60$}& \multicolumn{3}{c}{$T=100$} & \multicolumn{3}{c}{$T=200$}\\
               \cmidrule(r){3-5}  \cmidrule(r){6-8} \cmidrule(r){9-11}
                  & &  \multicolumn{3}{c}{$N$} & \multicolumn{3}{c}{$N$} &\multicolumn{3}{c}{$N$}\\
                \cmidrule(r){3-5} \cmidrule(r){6-8} \cmidrule(r){9-11}
              &  & $60$    & $100$   & $200$   & $60$    & $100$   & $200$    & $60$    & $100$   & $200$  \\
                    \midrule
\multirow{5}{*}{$\sigma=2.6$} & $\operatorname{tr}(\hat{\bf{F}}_M) \; c=3$ &  0.940 & 0.978 & 0.989 & 0.956 & 0.965 & 0.991 & 0.962 & 0.981 & 0.991 \\
                   & $\operatorname{tr}(\hat{\bf{F}}_M) \; c=5$ &0.926 & 0.964 & 0.981 & 0.939 & 0.969 & 0.984 & 0.945 & 0.972 & 0.987 \\
                   & $\operatorname{tr}(\hat{\bf{F}}_M) \; c=7$ & 0.892 & 0.947 & 0.969 & 0.909 & 0.955 & 0.975 & 0.923 & 0.963 & 0.979 \\
                   & $\operatorname{tr}(\hat{\bf{F}}_P)$ & 0.918 & 0.954 & 0.977 & 0.923 & 0.955 & 0.978 & 0.927 & 0.955 & 0.977 \\
                   & $\operatorname{tr}(\hat{\bf{F}}_Q^{0.5})$ &  0.951 & 0.977 & 0.988 & 0.956 & 0.977 & 0.989 & 0.960 & 0.978 & 0.989  \\
                   \midrule
\multirow{5}{*}{$\sigma=3$} & $\operatorname{tr}(\hat{\bf{F}}_M) \; c=3$ & 0.947 & 0.977 & 0.989 & 0.956 & 0.980 & 0.991 & 0.959 & 0.981 & 0.992 \\
                   & $\operatorname{tr}(\hat{\bf{F}}_M) \; c=5$ & 0.932 & 0.961 & 0.982 & 0.935 & 0.969 & 0.985 & 0.943 & 0.971 & 0.987  \\
                   & $\operatorname{tr}(\hat{\bf{F}}_M) \; c=7$ & 0.894 & 0.944 & 0.972 & 0.909 & 0.956 & 0.977 & 0.923 & 0.960 & 0.980 \\
                   & $\operatorname{tr}(\hat{\bf{F}}_P)$ & 0.872 & 0.931 & 0.960 & 0.892 & 0.947 & 0.972 & 0.898 & 0.943 & 0.971 \\
                   & $\operatorname{tr}(\hat{\bf{F}}_Q^{0.5})$ & 0.947 & 0.975 & 0.987 & 0.952 & 0.978 & 0.989 & 0.959 & 0.977 & 0.989\\
                   \midrule
\multirow{5}{*}{$\sigma=3.4$} & $\operatorname{tr}(\hat{\bf{F}}_M) \; c=3$ & 0.947 & 0.976 & 0.978 & 0.951 & 0.980 & 0.991 & 0.957 & 0.981 & 0.992 \\
                   & $\operatorname{tr}(\hat{\bf{F}}_M) \; c=5$ & 0.927 & 0.960 & 0.981 & 0.921 & 0.967 & 0.984 & 0.938 & 0.969 & 0.986  \\
                   & $\operatorname{tr}(\hat{\bf{F}}_M) \; c=7$ & 0.889 & 0.945 & 0.971 & 0.909 & 0.954 & 0.976 & 0.919 & 0.958 & 0.980 \\
                   & $\operatorname{tr}(\hat{\bf{F}}_P)$ & 0.846 & 0.907 & 0.951 & 0.871 & 0.932 & 0.964 & 0.879 & 0.932 & 0.965\\
                   & $\operatorname{tr}(\hat{\bf{F}}^{0.5}_Q)$ & 0.943 & 0.974 & 0.987 & 0.951 & 0.976 & 0.988 & 0.956 & 0.977 & 0.989 \\
                    \bottomrule
\end{tabular}}
\begin{tablenotes}
\item Simulation results over 500 repetitions. The DGP considered in this tables: $X_{it}=\sum_{j=1}^3{\lambda}_{ji}f_{jt}+ e_{it}$, ${\lambda}_{ji},f_{jt}\sim i.i.d \; \mathcal{N}(0,1),e_{it}\sim i.i.d \; 0.5 \mathcal{N}(0.8,0.6^2)+0.5 \mathcal{N}(-0.8,\sigma^2)$, for $\sigma=2.6,3,3.4$.
\end{tablenotes}
\end{table*}

\begin{table*}[t]
	\caption{Factor number estimation results for S2.}
	\centering
		\resizebox{1\linewidth}{!}{
			\begin{tabular}{l c c c c c c c c c c c c c c}
				\toprule
				&  &    &  \multicolumn{4}{c}{(D1)} &  \multicolumn{4}{c}{(D2)} & \multicolumn{4}{c}{(D3)}  \\
				\cmidrule(r){4-7}  \cmidrule(r){8-11} \cmidrule(r){12-15}
				\specialrule{0em}{1pt}{1pt}
				 & $T$ & $N$   & $\bar{\hat{r}}_{\mathrm{rank}}$ & $\mathit{Freq}_1$  & $\bar{\hat{r}}_{\mathrm{IC}}$ & $\mathit{Freq}_2$  & $\bar{\hat{r}}_{\mathrm{rank}}$ & $\mathit{Freq}_1$ & $\bar{\hat{r}}_{\mathrm{IC}}$ & $\mathit{Freq}_2$ & $\bar{\hat{r}}_{\mathrm{rank}}$ & $\mathit{Freq}_1$&  $\bar{\hat{r}}_{\mathrm{IC}}$ & $\mathit{Freq}_2$  \\
				\midrule
	\multirow{9}{*}{$c=3$} & 60  & 60  & 2.55 & 0.58 & 2.94 & 0.92 & 2.61 & 0.65 & 2.86 & 0.83 & 2.68 & 0.52 & 2.71 & 0.64 \\
                     & 60  & 100 & 2.55 & 0.55 & 2.96 & 0.96 & 2.81 & 0.81 & 2.94 & 0.94 & 2.90 & 0.77 & 2.89 & 0.85 \\
                     & 60  & 200 & 2.81 & 0.81 & 2.98 & 0.98 & 2.84 & 0.84 & 2.97 & 0.97 & 2.90 & 0.84 & 2.94 & 0.91 \\
                     & 100 & 60  & 2.81 & 0.81 & 2.94 & 0.94 & 2.90 & 0.90 & 2.95 & 0.95 & 2.87 & 0.87 & 2.96 & 0.94 \\
                     & 100 & 100 & 2.94 & 0.94 & 3.00 & 1.00 & 3.00 & 1.00 & 3.00 & 1.00 & 3.00 & 1.00 & 3.00 & 1.00 \\
                     & 100 & 200 & 3.00 & 1.00 & 3.00 & 1.00 & 3.00 & 1.00 & 3.00 & 1.00 & 3.00 & 1.00 & 3.00 & 1.00 \\
                     & 200 & 60  & 2.81 & 0.81 & 3.00 & 1.00 & 2.87 & 0.87 & 2.99 & 0.99 & 2.94 & 0.94 & 2.98 & 0.98 \\
                     & 200 & 100 & 3.00 & 1.00 & 3.00 & 1.00 & 2.94 & 0.94 & 3.00 & 1.00 & 2.94 & 0.94 & 3.00 & 1.00 \\
                     & 200 & 200 & 3.00 & 1.00 & 3.00 & 1.00 & 3.00 & 1.00 & 3.00 & 1.00 & 3.00 & 1.00 & 3.00 & 1.00 \\
                     \midrule
\multirow{9}{*}{$c=5$} & 60  & 60  & 2.58 & 0.61 & 2.88 & 0.89 & 2.66 & 0.70 & 2.77 & 0.80 & 2.55 & 0.58 & 2.63 & 0.59 \\
                     & 60  & 100 & 2.61 & 0.61 & 2.95 & 0.95 & 2.78 & 0.80 & 2.91 & 0.92 & 2.74 & 0.74 & 2.80 & 0.77 \\
                     & 60  & 200 & 2.81 & 0.81 & 2.97 & 0.97 & 2.85 & 0.85 & 2.89 & 0.89 & 2.87 & 0.87 & 2.81 & 0.81 \\
                     & 100 & 60  & 2.78 & 0.78 & 2.92 & 0.92 & 2.87 & 0.87 & 2.93 & 0.93 & 2.87 & 0.87 & 2.88 & 0.88 \\
                     & 100 & 100 & 2.94 & 0.94 & 3.00 & 1.00 & 3.00 & 1.00 & 3.00 & 1.00 & 2.94 & 0.94 & 3.00 & 1.00 \\
                     & 100 & 200 & 2.94 & 0.94 & 3.00 & 1.00 & 3.00 & 1.00 & 3.00 & 1.00 & 3.00 & 1.00 & 3.00 & 1.00 \\
                     & 200 & 60  & 2.90 & 0.90 & 2.97 & 0.97 & 2.87 & 0.87 & 2.98 & 0.98 & 2.94 & 0.94 & 2.97 & 0.97 \\
                     & 200 & 100 & 3.00 & 1.00 & 3.00 & 1.00 & 2.96 & 0.96 & 3.00 & 1.00 & 3.00 & 1.00 & 3.00 & 1.00 \\
                     & 200 & 200 & 3.00 & 1.00 & 3.00 & 1.00 & 3.00 & 1.00 & 3.00 & 1.00 & 3.00 & 1.00 & 3.00 & 1.00 \\
                     \midrule
\multirow{9}{*}{$c=7$} & 60  & 60  & 2.60 & 0.58 & 2.80 & 0.84 & 2.68 & 0.71 & 2.77 & 0.79 & 2.58 & 0.55 & 2.51 & 0.55 \\
                     & 60  & 100 & 2.61 & 0.61 & 2.84 & 0.84 & 2.81 & 0.81 & 2.84 & 0.84 & 2.74 & 0.74 & 2.72 & 0.73 \\
                     & 60  & 200 & 2.83 & 0.83 & 2.94 & 0.94 & 2.87 & 0.87 & 2.82 & 0.82 & 2.87 & 0.87 & 2.75 & 0.76 \\
                     & 100 & 60  & 2.77 & 0.77 & 2.84 & 0.88 & 2.87 & 0.87 & 2.85 & 0.87 & 2.87 & 0.87 & 2.80 & 0.82 \\
                     & 100 & 100 & 2.94 & 0.94 & 3.00 & 1.00 & 3.00 & 1.00 & 3.00 & 1.00 & 2.94 & 0.94 & 3.00 & 1.00 \\
                     & 100 & 200 & 2.94 & 0.94 & 3.00 & 1.00 & 3.00 & 1.00 & 3.00 & 1.00 & 3.00 & 1.00 & 3.00 & 1.00 \\
                     & 200 & 60  & 2.90 & 0.90 & 2.92 & 0.92 & 2.87 & 0.87 & 2.88 & 0.88 & 2.90 & 0.90 & 2.86 & 0.86 \\
                     & 200 & 100 & 3.00 & 1.00 & 3.00 & 1.00 & 2.94 & 0.94 & 3.00 & 1.00 & 3.00 & 1.00 & 3.00 & 1.00 \\
                     & 200 & 200 & 3.00 & 1.00 & 3.00 & 1.00 & 3.00 & 1.00 & 3.00 & 1.00 & 3.00 & 1.00 & 3.00 & 1.00 \\
				\bottomrule
		\end{tabular}}		
 \begin{tablenotes}
\item Simulation results over 500 repetitions. The DGP considered in this tables: $X_{it}=\sum_{j=1}^3{\lambda}_{ji}f_{jt}+ e_{it}$, ${\lambda}_{ji},f_{jt}\sim i.i.d \; \mathcal{N}(0,1),e_{it}=\rho e_{it-1}+v_{it}+\sum_{j\neq 0,j=-J}^{J}\beta v_{i-jt}, v_{it}\sim i.i.d\; t_3$. $(D1):\rho=0.2,\beta=0; (D2):\rho=0,\beta=0.2,J=3; (D3):\rho=0.2,\beta=0.2,J=3.$   
 \end{tablenotes}
\end{table*}

\begin{table}[ht]
	\caption{Factor number estimation results for S3.}
	\centering
		\resizebox{1\linewidth}{!}{
			\begin{tabular}{l c c c c c c c c c c c c c c}
				\toprule 
				&  &   &  \multicolumn{4}{c}{$\sigma=2.6$} &  \multicolumn{4}{c}{$\sigma=3$} & \multicolumn{4}{c}{$\sigma=3.4$}  \\
				\cmidrule(r){4-7}  \cmidrule(r){8-11} \cmidrule(r){12-15}
				 & $T$ & $N$   & $\bar{\hat{r}}_{\mathrm{rank}}$ & $\mathit{Freq}_1$  & $\bar{\hat{r}}_{\mathrm{IC}}$ & $\mathit{Freq}_2$  & $\bar{\hat{r}}_{\mathrm{rank}}$ & $\mathit{Freq}_1$ & $\bar{\hat{r}}_{\mathrm{IC}}$ & $\mathit{Freq}_2$ & $\bar{\hat{r}}_{\mathrm{rank}}$ & $\mathit{Freq}_1$&  $\bar{\hat{r}}_{\mathrm{IC}}$ & $\mathit{Freq}_2$  \\
				\midrule
				\multirow{9}{*}{$c=3$} & 60  & 60  & 2.84 & 0.58 & 2.88 & 0.84 & 2.87 & 0.58 & 2.81 & 0.77 & 2.90 & 0.47 & 2.96 & 0.90 \\
                     & 60  & 100 & 2.61 & 0.61 & 3.02 & 0.97 & 2.84 & 0.68 & 2.97 & 0.97 & 2.88 & 0.42 & 2.93 & 0.90 \\
                     & 60  & 200 & 2.94 & 0.87 & 3.00 & 1.00 & 2.71 & 0.68 & 2.97 & 0.97 & 3.10 & 0.45 & 2.97 & 0.97 \\
                     & 100 & 60  & 2.68 & 0.68 & 2.98 & 0.97 & 2.84 & 0.65 & 3.00 & 1.00 & 2.94 & 0.51 & 3.00 & 1.00 \\
                     & 100 & 100 & 2.95 & 0.95 & 3.00 & 1.00 & 3.10 & 0.71 & 3.00 & 1.00 & 3.29 & 0.67 & 3.00 & 1.00 \\
                     & 100 & 200 & 3.00 & 1.00 & 3.00 & 1.00 & 3.05 & 0.90 & 3.00 & 1.00 & 3.05 & 0.86 & 3.00 & 1.00 \\
                     & 200 & 60  & 2.90 & 0.84 & 3.00 & 1.00 & 3.03 & 0.68 & 3.00 & 1.00 & 3.12 & 0.77 & 3.00 & 1.00 \\
                     & 200 & 100 & 3.05 & 0.95 & 3.00 & 1.00 & 2.90 & 0.90 & 3.00 & 1.00 & 3.18 & 0.72 & 3.00 & 1.00 \\
                     & 200 & 200 & 3.00 & 1.00 & 3.00 & 1.00 & 3.00 & 1.00 & 3.00 & 1.00 & 3.05 & 0.95 & 3.00 & 1.00 \\
                     \midrule
\multirow{9}{*}{$c=5$} & 60  & 60  & 2.61 & 0.55 & 2.76 & 0.74 & 2.65 & 0.65 & 2.71 & 0.65 & 2.71 & 0.52 & 2.58 & 0.55 \\
                     & 60  & 100 & 2.68 & 0.68 & 2.81 & 0.81 & 2.55 & 0.55 & 2.87 & 0.84 & 2.58 & 0.65 & 2.90 & 0.87 \\
                     & 60  & 200 & 2.90 & 0.90 & 2.97 & 0.97 & 2.74 & 0.74 & 2.90 & 0.90 & 2.81 & 0.84 & 2.94 & 0.94 \\
                     & 100 & 60  & 2.77 & 0.77 & 2.87 & 0.87 & 2.65 & 0.68 & 2.90 & 0.90 & 2.81 & 0.68 & 2.88 & 0.87 \\
                     & 100 & 100 & 2.86 & 0.86 & 3.00 & 1.00 & 2.90 & 0.90 & 3.00 & 1.00 & 2.90 & 0.81 & 3.00 & 1.00 \\
                     & 100 & 200 & 3.00 & 1.00 & 3.00 & 1.00 & 2.95 & 0.95 & 3.00 & 1.00 & 2.95 & 0.95 & 3.00 & 1.00 \\
                     & 200 & 60  & 2.84 & 0.84 & 2.97 & 0.97 & 2.83 & 0.83 & 3.00 & 1.00 & 2.94 & 0.87 & 2.97 & 0.97 \\
                     & 200 & 100 & 3.00 & 1.00 & 3.00 & 1.00 & 2.90 & 0.90 & 3.00 & 1.00 & 3.00 & 1.00 & 3.00 & 1.00 \\
                     & 200 & 200 & 3.00 & 1.00 & 3.00 & 1.00 & 3.00 & 1.00 & 3.00 & 1.00 & 3.00 & 1.00 & 3.00 & 1.00 \\
                     \midrule
\multirow{9}{*}{$c=7$} & 60  & 60  & 2.71 & 0.58 & 2.66 & 0.68 & 2.74 & 0.68 & 2.62 & 0.65 & 2.77 & 0.74 & 2.46 & 0.48 \\
                     & 60  & 100 & 2.71 & 0.71 & 2.79 & 0.81 & 2.48 & 0.48 & 2.64 & 0.68 & 2.58 & 0.65 & 2.69 & 0.71 \\
                     & 60  & 200 & 2.90 & 0.90 & 2.93 & 0.94 & 2.68 & 0.68 & 2.87 & 0.87 & 2.81 & 0.84 & 2.84 & 0.84 \\
                     & 100 & 60  & 2.74 & 0.71 & 2.78 & 0.81 & 2.74 & 0.68 & 2.75 & 0.77 & 2.71 & 0.58 & 2.66 & 0.71 \\
                     & 100 & 100 & 2.95 & 0.95 & 3.00 & 1.00 & 3.00 & 0.90 & 3.00 & 1.00 & 2.86 & 0.86 & 3.00 & 1.00 \\
                     & 100 & 200 & 3.00 & 1.00 & 3.00 & 1.00 & 2.95 & 0.95 & 3.00 & 1.00 & 2.95 & 0.95 & 3.00 & 1.00 \\
                     & 200 & 60  & 2.84 & 0.84 & 2.90 & 0.90 & 2.81 & 0.81 & 2.94 & 0.94 & 2.90 & 0.90 & 2.97 & 0.97 \\
                     & 200 & 100 & 3.00 & 1.00 & 3.00 & 1.00 & 2.94 & 0.94 & 3.00 & 1.00 & 3.00 & 1.00 & 3.00 & 1.00 \\
                     & 200 & 200 & 3.00 & 1.00 & 3.00 & 1.00 & 3.00 & 1.00 & 3.00 & 1.00 & 3.00 & 1.00 & 3.00 & 1.00\\
				\bottomrule
\end{tabular}}		
\begin{tablenotes}
 \item Simulation results over 500 repetitions. The DGP considered in this tables: $X_{it}=\sum_{j=1}^3{\lambda}_{ji}f_{jt}+ e_{it}$, ${\lambda}_{ji},f_{jt}\sim i.i.d \; \mathcal{N}(0,1),e_{it}\sim i.i.d \; 0.5 \mathcal{N}(0.8,0.6^2)+0.5 \mathcal{N}(-0.8,\sigma^2)$, for $\sigma=2.6, 3, 3.4$.   
\end{tablenotes}
\end{table}

\begin{figure}[t]
\begin{minipage}[t]{0.3\linewidth}
        \centering
		\subfigure[$N=T=60,c=3$] {\includegraphics[width=1\textwidth]{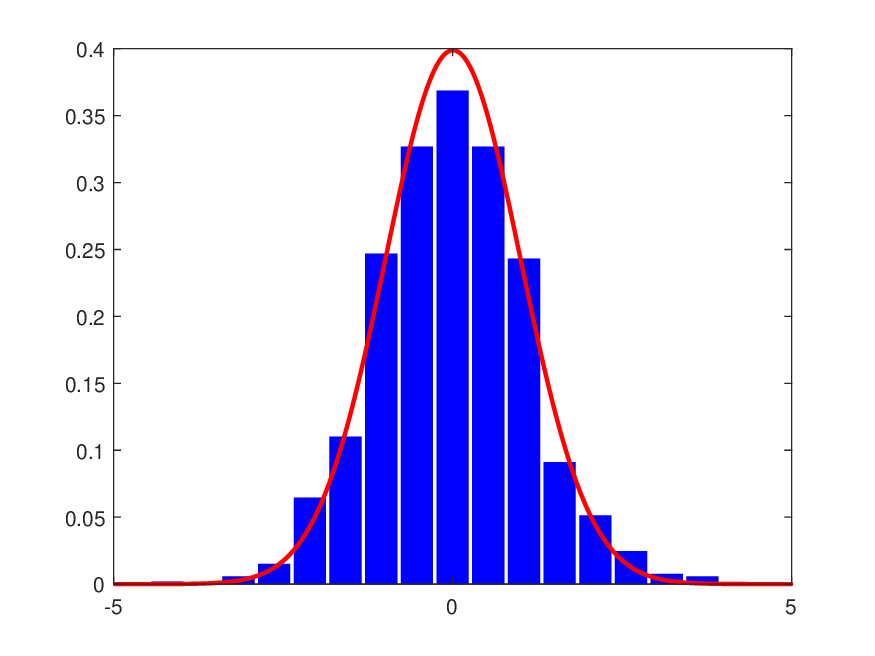}}
\vfill
\subfigure[$N=T=100,c=3$]{\includegraphics[width=1\textwidth]{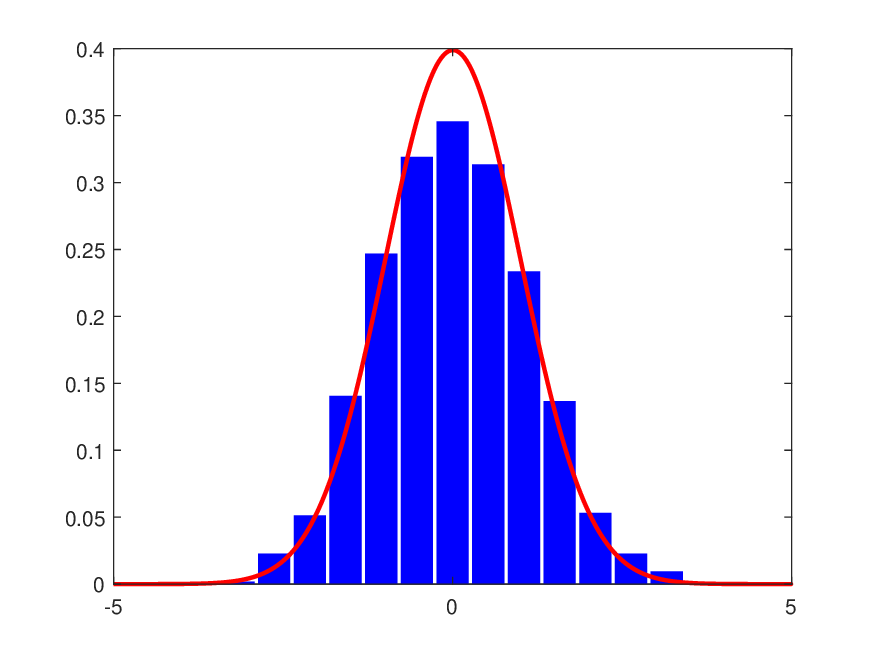}}
\vfill
\subfigure[$N=T=200,c=3$]{\includegraphics[width=1\textwidth]{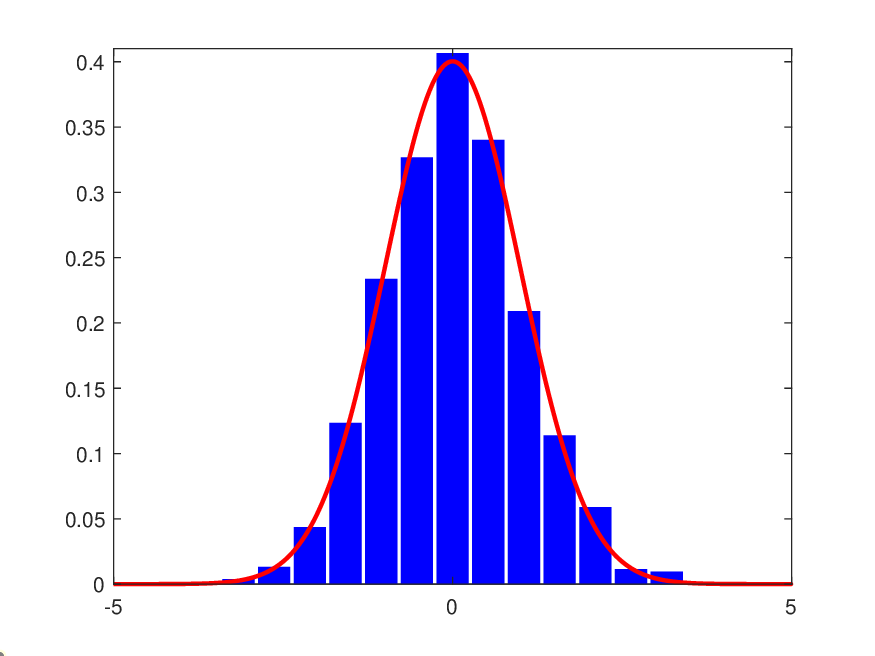}}
\end{minipage}
    \hfill
    \begin{minipage}[t]{0.3\linewidth}
    \centering
		\subfigure[$N=T=60,c=5$] {\includegraphics[width=1\textwidth]{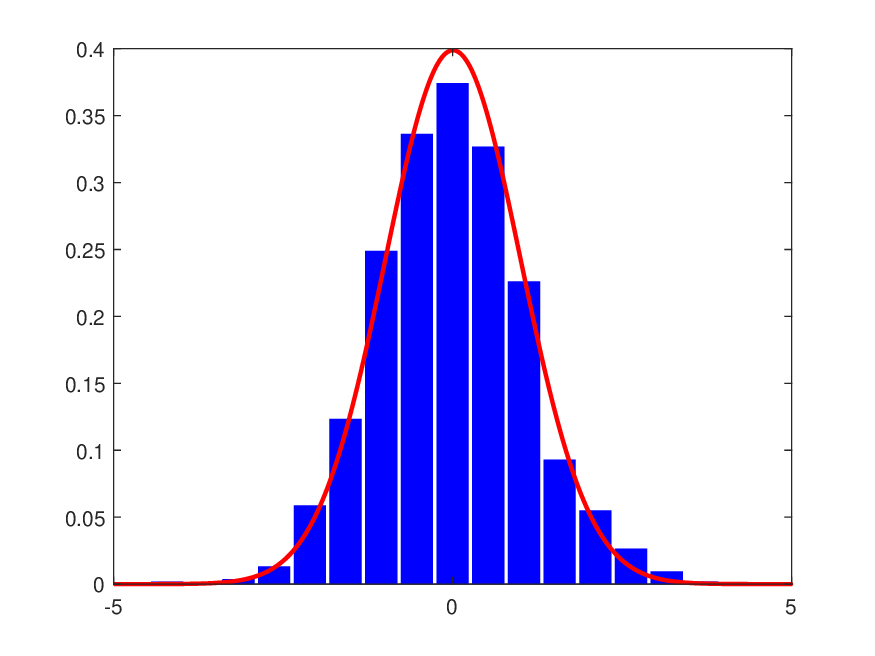}}
\vfill
\subfigure[$N=T=100,c=5$]{\includegraphics[width=1\textwidth]{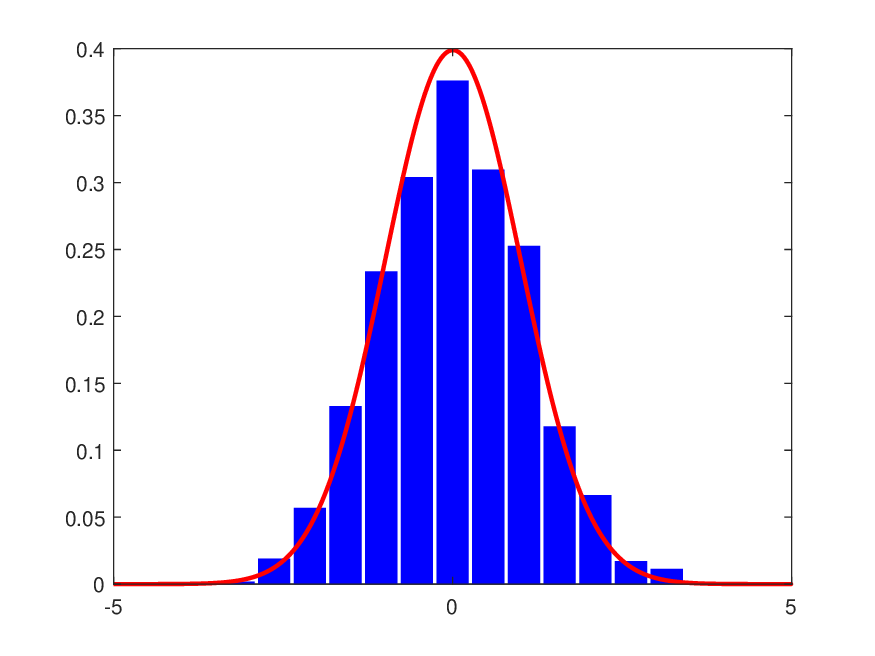}}
\vfill
\subfigure[$N=T=200,c=5$]{\includegraphics[width=1\textwidth]{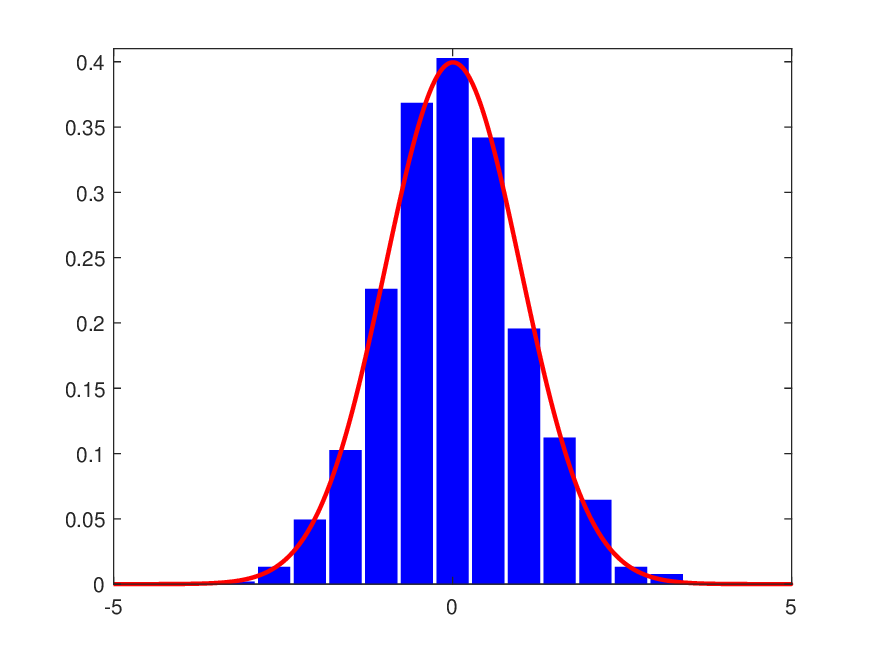}}
\end{minipage}
\hfill
    \begin{minipage}[t]{0.3\linewidth}
    \centering
		\subfigure[$N=T=60,c=7$] {\includegraphics[width=1\textwidth]{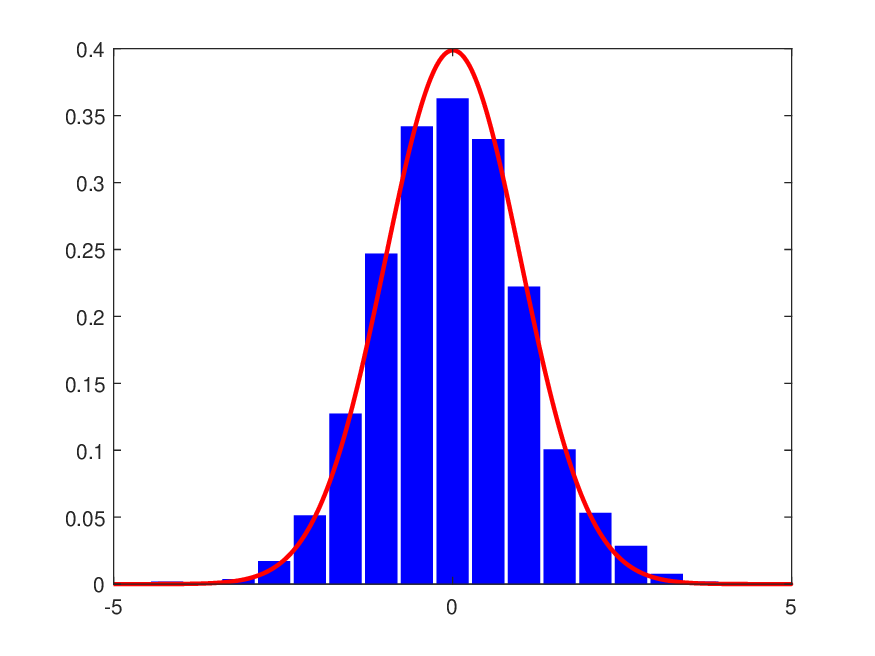}}
\vfill
\subfigure[$N=T=100,c=7$]{\includegraphics[width=1\textwidth]{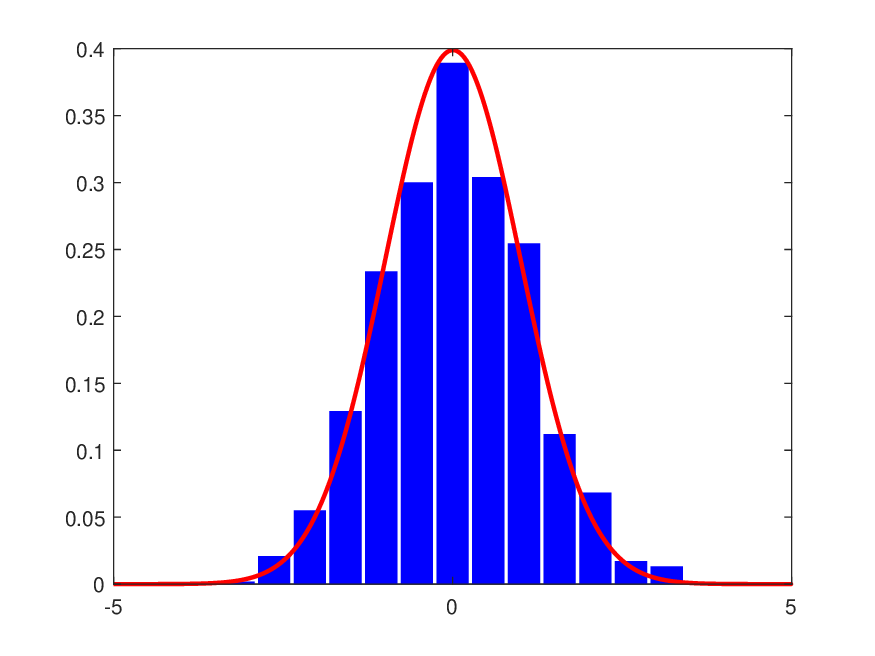}}
\vfill
\subfigure[$N=T=200,c=7$]{\includegraphics[width=1\textwidth]{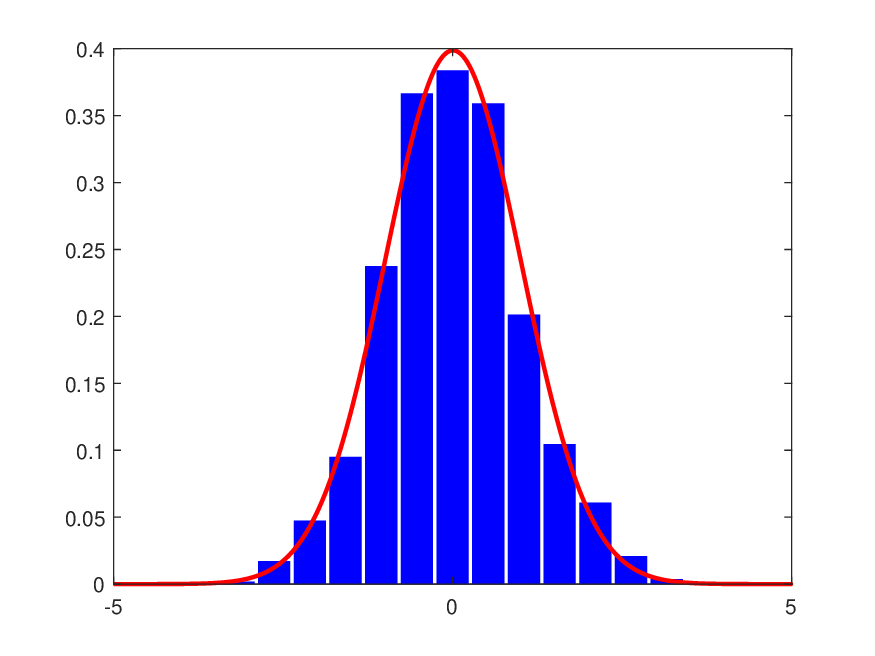}}
\end{minipage}
\caption{Normal approximations of the estimated MFA factors. Notes: Simulation results over 1000 repetitions. The DGP considered in this figure: $X_{it}=\lambda_if_t+e_{it}$, $\lambda_i, f_t, e_{it} \sim i.i.d \; \mathcal{N}(0,1)$. For each subfigure, the red curve plots the standard normal density, and the blue bars plot the histogram for the estimates $\sqrt{Nh^3}[\hat{\bm{\Psi}}_t^{-1}\hat{\bm{\Omega}}_t\hat{\bm{\Psi}}^{-1}_t]^{-1/2}(\hat{f}_t-f_{0t})$, where $t=T/2$. The estimates are obtained with the bandwidth $h=c\cdot L_{NT}^{-1/12}$, for $c=3$ (first column), $c=5$ (second column), $c=7$ (third column).}
\end{figure}

\begin{figure}[ht]
\begin{minipage}[t]{0.3\linewidth}
        \centering
		\subfigure[$N=T=60,c=3$] {\includegraphics[width=1\textwidth]{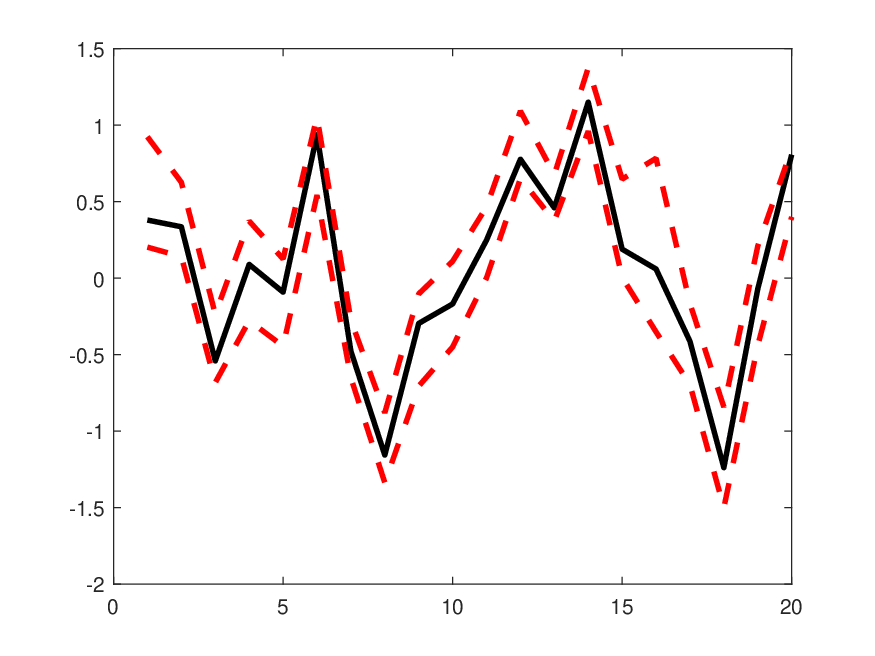}}
\vfill
\subfigure[$N=T=100,c=3$]{\includegraphics[width=1\textwidth]{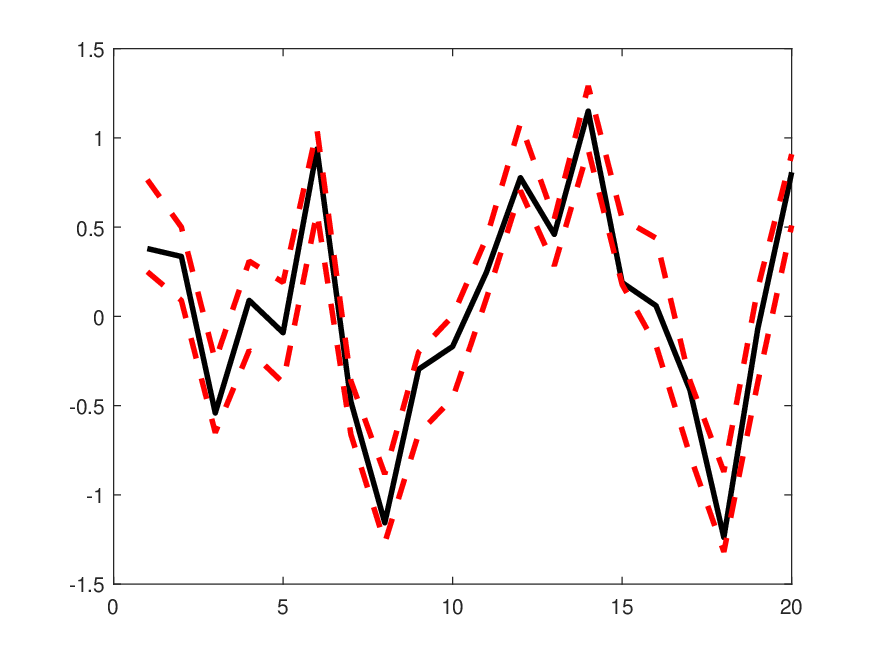}}
\vfill
\subfigure[$N=T=200,c=3$]{\includegraphics[width=1\textwidth]{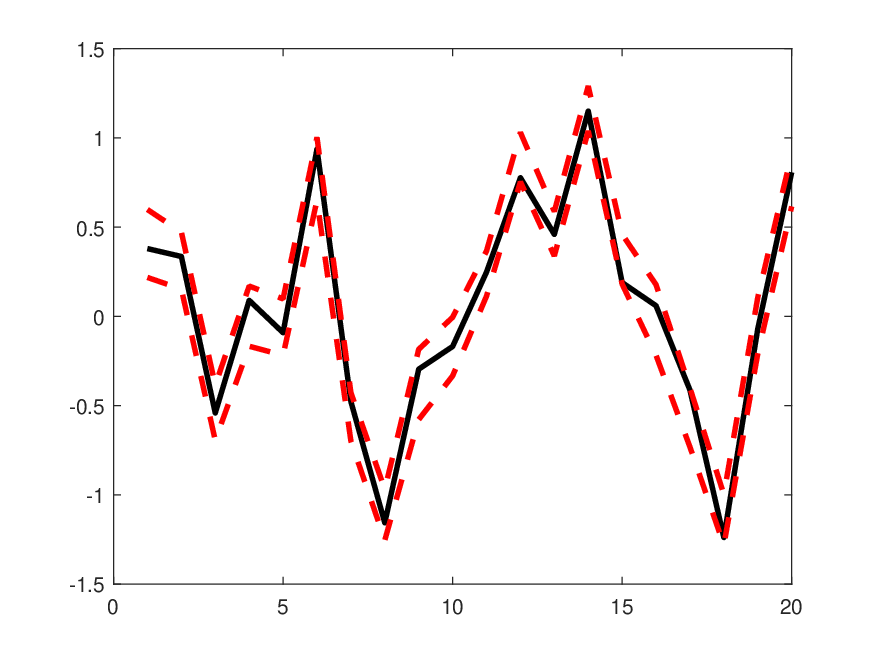}}
\end{minipage}
    \hfill
    \begin{minipage}[t]{0.3\linewidth}
    \centering
		\subfigure[$N=T=60,c=5$] {\includegraphics[width=0.98\textwidth]{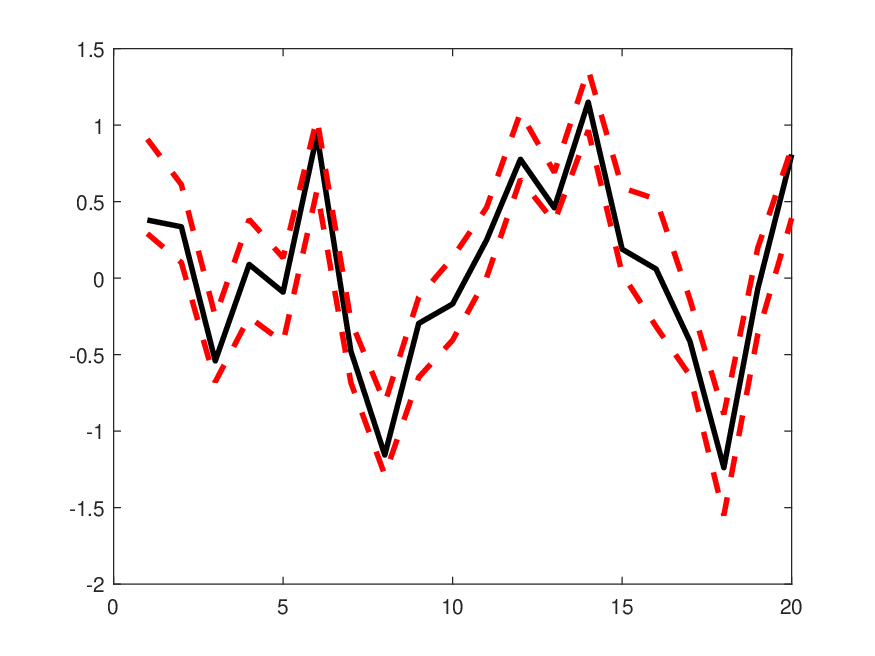}}
\vfill
\subfigure[$N=T=100,c=5$]{\includegraphics[width=1\textwidth]{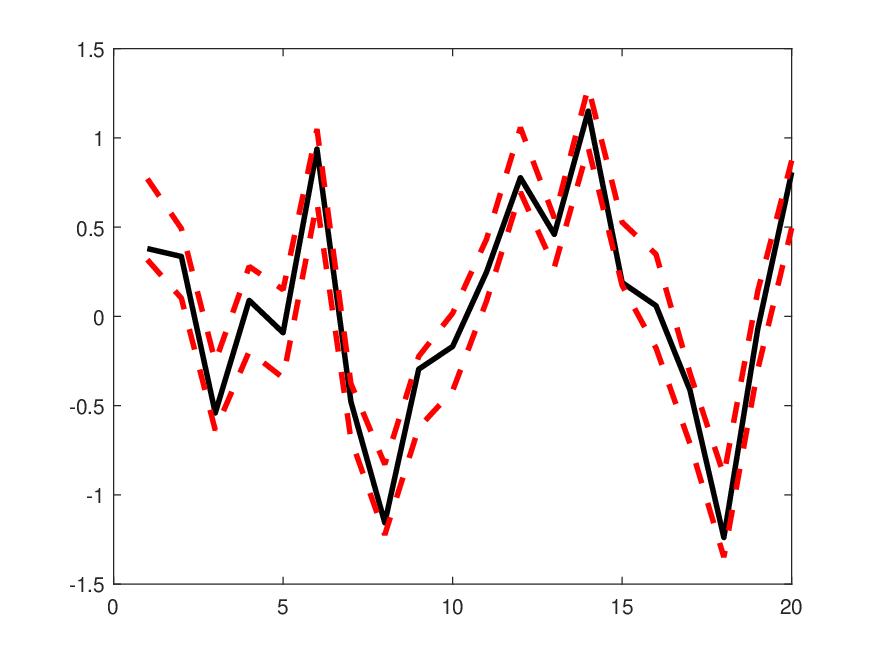}}
\vfill
\subfigure[$N=T=200,c=5$]{\includegraphics[width=1\textwidth]{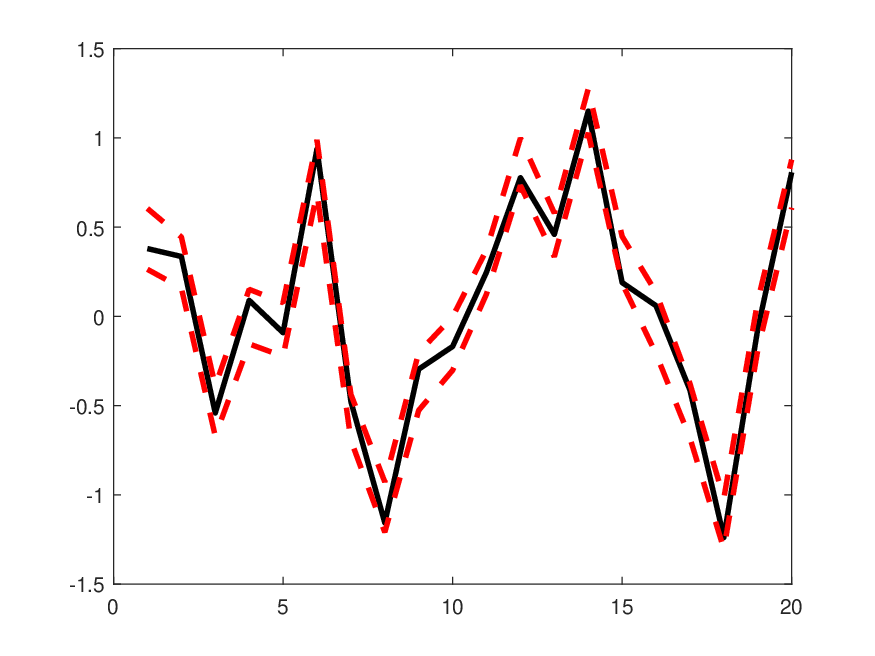}}
\end{minipage}
\hfill
    \begin{minipage}[t]{0.3\linewidth}
    \centering
		\subfigure[$N=T=60,c=7$] {\includegraphics[width=1\textwidth]{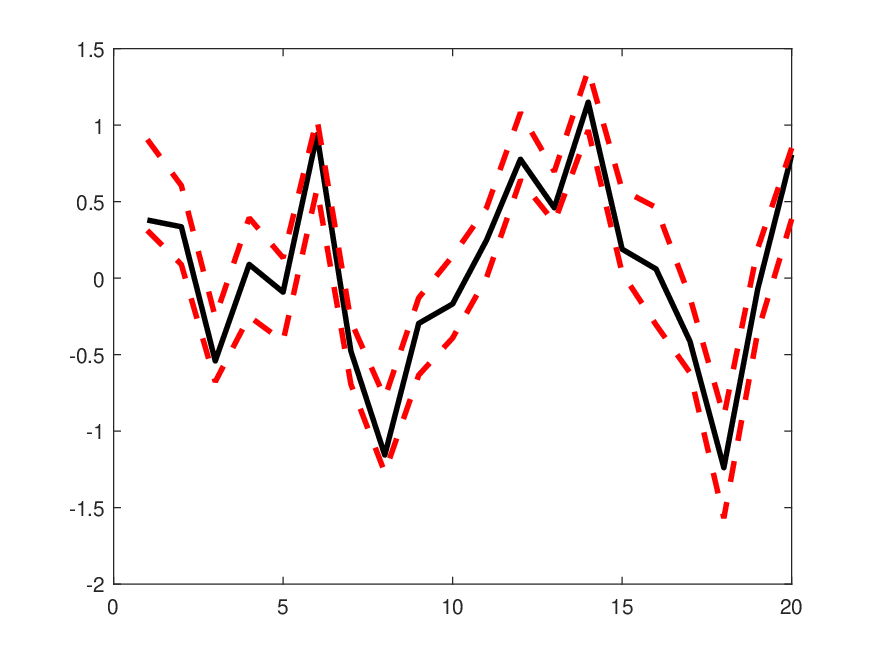}}
\vfill
\subfigure[$N=T=100,c=7$]{\includegraphics[width=1\textwidth]{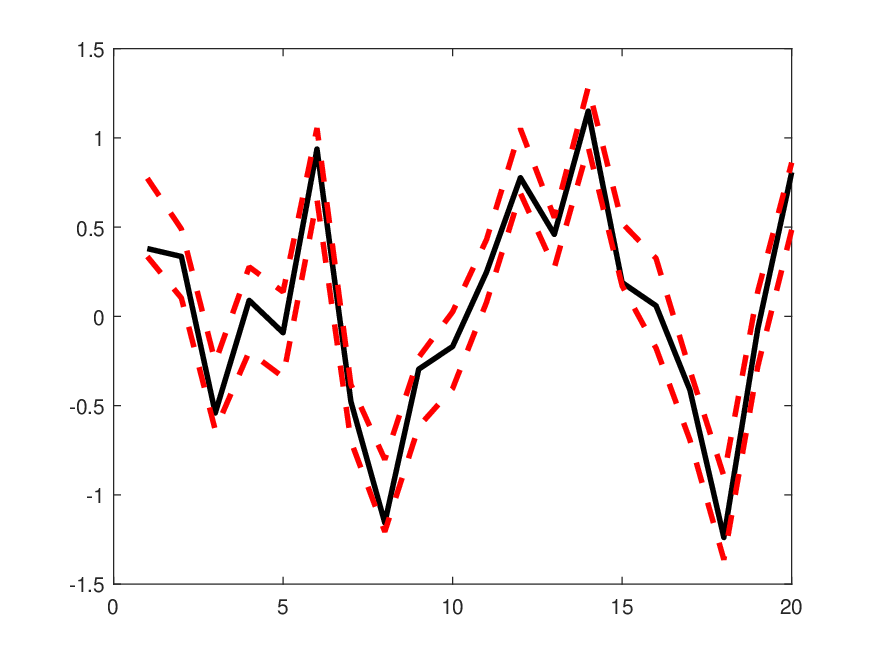}}
\vfill
\subfigure[$N=T=200,c=7$]{\includegraphics[width=1\textwidth]{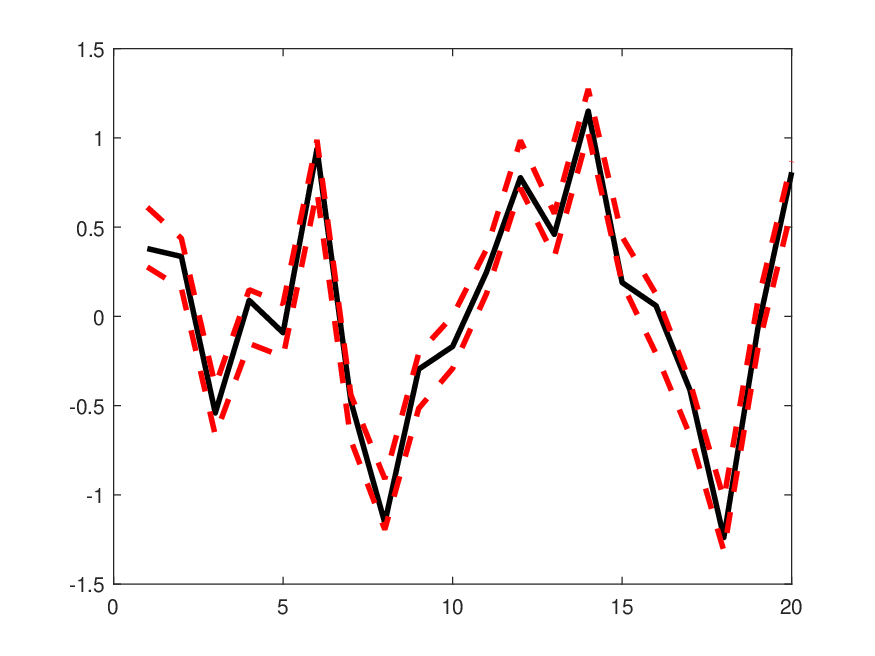}}
\end{minipage}
\caption{$95\%$ Confidence intervals for the ture factor process.
The DGP considered in this figure: $X_{it}=\lambda_if_t+e_{it}$, $\lambda_i, f_t, e_{it} \sim i.i.d \; \mathcal{N}(0,1)$. For each subfigure,
the black sold curve in the middle plots the true factor process $\{f_{0t},t=1,2,\cdots,20\}$, and two red dashed curves plot the estimated $95\%$ confidence intervals for $\{f_{0t}, t=1,2,\cdots,20\}$. The estimates are obtained with the bandwidth $h=c\cdot L_{NT}^{-1/12}$, for $c=3$ (first column), $c=5$ (second column), $c=7$ (third column).}
\end{figure}

\bibliographystyle{apalike}
\bibliography{Ref_MFM20240927}

\end{document}